\documentclass[12pt]{My_preprint}
\title{
Inertial effects on the interphase drag force and rheology of dilute suspensions of buoyant droplets at low Reynolds number
    }

\author[1]{Nicolas Fintzi}\author[1]{Jean-Lou Pierson}
\affil[1]{IFP Energies Nouvelles, Rond-point de l'echangeur de Solaize, 69360 Solaize, France}
\normalmarginpar

\begin{document}

\maketitle

\begin{abstract}
    In this work, we compute the hydrodynamic force and the first and second moments of force acting on a translating spherical droplet immersed in a uniform flow using the reciprocal theorem. 
    We consider the low but finite Reynolds number regime, $Re = a U \rho_f / \mu_f$, and the dilute limit of small droplet volume fraction $\phi$. 
    Here, $U$ denotes the magnitude of the relative velocity between the phases, $a$ the droplet radius, and $\rho_f$ and $\mu_f$ the density and viscosity of the continuous phase, respectively.
    We show that the $O(Re)$ inertial corrections to the first and second moments of force scale as $O(\rho_f \phi U^2)$ and $O(a\rho_f\phi U^2)$, respectively.
    Equivalently, in dimensionless form, these corrections scale as $O(\phi Re)$ and $O(\frac{a}{R}\phi Re)$, where $R$ denotes the macroscopic length scale. 
    Moreover, the ensemble average of the drag force and the higher-order force moments over the distribution of droplet velocities introduces additional contributions proportional to the velocity variance of the dispersed phase, both in the interphase momentum exchange and in the effective stress of the continuous phase.
    As a consequence, in dilute emulsions of buoyant droplets, the effective stress depends quadratically on the relative velocity between the phases, on the velocity variance of the dispersed phase, and on the spatial gradients of these quantities. 
    Finally, we consider a steady-state pipe flow to examine how the force moments derived in this study affect the behaviour of the two-phase flow in this particular configuration. 
\end{abstract}
 
\section{Introduction}

Dispersed two-phase flows composed of droplets or bubbles are ubiquitous in processes developed at IFP Energies Nouvelles. 
Separation processes such as flotation rely on the injection of small bubbles to recover contaminants or dispersed droplets \citep{nguyen2003colloidal}. 
Dispersed phases are also widely used to enhance mass transfer, for instance in bubble columns \citep{kantarci2005} and liquid-liquid extractors \citep{weatherley2020}, which can take various forms, including mixer-settlers and agitated columns. 
In all these processes, the flow is both turbulent and buoyancy-driven.
Buoyancy plays an important role in separating the dispersed phase from the continuous phase. 
When buoyancy effects are too small, operational issues may arise, such as flooding in liquid-liquid extraction columns or the need for excessively large tanks in flotation systems. 
It is therefore essential to accurately predict the effects of relative motion between the phases in order to better design and optimize these processes.

At the process scale (typically on the order of meters), engineers cannot rely on direct numerical simulations, as their computational cost far exceeds the capabilities of current supercomputers and is incompatible with engineering timescales. 
Instead, modeling efforts rely on averaged equations of motion, which have proven accurate in predicting many two-phase flow phenomena. 
Even the simplest two-fluid models provide valuable insights into various aspects of two-phase flows, including the prediction of flooding in liquid-liquid extraction columns \citep{wallis2020}. 
Since the work of \citet{wallis2020} and references therein, numerous refinements of two-fluid models have been proposed \citep{buyevich1979flow,drew1983mathematical,lhuillier1992ensemble,zhang1994ensemble,jackson1997locally}.
Because the present study focuses on dispersed two-phase flows, we adopt the hybrid formalism originally developed by \citet{buyevich1979flow,lhuillier1992ensemble,zhang1994ensemble,jackson1997locally}. 
This formalism enables the derivation of physically grounded closure laws for both the continuous and dispersed phases. 
However, even for dilute buoyancy dominated flows, only a limited number of closure models are currently available in the literature, motivating the present work.
Most previous studies on dilute suspensions have focused either on the Stokes flow regime or on the potential flow limit, and have examined how momentum exchange terms in two-fluid equations ---namely, the drag force and higher-order moments of the hydrodynamic forces--- affect the behaviour of the two phases.

The drag force acting on a droplet immersed in a uniform, steady Stokes flow was first derived by \citet{hadamard1911} and \citet{rybczynski1911}. 
Its extension to steady but non-uniform flows, commonly referred to as the Fax\'en correction, was obtained by \citet{hetsroni1970} (see also \citet{zhang1997momentum} and \citet{fintzi2025averaged} for derivations within a two-fluid framework).
Einstein \citep{einstein1905neue} was the first to show that the effective viscosity of a dilute suspension of solid spheres immersed in a viscous fluid is increased. 
This result was later extended by \citet{taylor1932viscosity} to suspensions of spherical droplets, demonstrating that the viscosity increase depends on the viscosity ratio between the phases.
At the time, the connection between suspension viscosity and the first moment of the hydrodynamic force ---whose deviatoric part for a solid particle is now known as the stresslet--- had not yet been established. 
This link was later clarified by \citet{batchelor1970stress}, who showed that the averaged symmetric part of the first force moment accounts for the increase in effective viscosity. 
It is worth noting that a spherical inclusion translating in a uniform flow does not generate a stresslet, which vanishes in this configuration.
Subsequently, \citet{lhuillier1996contribution}, \citet{jackson1997locally}, and \citet{zhang1997momentum}, still within the framework of viscous and dilute suspensions, demonstrated that the second moment of the hydrodynamic force also contributes to the suspension rheology, leading to non-Newtonian behaviour. 
Because the second force moment appears under a divergence operator in the effective stress, it induces a non-homogeneous rheology. 
As a result, such suspensions may also be described as second-gradient fluids \citep{gatignol2023thermomecanique}. 
Higher-order force moments were computed by \citet{zhou2020lamb}, although their relevance in practical two-fluid models remains questionable, since their magnitude decreases rapidly with increasing order under the assumption of scale separation \citep{jackson1997locally,fintzi2025averaged}.

In the opposite limit of potential flow, several studies have investigated momentum transfer between the two phases. 
It is well known that the force acting on a spherical particle immersed in a uniform inviscid potential flow reduces to the added-mass force \citep{pozrikidis2011introduction}. 
The dissipation method originally developed by Levich \citep{levich1962} allows to determine the drag force on a steadily moving bubble by considering a kinetic-energy budget over the entire fluid domain. 
\citet{biesheuvel1984two} computed both the force and the first moment for a dilute suspension of translating spherical bubbles in potential flow, and additionally derived the Reynolds stresses induced by bubble motion. 
This contrasts with the Stokes-flow regime, in which the slow decay of the disturbance fields generated by particle motion leads to divergent integrals for the Reynolds stresses \citep{caflisch1985variance,fintzi2025statistical}.
At the same time, \citet{ryskin1980}, using an extension of Levich dissipation method, derived the increase in suspension viscosity due to bubbles immersed in a purely extensional flow at very high Reynolds numbers. 
Later, \citet{zhang1994ensemble}, within the hybrid formalism, extended the work of \citet{biesheuvel1984two} by accounting for a finite variance of the particle centre-of-mass velocity. 
In parallel, several authors focused on closures for the dispersed-phase equations, including particle velocity fluctuations and collisional contributions \citep{sangani1993,bulthuis1995}.

A large number of results are available in the literature for the force and torque acting on a spherical particle immersed in a flow at low but finite Reynolds number (see \citet{leal1980}, \citet{candelier2023second}, and references therein).
In contrast, still within the low but finite Reynolds-number regime, very few results exist for higher-order moments of the hydrodynamic force.
Notable exceptions are the studies of \citet{lin1970}, \citet{stone2001inertial}, and later \citet{raja2010inertial,subramanian2011influence}, who investigated the small effects of fluid inertia on suspensions of neutrally buoyant spheres or droplets, respectively, immersed in viscous linear flows. 
These authors computed the first inertial correction to the first moment of the hydrodynamic force \citep{lin1970}, as well as the pseudo-turbulent stress \citep{stone2001inertial,raja2010inertial}.   
In this framework, the resulting mixture can be assimilated to a Reiner-Rivlin fluid, since the effective stress was found to be proportional to the square of the mean shear rate in the steady-state regime.
Then \citet{subramanian2011influence} extended~\citet{raja2010inertial}'s result to higher orders in shear Reynolds number by taking into account the outer region contribution to the stresses. In this last formulation the fluid mixture exhibits non-trivial behaviour with respect to the shear rate since it is configuration dependent, and therefore it seems that the idea of Reiner-Rivlin fluids is not even applicable for this case. 
It is worth noting, however, that the closures proposed in these works are not material-frame indifferent, a limitation that is still the subject of discussion in the modeling of inertial suspension  \citep{ryskin1980}. 

The originality of this work lies in our focus on buoyant droplets translating at low but finite Reynolds number, a configuration that has not previously been investigated within the framework of averaged two-phase flow equations. 
Where previous studies~\citep{lin1970,stone2001inertial,raja2010inertial,subramanian2011influence} focused on the response of the mixture stress to imposed shear at order $O(Re)$, we instead study the $O(Re)$ response of the continuous phase stress to slip velocity between the dispersed and continuous phases. 
The latter configuration is fundamentally different from the former as droplet motions can be independent of continuous phase motions, and may induce stresses in the latter without the continuous phase necessarily being in motion (from an average point of view).
The resulting stress will therefore be radically different from a Newtonian behavior, or a shear-thinning or -thickening behavior, because it is not just the mixture flow interacting with itself via a given constitutive law, but rather the dispersed phase inducing stresses in the continuous phase due to the slip velocity between phases. 
Additionally, we compute both the first and second moments of the hydrodynamic force, thereby including non-homogeneous rheological contributions to the effective stress of a buoyant suspension. 
A second contribution of this study is the formulation of a general reciprocal relation that enables the computation of all force moments acting on droplets with surface properties, including the drag force and the first and second moments. 
In this respect, the methodology adopted here is closely related to that employed in \citet{stone2001inertial}, and \citet{raja2010inertial} despite the fundamentally different setup studied.
Finally, by adopting a rigorous statistical averaging procedure, we show explicitly how velocity-variance terms arise in the ensemble-averaged drag force and in the effective stress of the continuous phase. 
It should be noted, however, that the system of equations is not fully closed, as no explicit closure is proposed for the velocity-variance terms appearing in the averaged momentum equations.
Although some hints on their modelling are provided in \citet{fintzi2025averaged}.

This manuscript is organized as follows.
In~\ref{sec:context} we recall the averaged mass and momentum equations, and identify the terms that require closure relations.
We then show in~\ref{sec:closure_problem} how the moments of hydrodynamic forces are related to conditionally averaged fields, which are governed by the conditionally averaged equations presented in~\ref{sec:governing_equation}. 
To avoid the complete resolution of these equations, we derive in~\ref{sec:reciprocal} a general reciprocal relation.
This relation provides a systematic method for computing force moments of arbitrary order acting on a droplet. 
Then, in~\ref{sec:compute_moments} we compute the $O(Re)$ corrections to the force moments acting on a droplet immersed in a uniform flow. 
These moments are then ensemble-averaged in~\ref{sec:averaged_moments}.
Finally, in~\ref{sec:averaged_equations} all results are incorporated into the averaged equations introduced in~\ref{sec:context}, and we discuss the rheological behaviors of the suspension. 
Particularly, we take the example of steady-state established pipe flow to better understand in which respect the new stresses derived in this study modify the two-phase flow behavior in this specific situation. \section{Problem formulation}

\subsection{Averaged equations for dispersed two-phase flows}
\label{sec:context}
To clarify the respective roles of forces and higher-order moments in dispersed two-phase flow modeling, we introduce the averaged mass and momentum equations for the dispersed and continuous phases, obtained through an ensemble-averaging procedure.
For a monodisperse suspension, the governing mass and momentum equations for each phase can be expressed as \citep{fintzi2025averaged}
\begin{align}
    \label{eq:dt_phif}
    \phi_f + \phi &\approx 1,\\
\div \textbf{U} &= 0 
    \label{eq:div_u},\\
    \label{eq:dt_phip}
(\pddt + \textbf{U}_p \cdot \grad)\phi
    &=
    - \phi \div \textbf{U}_p,\\
    \label{eq:dt_up}
    \rho_p \phi  (\pddt + \textbf{U}_p \cdot \grad)\textbf{U}_p
&=
     \phi (\div \bm\Sigma
    + \rho_p  \textbf{g})
    + \div \bm\Sigma^p
    + \textbf{F}
    ,\\
    \rho_f \phi_f (\pddt + \textbf{U}_f  \cdot \grad) \textbf{U}_f
    &= \phi_f  \left(\div \bm{\Sigma}
    + \rho_f \textbf{g}\right)
    + \div \bm\Sigma^f
    - \textbf{F},
    \label{eq:dt_uf2}
\end{align}
respectively. 
Note that in contrast to \citep{fintzi2025averaged}, all ensemble averaged quantities are written using capital letters.
The subscripts $_f$ and $_p$ refer to the continuous phase and the dispersed phase, respectively.
The vector $\textbf{g}$ is the acceleration of gravity, $\rho_k$ is the density, $\mu_k$ is the viscosity, and $\textbf{U}_k$ is the averaged velocity of phase $k$. 
We use the notation $\phi = n_p v_p$, where $n_p$ is the droplet number density and $v_p$ is the volume of a single droplet.
$\phi_f$ is the volume fraction of the continuous phase.
$\bm\Sigma = - P_f \bm\delta + 2 \mu_f \textbf{E}$, with $ \textbf{E} = \frac{1}{2}[\grad \textbf{U} +  (\grad \textbf{U})^\dagger ]$, is the \textit{mean Newtonian stress} of the mixture, with $P_f$ being the mean hydrodynamic pressure, and $\textbf{U} = \phi_f \textbf{U}_f + \phi \textbf{U}_p$ the averaged velocity of the mixture\footnote{
    Note that the definition of \textbf{U} is only exact in the homogeneous regime \citep{fintzi2025averaged}. 
} 
$, \bm{\Sigma}^p$ and $\bm{\Sigma}^f$ are the effective stresses of the dispersed and continuous phase, respectively.  
Finally, $\textbf{F}$ represents the interphase momentum exchange.

The number density, continuous phase volume fraction, interphase force, and effective stresses of the dispersed and continuous phase can be expressed formally as \citep{fintzi2025averaged},
\begin{align}
    n_p &= \pavg{}
    \label{eq:n_p},\\
    \phi_f &= \avg{\chi_f}
    \label{eq:chi_f},\\
    \textbf{F}  &= \pSavg{\bm\sigma^*_f\cdot \textbf{n}}
    \label{eq:f_alpha}
    ,\\
    \bm{\Sigma}^p &= -\rho_pv_p\pavg{ \textbf{u}_\alpha'\textbf{u}_\alpha'}
    \label{eq:def_uup}
    ,\\
    \bm{\Sigma}^f 
    &= 
    - \avg{\chi_f\rho_f \textbf{u}_f'\textbf{u}_f'} 
    + \pavg{\intS{\textbf{r}\bm\sigma^{*}_f\cdot \textbf{n}} - \delta_p\intO{2\mu_f\textbf{e}_d^*}}\nonumber\\
    &- \div
        \pavg{ \frac{1}{2}\intS{\textbf{rr}\bm\sigma^{*}_f\cdot \textbf{n}}
        - \delta_p\intO{2\mu_f \textbf{r} \textbf{e}_d^*}}
        + \grad\grad (\ldots). 
    \label{eq:def_sigma_eff_f}
\end{align}
respectively. 
Here the operator $\avg{\ldots}$ denotes ensemble averaging,  $\chi_f$ is the phase indicator function of the continuous phase, $\delta_p = \sum_\alpha \delta(\textbf{x}-\textbf{x}_\alpha)$ is the distribution pointing on the particles centre of mass (denoted $\textbf{x}_\alpha$), and \textbf{n} the unit vector normal (outward) to the droplets surface. 
$\textbf{u}_\alpha$ is the centre of mass velocity of a particle labelled $\alpha$. 
$\Omega_\alpha$ and $\Gamma_\alpha$, represent the volume and surface, respectively, of the droplets $\alpha$. 
The superscript $'$ indicates the relative values of a quantity with respect to its phase- or particle-averaged value. 
Specifically, $p_f' = p_f^0 - P_f$, $\textbf{u}_\alpha' = \textbf{u}_\alpha - \textbf{U}_p$, and $\textbf{u}_f' = \textbf{u}_f^0  -\textbf{U}_f$, with $\textbf{u}_f^0$ and $p_f^0$,  the local (non-averaged) velocity and pressure fields of the continuous phase. 
The superscript $^*$ represents the relative values of a quantity with respect to the mixture ensemble averaged value, such that $\bm{\sigma}_f^* = \bm{\sigma}_f^0  - \bm{\Sigma}$ and $\textbf{e}_d^* = \textbf{e}_d^0 - \textbf{E}$, where $\bm{\sigma}_f = -p_f^0 + \mu_f [\grad \textbf{u}_f^0 + (\grad \textbf{u}_f^0)^\dagger]$ is the local stress of the continuous phase, and $\textbf{e}_d^0 =\frac{1}{2}(\grad \textbf{u}_d^0+^\dagger\grad \textbf{u}_d^0)$, is the internal shear rate within the droplet phase.
Note that $\textbf{u}_d^0$ is the internal velocity field of the fluid within the droplets, in opposition to $\textbf{u}_\alpha$ which is the centre of mass velocity. 

In this work, our focus is specifically on $\textbf{F}$ and on the second and third contributions appearing in the effective stress tensor $\bm\Sigma^f$.

\subsection{The closure problem}
\label{sec:closure_problem}

In the present context, the closure problem consists in expressing all ensemble-averaged terms of the form $\avg{\ldots}$ in~\ref{eq:f_alpha,eq:def_uup,eq:def_sigma_eff_f} solely in terms of the macroscopic variables namely $n_p$, $\phi_f$, $P_f$, $\mathbf{U}_p$, and $\mathbf{U}_f$.
We now  assume that the droplets are spherical with radius $a$. 
The surface exchange terms may be rewritten in terms of integrals of conditional averaged quantities \citep{lhuillier1992ensemble,zhang1997momentum,fintzi2025statistical}. 
If we take the example of the drag force term, we obtain, 
\begin{equation}
    \pSavg{\bm\sigma^*_f\cdot \textbf{n}}
    =
    \int_{\mathbb{R}^3} P[\textbf{w}|\textbf{x},t] n_p[\textbf{x},t]\intS[p]{
        \left\{
            -p_f^1 \bm\delta 
    + \mu_f [\grad \textbf{u}^1 + ^\dagger\grad \textbf{u}^1]
        \right\}\cdot \textbf{n}
    }(\textbf{r})  
    d^3 \textbf{w},
    \label{eq:conditional_average2}
\end{equation}
where $p_f^1[\textbf{r}|\textbf{x},\textbf{w}]$ and $\textbf{u}^1[\textbf{r}|\textbf{x},\textbf{w}]$, are the pressure and velocity fields ensemble averaged on every configuration where a droplet is located at $\textbf{x}$ with centre of mass velocity \textbf{w}, minus the ensemble averaged pressure and velocity fields, i.e. $P_f[\textbf{r},t]$, and $\textbf{U}[\textbf{r},t]$, \textbf{r} being the points on the droplets surface.
$\Gamma _p$ denotes the domain of the surface of the test droplet, whose surface is described by the equation $|\textbf{x}- \textbf{y}| - a=r - a = 0$. 
Here, $ P[\textbf{w}|\textbf{x},t]$ is the probability of finding a droplet with centre of mass velocity \textbf{w} knowing that a droplet is present at \textbf{x} at time $t$. 
The equations governing $p_f^1$ and $\textbf{u}^1$ can be obtained by conditionally averaging the local mass and momentum equations \citep{fintzi2025statistical}.
The kinematics boundary conditions at the interface of the test droplet are well-defined only if the conditional average is performed on a sample of configurations where the droplets posses the same centre of mass velocity and shape.
Hence, conditionally averaging on the centre of mass position \textbf{x}, and on the centre of mass velocity \textbf{w}, is a necessary step to reduce the degree of freedom of the conditionally averaged problem.

In the following, we consider arbitrary values of density and viscosity ratios, defined as:
\begin{align}
    \lambda = \mu_p/\mu_f,  && \zeta =\rho_p /\rho_f,
\end{align}
respectively. 
Additionally, we limit this theoretical investigation to dilute suspensions ($\phi\ll 1$).
We defined the droplet Reynolds number as,
\begin{equation}
    Re  = \frac{U\rho_f a}{\mu_f}, 
\end{equation}
where $a$ is the radius of the droplets, and $U = |\textbf{w}_r[\textbf{y}]|$, with $\textbf{w}_r[\textbf{x}] = \textbf{U}[\textbf{x}] - \textbf{w}$, the magnitude of the relative velocity between the centre of mass velocity of the test droplet and the ensemble averaged velocity of the mixture evaluated at the centre of mass position \textbf{y}. 
Note that the value of the Reynolds number is considered arbitrary up to~\ref{sec:computation_forces} where we consider a small but finite Reynolds number. 
Finally, the \textit{capillary} number,
\begin{equation}
    Ca = \mu_f U/\gamma,
\end{equation}
is assumed to be small enough so that the droplets remain spherical. 

\subsection{Conditionally averaged equations}
\label{sec:governing_equation}

At the leading order in droplet volume fraction, the governing equations for $\textbf{u}^1$ and $p_f^1$ are equivalent to those of an isolated droplet immersed in an unbounded domain \citep{hinch1977averaged,koch1993hydrodynamic}. 
Hence, we consider the problem of an isolated test droplet immersed in an arbitrary flow where only the uniform relative motions induce inertia. 
The disturbances pressure and velocity fields ($\textbf{u}^1$ and $p_f^1$), relative to the position of a test droplet are noted $\textbf{u}_{o}$, $\textbf{u}_{i}$, $p_{o}$ and $p_{i}$, for the velocity outside, and inside, the pressure outside, and inside the volume of the test droplet, respectively. 
\ref{fig:disturbance} displays a schematic representation of the problem. 
\begin{figure}[h!]
    \centering
    \begin{tikzpicture}[scale= 1.5]
        \filldraw[ gray!50!white](0,0)circle (0.5);
        \draw(0,0)circle (0.5)node[right,below]{   $\;\;\;\;\;\;\;\;\;\;\;\Gamma$};
        \draw[->,blue!50!black,thick](0,0)--++(0,1)node[right]{$\textbf{w}$};
        \draw (0,0)node{$\bullet$}node[right]{$\textbf{y}$};
        \draw[->] (-0.45,0.1)--++(-1,0.3)node[above]{$\textbf{n}$};
        \draw[->](-0.5,1)--++(-0.5,1)node[midway,left]{$\textbf{u}_{i/o}[\textbf{x}|\textbf{y},\textbf{w}]$};
\draw (0,0)node[below]{$\Omega_i$};
        \draw (1.5,1.25)node{$\Omega_o$};
    \end{tikzpicture} 
    \caption{Sketch of the problem. $\textbf{u}_{o/i}[\textbf{x}|\textbf{y},\textbf{w}]$ is the disturbance velocity field evaluated at \textbf{x}, generated by a test droplet positioned at \textbf{y} with centre of mass velocity \textbf{w}. 
    \textbf{n} is the surface normal (outward) vector of the droplet. 
    The exterior domain of the test droplet is denoted $\Omega_o$, and the interior domain $\Omega_i$.}
    \label{fig:disturbance}
\end{figure}

Outside and inside the volume of the test droplet we can write the mass and momentum equations for the disturbances fields ($p_{o/i}$ and $\textbf{u}_{o/i}$) by conditionally averaging the Navier stokes equations \citep{koch1993hydrodynamic,fintzi2025statistical}.
This lead to the equations for the disturbances fields originally derived by~\cite{maxey1983equation} and \citet{gatignol1983}. 
However, it is important to emphasize that-even in dilute suspensions-conditional averaging introduces additional contributions in~\ref{eq:momentum_out,eq:momentum_in} (see~\cite[Eq. (9)]{koch1993hydrodynamic} or \citet{fintzi2025statistical}), which are not present in~\cite{maxey1983equation} and \citet{gatignol1983}'s equations. 
In particular, one extra term corresponds to the velocity-variance of the conditionally averaged velocity field (see \citet{koch1993hydrodynamic} or \citet[Chapter 4]{fintzi2025statistical}). 
In dilute flow, this term turns out to be negligible \citep[Appendix A]{koch1993hydrodynamic}.
Consequently, in the present context, we consider that the conditionally averaged equations are equivalent to the equations for an isolated droplet. 
The only difference is that conditionally averaged equations are explicitly related to ensemble-averaged fields (\textbf{U} and $P_f$) through their boundary conditions \citep{fintzi2025statistical}.
In dimensionless form, they read, 
\begin{align}
    \div \textbf{u}_{o} &= 0
    \\
    \div\bm\sigma_{o}
    &= 
    Re [
    \pddt \textbf{u}_{o}
    + \textbf{u}_{o}\cdot \grad \textbf{u}_{o}
    + \textbf{u}_{o}\cdot \grad \textbf{w}_r
    + \textbf{w}_r \cdot \grad \textbf{u}_{o}] \nonumber \\
    &= Re \textbf{f}_{o},
    \label{eq:momentum_out}
\end{align}
for all $|\textbf{x}- \textbf{y}| = |\textbf{r}| = r > 1$, and, 
\begin{align}
    \div \textbf{u}_{i} &= 0,
    \\
    \div\bm\sigma_{i}
    &= 
    (\zeta/\lambda -1)  \div\bm\Sigma
    + \frac{\zeta Re}{\lambda} [
    \pddt \textbf{u}_{i}
    + \textbf{u}_{i}\cdot \grad \textbf{u}_{i}
    + \textbf{u}_{i}\cdot \grad \textbf{w}_r 
    + \textbf{w}_r \cdot \grad \textbf{u}_{i}]\nonumber \\
    &= 
    (\zeta/\lambda -1)  \div\bm\Sigma
    +\frac{\zeta}{\lambda}Re \textbf{f}_{i}
    =  \textbf{f}_{i}^{\text{tot}}
    ,\label{eq:momentum_in}
\end{align}
for $r<1$. 
The terms  $\textbf{f}_{o/i}$ denote dimensionless effective body force incorporating the inertial terms from the Navier-Stokes equations.
Additionally, $\bm\sigma_{i/o} = -p_{i/o}\bm\delta + 2\textbf{e}_{i/o}$ with $\textbf{e}_{i/o} = \frac{1}{2}[\grad \textbf{u}_{i/o} + ^{\dagger}\grad \textbf{u}_{i/o}]$ is the dimensionless Newtonian stresses in the interior and exterior of the test droplet. 
Note that the distances have been made dimensionless using the radius $a$, and the velocities using the velocity scale $U$.
Likewise, the stresses in the exterior of the test droplet $\bm\sigma_o$, and the ensemble averaged stress $\bm\Sigma$, have been made dimensionless using the viscous scale $U \mu_f /a$, and the stress in the interior of the test droplet $\bm\sigma_i$, using the scale $U \mu_d /a$. 
Recall that $\bm\sigma_{i}$ is a disturbance field defined with respect to the mean velocity $\textbf{U}$, and the mean stress $\bm\Sigma$. 
Hence, because of the difference in viscosity and density between the dispersed and continuous phases (i.e. between $\bm\Sigma$ and $\bm\sigma_i$), $\textbf{f}_{i}^{\text{tot}}$ includes in addition the divergence of the ensemble-averaged stress times the ratio of the material properties: $(\zeta/\lambda - 1)$.
Note that in~\ref{eq:momentum_out,eq:momentum_in}, and in the following $\textbf{w}_r$, $\textbf{U}$ and $\textbf{E}$ are dimensionless.
In particular they correspond to $\textbf{w}_r/U$,$\textbf{U}/U$ and $\textbf{E}a/U$, respectively.

At the surface of the droplet ($r = 1$) the continuity of velocity and the non-deformability of the test droplet ($Ca \ll 1$), and the imposed velocity of the test droplet centre of mass, imposes, 
\begin{align}
    \textbf{u}_{i} - \textbf{u}_{o}= 0,
    && 
    (\textbf{u}_{i}+\textbf{w}_r) \cdot \textbf{n}
    =
    0.
    \label{eq:normal_vel}
\end{align} 
$\textbf{u}_{i}+\textbf{w}_r$ can be interpreted as the velocity observed in the reference frame moving with the test droplet. 
The correct boundary condition for the disturbance shear rate is (at $r=1$),
\begin{equation}
    \mathbf{n}\cdot 2 [\textbf{e}_{o} - \lambda \textbf{e}_{i} + (1-\lambda)\textbf{E}
]\cdot (\bm\delta - \textbf{nn})
    =
    \textbf{b}\cdot (\bm\delta - \textbf{nn}),
\label{eq:boundary_cdt_stress}
\end{equation}
with, $\textbf{b}[\textbf{r},t]$ a tangential stress jump at the interface normalized by $\mu_f U / a$.
Here we have considered a stress jump in order to stay general when deriving the reciprocal relation (see~\ref{sec:reciprocal}), nevertheless, $\textbf{b}$ will be taken to be zero in the final calculation of the moments of forces~\ref{sec:compute_moments}. 
In particular, the contribution of Marangoni stresses to the rheology of the droplet suspension is not included in the present analysis (see \citet{fintzi2025averaged} for further discussion).
Note that in general the ensemble averaged quantities $\textbf{U}$ and $\textbf{E}$ as well as $\textbf{b}$ depend on the position along the surface of the test droplet and time, so that $\textbf{w}_r[\textbf{r},t]$, $\textbf{E}[\textbf{r},t]$ and $\textbf{b}[\textbf{r},t]$ in~\ref{eq:boundary_cdt_stress,eq:normal_vel} are not spatially constant. 
Hence, it is important to note that although $\textbf{w}_r[\textbf{y}]$ is a unit vector,  $\textbf{w}_r[\textbf{x}]$ is not, because in the general case $\textbf{U}[\textbf{x}] \neq \textbf{U}[\textbf{y}]$.

For sufficiently large $|\textbf{y}-\textbf{x}|$, it is assumed that the  conditionally averaged velocity and pressure fields are not affected by the presence of the test droplet, hence they are equal to their respective ensemble averaged values. 
Hence, far from the test droplet the disturbance fields, which are defined as the difference between conditionally and ensemble averaged fields, vanish. 
It reads, \begin{align}
    \lim_{r \to\infty }\textbf{u}_{o}[\textbf{r}|\textbf{y},\textbf{w}] = 0,
    && \lim_{r \to\infty }p_{o}[\textbf{r}|\textbf{y},\textbf{w}]= 0. 
    \label{eq:BC_r_infty_1}
\end{align}

\section{Reciprocal theorem for droplets}\label{sec:reciprocal}

The reciprocal theorem for a droplet already exists in the literature; see, for example, \citet{lovalenti1993force, raja2010inertial}. 
However, \citet{lovalenti1993force} focuses on the force, while \citet{raja2010inertial} focuses on the stresslet. 
Here, we provide a detailed derivation of the reciprocal theorem formulated to compute moments of any order for a droplet, thereby encompassing both the force and the stresslet.

\subsection{Auxiliary solutions}
The reciprocal theorem requires the use of a known solution.
We call that solution the \textit{auxiliary} solution, since its only purpose is to compute the \textit{real} solution of the equations introduced above. 
The \textit{auxiliary} velocity and stress fields are denoted by $\hat{\textbf{u}}_{i/o}$ and $\hat{\bm\sigma}_{i/o} = - \hat{p}_{i/o}\bm\delta + \grad \hat{\textbf{u}}_{i/o} + ^\dagger\grad \hat{\textbf{u}}_{i/o}$, respectively. 
In general, all \textit{auxiliary} quantities are written with a hat. 

In the \textit{auxiliary} problem, we neglect all inertial effects and consider an arbitrary quadratic flow; thus $\hat{p}_{o/i}$ and $\hat{\textbf{u}}_{o/i}$ satisfy,
\begin{align}
    \div \hat{\textbf{u}}_{o} = 0 ,
    &&\div\hat{\bm\sigma}_{o}  = 0 ,
    \label{eq:momentum_out_s}\\
    \div \hat{\textbf{u}}_{i} = 0 ,
    && \div\hat{\bm\sigma}_{i}  = 0 ,
    \label{eq:momentum_in_s}
\end{align}
for $r>1$ and $r<1$, respectively. 
The boundary conditions at $r\to\infty$ read, 
\begin{align}
    \lim_{r \to\infty }\hat{\textbf{u}}_{o}[\textbf{r}|\textbf{y},\textbf{w}] = 0,
    && \lim_{r \to\infty }\hat{p}_{o}[\textbf{r}|\textbf{y},\textbf{w}]= 0. 
    \label{eq:BC_r_infty_2}
\end{align}
and at $r=1$,
\begin{align}
    \label{eq:normal_vel2_s}
    \hat{\textbf{u}}_{i} &= \hat{\textbf{u}}_{o}\\
    \hat{\textbf{u}}_{i} \cdot \textbf{n} &= - \hat{\textbf{w}}_r \cdot \textbf{n}
    \label{eq:normal_vel_s}
    \\
    \label{eq:normal_stres_s}
    \textbf{n}\cdot 2[\hat{\textbf{e}}_{o} - \lambda \hat{\textbf{e}}_{i} + (1-\lambda) \hat{\textbf{E}}
]\cdot (\bm\delta - \textbf{nn})
    &=
    \hat{\textbf{b}}\cdot (\bm\delta - \textbf{nn})
\end{align}
where we introduced $\hat{\textbf{w}}_r = \hat{\textbf{U}} - \textbf{w}$ as the undisturbed relative velocity field of the \textit{auxiliary} problem, similarly $\hat{\textbf{E}}$ is the undisturbed shear rate of the \textit{auxiliary} problem. 
Note that for the \textit{test} problem it is important that $\hat{\textbf{b}}\neq 0$, this will be useful to write a reciprocal relation in~\ref{sec:reciprocal}. 
Indeed, this stress discontinuity is a mathematical artefact in order to evaluate the surface integral of the shear stress at the interface of the test droplet \citep{raja2010inertial} but have no physical meaning here.
For our purposes, we only need $\hat{\textbf{b}}$ and $\hat{\textbf{U}}$ to be quadratic fields.
Hence, we consider the relations, 
\begin{align}
    \hat{\textbf{w}}_r(\textbf{y} + \textbf{r}) 
    &=  \hat{\textbf{U}} - \textbf{w}
    +  \textbf{r} \cdot  \grad\hat{\textbf{U}}|_{\textbf{r}=0}
    +  \frac{1}{2}\textbf{rr} :  \grad\grad\hat{\textbf{U}}|_{\textbf{r}=0} + ...,
    \label{eq:w_r_expand}\\
     \hat{\textbf{E}}(\textbf{y} + \textbf{r}) 
    &=   \hat{\textbf{E}}|_{\textbf{r}=0}
    + \textbf{r} \cdot  \grad \hat{\textbf{E}}|_{\textbf{r}=0}+ ...,
\\
     \hat{\textbf{b}}(\textbf{y} + \textbf{r}) 
    &=   \hat{\textbf{b}}|_{\textbf{r}=0}
    + \textbf{r} \cdot  \grad \hat{\textbf{b}}|_{\textbf{r}=0}
    + \frac{1}{2}\textbf{rr} :  \grad\grad \hat{\textbf{b}}|_{\textbf{r}=0}+ .... 
\end{align}

According to the linearity of the Stokes equations~\ref{eq:momentum_out_s,eq:momentum_in_s}, and of the boundary conditions~\ref{eq:normal_vel2_s,eq:normal_vel_s,eq:normal_stres_s} we deduce that $\hat{\textbf{u}}_{i/o}$ and $\hat{p}_{i/o}$  must be linear combination of spherical harmonics proportional to $\hat{\textbf{w}}_r|_{\textbf{r}=0}$, $\hat{\textbf{b}}|_{\textbf{r}=0}$, and their derivatives \citep{brenner1963stokes}.
Hence, we can write the general solution of $\textbf{u}_{i/o}$ and $p_{i/o}$ as,
\begin{align}
    \begin{pmatrix}
        \hat{\textbf{u}}_{o}\\
        \hat{p}_{o}\\
        \hat{\textbf{u}}_{i}\\
        \hat{p}_{i}
    \end{pmatrix}
    =
    \begin{pmatrix}
        \mathbb{U}_{o}^{(1)} + \mathbb{U}_{o}^{(2)}\cdot \grad + \mathbb{U}_{o}^{(3)} :\grad\grad &
        \mathbb{U}_{o}^\text{(1-b)} + \mathbb{U}_{o}^\text{(2-b)}\cdot \grad + \mathbb{U}_{o}^\text{(3-b)} :\grad\grad \\
        \mathbb{P}_{o}^{(1)} + \mathbb{P}_{o}^{(2)}\cdot \grad + \mathbb{P}_{o}^{(3)} :\grad\grad &
        \mathbb{P}_{o}^\text{(1-b)} + \mathbb{P}_{o}^\text{(2-b)}\cdot \grad + \mathbb{P}_{o}^\text{(3-b)} :\grad\grad \\
        \mathbb{U}_{i}^{(1)} + \mathbb{U}_{i}^{(2)}\cdot \grad + \mathbb{U}_{i}^{(3)} :\grad\grad &
        \mathbb{U}_{i}^\text{(1-b)} + \mathbb{U}_{i}^\text{(2-b)}\cdot \grad + \mathbb{U}_{i}^\text{(3-b)} :\grad\grad \\
        \mathbb{P}_{i}^{(1)} + \mathbb{P}_{i}^{(2)}\cdot \grad + \mathbb{P}_{i}^{(3)} :\grad\grad &
        \mathbb{P}_{i}^\text{(1-b)} + \mathbb{P}_{i}^\text{(2-b)}\cdot \grad + \mathbb{P}_{i}^\text{(3-b)} :\grad\grad \\
    \end{pmatrix}
    \cdot 
    \begin{pmatrix}
        \hat{\textbf{w}}_r\\
        \hat{\textbf{b}}
    \end{pmatrix}
    \label{eq:big_solution}
\end{align}
The tensor $\mathbb{U}_{i/o}^{(n)}$ and $\mathbb{P}_{i/o}^{(n)}$ are $n+1$ and $n$ order tensors given in \citep{leal2007advanced} and in \citet[Appendix F]{fintzi2025averaged}, that depend only on the relative coordinate $\textbf{r} = \textbf{x}-\textbf{y}$ and the viscosity ratio $\lambda$.
We also define the $n+2$ order tensors, 
\begin{equation}
    \mathbb{S}_{i/o}^{(n)} = 
    - \bm \delta \mathbb{P}_{i/o}^{(n)}
    + \grad \mathbb{U}_{i/o}^{(n)}
    + ^\dagger\grad \mathbb{U}_{i/o}^{(n)},
    \label{eq:big_S}
\end{equation}
\footnote{The $^\dagger$ used in~\ref{eq:big_S} represents the transpose operator which acts on the two closest indices.
For example, the transpose of the triadic $\textbf{abc}$, may be written:  $^\dagger\textbf{abc} = \textbf{bac}$ or $\textbf{abc}^\dagger = \textbf{acb}$. }
where $n =1,2,3$ or $1\text{-}b,2\text{-}b,3\text{-}b$. 
\ref{eq:big_S} provides the stresses fields of the \textit{auxiliary} problem upon contraction with $\hat{\textbf{w}}_r$, and $\hat{\textbf{b}}$, and the higher derivatives. 

\subsection{Singularity solutions}

Following \citet{stone2001inertial,raja2010inertial}, we also consider the \textit{auxiliary} problem of a point source located at the origin.
The reason why this solution is necessary is that, in the previous \textit{auxiliary} problem $\div \hat{\textbf{U}}= 0$.
Hence, as demonstrated below the previous \textit{auxiliary} solutions are unable to provide a formula for the trace of the first moment using the reciprocal theorem.
Because we also aim to compute the traces of the second moment of forces, one also need the solutions of a point force and a potential dipole.  
These two classes of solutions, i.e., point source and point force (and their derivatives), obey the non-homogeneous Stokes equations, 
\begin{align}
    \grad^2 \hat{\textbf{u}}_o = \grad \hat{p}_o + (\textbf{Q}^{(n)}\odot \grad^{(n)})\grad \delta(\textbf{r}),
    &&
    \grad^2 \hat{\textbf{u}}_o = \grad \hat{p}_o + (\textbf{R}^{(n+1)}\odot \grad^{(n)}) \delta(\textbf{r}),
\end{align}
respectively. 
Where $\textbf{Q}^{(n)}$ and $\textbf{R}^{(n+1)}$ are two arbitrary $n^{th}$ and $(n+1)^{th}$ order tensors, respectively. 
Similarly, the $\grad^{(n)}$ denote $n^{th}$ order outter product of gradients vectors.
The operator $\odot$ represents the contraction product on the $n^{th}$ common indices. 
Both equations are completed by the continuity equation $\div \textbf{u}_{o}$. 
The solution of the point source and point force read, 
\begin{align}
    \hat{\textbf{u}}_o = -\frac{1}{4\pi}\grad^{(n+1)}(1/r)\odot \textbf{Q}^{(n)},
    \label{eq:pts_source_n}
    &&
    \hat{\textbf{u}}_o = \frac{1}{8\pi}\grad^{(n)}[\grad^{(2)} - \bm\delta \grad^2]r
    \odot \textbf{R}^{(n+1)},
\end{align}
respectively. 
Note that $\frac{1}{8\pi}[\grad^{(2)} - \bm\delta \grad^2]r = - (\textbf{nn} + \bm\delta)/(8\pi r) $ is the free-space green function of Stokes equations~\citep{pozrikidis1992boundary}. 
The stress tensors read, 
\begin{align}
    \hat{\bm\sigma}_o &= -\frac{1}{2\pi}\grad^{(n+2)}(1/r)\odot \textbf{Q}^{(n)}, 
    \\
    \hat{\bm\sigma}_o &= \frac{1}{8\pi}\grad^{(n)}[
        2\grad^{(3)}
        -(\grad \bm\delta + ^\dagger\grad\bm\delta+\bm\delta \grad)\grad^2
    ]r\odot \textbf{R}^{(n+1)},
\end{align}
for the point source and point forces solutions, respectively. 
More details on how to obtain those solutions are given in~\citet[Chapter 6]{pozrikidis2011introduction} or in~\ref{ap:singularity_sol}. 

As an example, the solution of a point source (i.e.,~\ref{eq:pts_source_n}~(left) with $n=0$) used in the calculation of the first moment of force reads, 
\begin{align}
    \hat{\textbf{u}}_{o} = \frac{Q^{(0)}}{4\pi} \textbf{n}r^{-2},
    && \hat{\bm\sigma}_{o} = \mu_f \frac{Q^{(0)}}{2\pi}\left(
        \bm\delta
        - 3 \textbf{nn}
    \right)r^{-3}
    \label{eq:point_source}
\end{align}
Note that this expression is valid throughout the domain excluding the point $\textbf{r} =  \textbf{x} -  \textbf{y} = 0$, hence we may use either the subscript $o$ or $i$. 
The solution for a point source dipole $(\textbf{Q}^{(1)}$) and point force $(\textbf{R}^{(1)})$ will also be used to derive relations for the traces of the second moment of force. 
In summary, as the values of $\hat{\textbf{w}}_r$, $\hat{\textbf{b}}$, $\textbf{R}^{(n)}$, and $\textbf{Q}^{(n)}$, are entirely arbitrary, the solutions given by~\ref{eq:big_solution,eq:pts_source_n,eq:point_source} can be used as a tool to derive expressions for the first and second moments of the forces by means of the reciprocal theorem derived in the next section.

\subsection{Reciprocal theorem}
The derivation of the general formula proceeds in three main steps: (1) we write a reciprocal relation in the exterior of the test droplet, (2) we then write the corresponding relation in the interior of the test droplet, and (3) finally, using the boundary conditions at the droplet interface, we combine the two expressions to obtain the final reciprocal relation.

\subsubsection{First steps}
We first take the dot product of~\ref{eq:momentum_out} with $\hat{\textbf{u}}_{o}$, and the dot product of~\ref{eq:momentum_out_s} with $\textbf{u}_{o}$, subtracting both expression gives, 
\begin{equation}
    \div (\bm\sigma_{o}\cdot \hat{\textbf{u}}_o)
=
    \div (\hat{\bm\sigma}_{o}\cdot \textbf{u}_o)
+ Re (\hat{\textbf{u}}_{o}\cdot \textbf{f}_{o}). 
    \label{eq:first_step_out}
\end{equation}
To derive this relation we used the identities $\bm\sigma_o :\grad \hat{\textbf{u}}_o = 2\textbf{e}_o :\grad \hat{\textbf{u}}_o = 2{\textbf e}_o : \hat{\textbf e}_o$, and  $\hat{\bm\sigma}_o :\grad {\textbf{u}}_o = 2\hat{\textbf{e}}_o : \textbf{e}_o$, which imply $\bm\sigma_o :\grad \hat{\textbf{u}}_o = \hat{\bm\sigma}_o :\grad {\textbf{u}}_o$. 
This simplification explains why complete knowledge of the velocity field in the exterior of the droplet becomes unnecessary when $\textbf{f}_o = 0$.
Consequently, the power of the reciprocal theorem is rooted in the commutativity of the operation $\textbf e_o:\hat{\textbf e}_o$. 
Integrating this expression on the whole domain $\Omega_{o}$, and using the divergence theorem we obtain,

\begin{equation}
   - \intS[p]{\hat{\textbf{u}}_{o} \cdot  \bm\sigma_{o}\cdot \textbf{n}}
    =
    -\intS[p]{\textbf{u}_{o}\cdot \hat{\bm\sigma}_{o}\cdot \textbf{n}}
    + Re\intO[o]{\hat{\textbf{u}}_{o}\cdot \textbf{f}_{o}}.
    \label{eq:int_first_step0}
\end{equation}
The surface integrals appears only on the boundary of the test droplet because far from the test droplet ($r\to\infty$), the fields $\hat{\textbf{u}}_o$, $\hat{\bm\sigma}_o$, $\textbf{u}_o$ and $\bm\sigma_o$ vanish (see~\ref{eq:BC_r_infty_1} and~\ref{eq:BC_r_infty_2}). 
\ref{eq:int_first_step0} already provides a sufficient reciprocal relation when considering solid particles \citep{masoud2019reciprocal}.
Indeed, in that case $\textbf{u}_{o} = -\textbf{w}_r$, and $\hat{\textbf{u}}_{o} = -\hat{\textbf{w}}_r$ at $r=1$, because of the no-slip boundary condition at the particle interface, hence leading directly to a formula for the force traction on the surface of the solid particle (first term on the left-hand side of~\ref{eq:int_first_step}).
Of course, one still needs to provide an approximation for $\textbf{f}_o$ if inertia is considered. For fluid particles, a few more steps are necessary. Inserting the relation $\textbf{u}_o = \textbf{u}_o+\textbf{w}_r-\textbf{w}_r$ and $\hat{\textbf{u}}_o = \hat{\textbf{u}}_o+\hat{\textbf{w}}_r-\hat{\textbf{w}}_r$ gives, 
\begin{align}
    &\intS[p]{\hat{\textbf{w}}_{r} \cdot  \bm\sigma_{o}\cdot \textbf{n}}
    - 2\intS[p]{(\hat{\textbf{u}}_{o} + \hat{\textbf{w}}_r) \cdot  \textbf e_{o}\cdot \textbf{n}}
     \nonumber\\
    &= \intS[p]{\textbf{w}_{r}\cdot \hat{\bm\sigma}_{o}\cdot \textbf{n}}
    - 2\intS[p]{(\textbf{u}_{o} + \textbf{w}_r)\cdot \hat{\textbf e}_{o}\cdot \textbf{n}}
    + 
    Re\intO[o]{\hat{\textbf{u}}_{o}\cdot \textbf{f}_{o}}.
    \label{eq:int_first_step}
\end{align}
Where we used the relation, $(\hat{\textbf{u}}_{o} + \hat{\textbf{w}}_r) \cdot  \bm\sigma_{o}\cdot \textbf{n} = 2 (\hat{\textbf{u}}_{o} + \hat{\textbf{w}}_r) \cdot  \textbf e_{o}\cdot \textbf{n}$, allowed  because $(\hat{\textbf{u}}_{o} + \hat{\textbf{w}}_r) = (\bm\delta - \textbf{nn})\cdot (\hat{\textbf{u}}_{o} + \hat{\textbf{w}}_r)$ on the surface of the test droplet, see~\ref{eq:normal_vel_s}. 
Similar considerations lead to $({\textbf{u}}_{o} + {\textbf{w}_r}) \cdot  \hat{\bm\sigma}_{o}\cdot \textbf{n} =2  ({\textbf{u}}_{o} + {\textbf{w}_r}) \cdot  \hat{\textbf e}_{o}\cdot \textbf{n}$. 
\subsubsection{Intermediate steps}

The equations governing the flow inside the droplets have not yet been considered.
We now take the dot product of~\ref{eq:momentum_in} with $(\hat{\textbf{u}}_i + \hat{\textbf{w}}_r)$ and of~\ref{eq:momentum_in_s} with $(\textbf{u}_i+ \textbf{w}_r)$, subtracting both expression leads to, 
\begin{equation}
    \div [\bm\sigma_{i}\cdot (\hat{\textbf{u}}_i+\hat{\textbf{w}}_r)]
    - 2\textbf{e}_i : \hat{\textbf{E}}
=
    \div [\hat{\bm\sigma}_{i}\cdot (\textbf{u}_i+\textbf{w}_r)]
    - 2\hat{\textbf{e}}_i :\textbf{E}
    + (\hat{\textbf{u}}_i+\hat{\textbf{w}}_r)\cdot \textbf{f}_{i}^{tot},
    \label{eq:second_step_out}
\end{equation}
where we used the relation, $- \bm\sigma_i :\grad (\hat{\textbf{u}}_i+\hat{\textbf{w}}_r)+ \hat{\bm\sigma}_i :\grad ({\textbf{u}}_i+{\textbf{w}}_r) =
- 2\textbf{e}_i :\hat{\textbf{E}} + 2\hat{\textbf{e}}_i :\textbf{E}$. 
Integrating this relation over the domain $\Omega_i$, and using the divergence theorem, as well as similar simplifications used in~\ref{eq:int_first_step}, gives, 
\begin{align}
    &\intS[p]{ (\hat{\textbf{u}}_i+\hat{\textbf{w}}_r)\cdot \textbf{e}_{i}\cdot\textbf{n}}
    - \intO[i]{\textbf{e}_i : \hat{\textbf{E}}} \nonumber \\
    &=
    \intS[p]{ (\textbf{u}_i+\textbf{w}_r)\cdot\hat{\textbf{e}}_{i}\cdot \textbf{n}}
    - \intO[i]{\hat{\textbf{e}}_i :\textbf{E}}
    + \frac{1}{2} \intO[i]{(\hat{\textbf{u}}_i+\hat{\textbf{w}}_r)\cdot \textbf{f}_{i}^{tot}} 
    \label{eq:second_step_int}
\end{align}
The first terms on both the left- and right-hand sides of \ref{eq:second_step_int} may be obtained using \ref{eq:boundary_cdt_stress} and ~\ref{eq:second_step_int}.
Indeed, since $(\hat{\textbf{u}}_i+\hat{\textbf{w}}_r)\cdot (\bm\delta-\textbf{nn}) = (\hat{\textbf{u}}_i+\hat{\textbf{w}}_r)$, multiplying~\ref{eq:boundary_cdt_stress} by $(\hat{\textbf{u}}_i+\hat{\textbf{w}}_r)$ yields
\begin{equation} 
         -2 \lambda (\hat{\textbf{u}}_{i/o}+\hat{\textbf{w}}_r) \cdot \textbf{e}_i \cdot \textbf{n} 
        =
        \textbf{b}\cdot (\hat{\textbf{u}}_{i/o}+\hat{\textbf{w}}_r) -2 (\hat{\textbf{u}}_{i/o}+\hat{\textbf{w}}_r)\cdot\textbf{e}_{o}\cdot \textbf{n} -2 (1 -\lambda) (\hat{\textbf{u}}_{i/o}+\hat{\textbf{w}}_r)\cdot\textbf{E} \cdot \textbf{n}
    \label{eq:boundary_with_the_velocity}
\end{equation}
with a similar relation for the \textit{test} problem boundary condition. Adding~\ref{eq:int_first_step} and~\ref{eq:second_step_int} times $2\lambda$, while using the boundary condition given by~\ref{eq:boundary_with_the_velocity} leads to the expression,
\begin{multline}
    \intS[p]{\hat{\textbf{w}}_{r} \cdot  \bm\sigma_{o}\cdot \textbf{n}}
    -\lambda \intO[i]{2\textbf{e}_i : \hat{\textbf{E}}}
    - (1-\lambda) \intS[p]{2(\textbf{u}_{i} + \textbf{w}_r)\cdot \hat{\textbf{E}}\cdot \textbf{n}}
    + \intS[p]{(\textbf{u}_{i} + \textbf{w}_r)\cdot \hat{\textbf{b}}}
    \\
    =
    \intS[p]{\textbf{w}_{r}\cdot \hat{\bm\sigma}_{o}\cdot \textbf{n}}
    - \lambda \intO[i]{2\hat{\textbf{e}}_i :\textbf{E}}
    - (1-\lambda) \intS[p]{2(\hat{\textbf{u}}_{i} + \hat{\textbf{w}}_r) \cdot  \textbf{E} \cdot \textbf{n}}\\ 
    + \intS[p]{(\hat{\textbf{u}}_{i} + \hat{\textbf{w}}_r) \cdot  \textbf{b}}
    + \lambda \intO[i]{(\hat{\textbf{u}}_i+\hat{\textbf{w}}_r)\cdot \textbf{f}_{i}^{tot}} 
    + Re\intO[o]{\hat{\textbf{u}}_{o}\cdot \textbf{f}_{o}}.
    \label{eq:int_third_step}
\end{multline}

\subsection{Final steps}

To obtain the final form of the reciprocal theorem, we proceed by noting that the second and third terms on the left-hand side  of~\ref{eq:int_third_step} may be combined upon using the divergence theorem on the third term, and using the relation, 
\begin{equation}
    - \lambda \textbf{e}_i : \hat{\textbf{E}}
    - (1-\lambda)\div [\hat{\textbf{E}} \cdot (\textbf{u}_i + \textbf{w}_r)]
    =
- \textbf{e}_i : \hat{\textbf{E}}
    - (1-\lambda) \hat{\textbf{E}}:\textbf{E}
    - (1-\lambda)\div \hat{\textbf{E}} \cdot (\textbf{u}_i + \textbf{w}_r).
    \label{eq:tricks_two}
\end{equation}
One can also derive the same manipulation on the second and third term on the right-hand side of~\ref{eq:int_third_step}, namely,
\begin{equation}
    - \lambda \hat{\textbf{e}}_i : {\textbf{E}}
    - (1-\lambda)\div [{\textbf{E}} \cdot (\hat{\textbf{u}}_i + \hat{\textbf{w}}_r)]
    =
- \hat{\textbf{e}_i} : {\textbf{E}}
    - (1-\lambda) \hat{\textbf{E}}:\textbf{E}
    - (1-\lambda)\div {\textbf{E}} \cdot (\hat{\textbf{u}}_i + \hat{\textbf{w}}_r).
    \label{eq:tricks_one}
\end{equation}
Because the product $\hat{\textbf{E}} : {\textbf{E}}$ is commutative, the second term on the right-hand side of~\ref{eq:tricks_two,eq:tricks_one} cancel each other in the final expression.  
Injecting both~\ref{eq:tricks_one,eq:tricks_two} in~\ref{eq:int_third_step} leads to the final form of the reciprocal theorem for droplets, namely
\begin{multline}
    \intS[p]{\hat{\textbf{w}}_{r} \cdot  \bm\sigma_{o}\cdot \textbf{n}}
    - \intO[i]{2\textbf{e}_i : \grad\hat{\textbf{U}}}
    - (1-\lambda) \intO[i]{(\textbf{u}_{i} + \textbf{w}_r)\cdot \grad^2 \hat{\textbf{U}}}
    + \intS[p]{(\textbf{u}_{i} + \textbf{w}_r)\cdot \hat{\textbf{b}}}
    \\
    =
    \intS[p]{\textbf{w}_{r}\cdot \hat{\bm\sigma}_{o}\cdot \textbf{n}}
    - \intO[i]{2\hat{\textbf{e}}_i :\grad\textbf{U}}
    - (1-\lambda) \intO[i]{(\hat{\textbf{u}}_{i} + \hat{\textbf{w}}_r) \cdot \grad^2 \textbf{U} }\\ 
    + \intS[p]{(\hat{\textbf{u}}_{i} + \hat{\textbf{w}}_r) \cdot  \textbf{b}}
+  \lambda \intO[i]{(\hat{\textbf{u}}_i+\hat{\textbf{w}}_r)\cdot \textbf{f}_i^{tot}} 
    + Re\intO[o]{\hat{\textbf{u}}_{o}\cdot \textbf{f}_{o}}.
    \label{eq:int_final_step}
\end{multline}
Note that we used the relations, $2{\textbf{e}_i} : \hat{\textbf{E}} = 2{\textbf{e}_i} : \hat{\grad\textbf{U}}$ and $\div\textbf{E} = \frac{1}{2}\grad^2 \textbf{U}$ because the background flow is divergence free~\ref{eq:div_u}. 
Upon making the good choice for $\hat{\textbf{w}}_r$ and $\hat{\textbf{b}}$ (i.e. whether it is constant, linear, or quadratic, see~\ref{eq:w_r_expand}), one recovers on the left-hand side of~\ref{eq:int_final_step} the expression of the force, first, and second moments applied on the droplet in a yet arbitrary flow described by $\textbf{w}_r$ and $\textbf{b}$, and their gradients. 

In~\ref{eq:int_final_step} recall that $\textbf{n}$ is the outward normal to the test-droplet surface ($\Gamma_p$), while $\Omega_o$ and $\Omega_i$ represent the domains outside and inside the test droplet, respectively.  
On the right-hand side of~\ref{eq:int_final_step} all the terms are either known quantities from the \textit{auxiliary} problem ($\hat{\bm\sigma}_{o},\hat{\textbf{e}}_i, \hat{\textbf{u}}_{i/o/r}$), or boundary conditions of the ``real'' problem ($\textbf{U},\textbf{w}_r$ and $\textbf{b}$). 
However, one exception remains, that is the inertial term $\textbf{f}_{o}$ (resp. $\textbf{f}_i^{tot}$), which is a function of the unknown velocity field $\textbf{u}_{o}$ (resp. $\textbf{u}_i$). 
Hence, to compute the right-hand side of~\ref{eq:int_final_step}, we must use some approximations allowing us to compute the last two terms of~\ref{eq:int_final_step} without the knowledge of the complete solution. 
This is the subject of the next section.

\section{Moments of forces on a translating droplet}\label{sec:compute_moments}
In this section, we provide the general formulation for the drag force, first moments of forces, and second moments of forces, as appearing in~\ref{eq:f_alpha,eq:def_sigma_eff_f}. 
Once these formulas are properly established, we will consider the case of a droplet embedded in a steady uniform flow (i.e., when $\textbf{w}_r[\textbf{r},t]=\textbf{w}_r$ is a steady and uniform vector field) at a small but finite Reynolds number.

\subsection{Formulas for the force, first, and second moments}
Let us consider the configurations of the \textit{auxiliary} problem listed~\ref{tab:textproblem}.
\begin{table}
    \caption{Configurations of the \textit{auxiliary} problem. Configuration 1 is used to compute the force; 
    Configurations 2 and 3 to compute the first moment of the force; 
    Configurations 4-6 to compute the second moment of the force;
    and Configuration 4 to compute the droplet internal shear rate.
    }
    \label{tab:textproblem}
    \begin{center}        
    \begin{tabular}{|cccc|}\hline
        Configurations & $\hat{\textbf{w}}_r(\textbf{x})$&$\hat{\textbf{b}}(\textbf{x})$ & Point source / force solutions\\ \hline
        1 & $\hat{\textbf{w}}_r[\textbf{r}] = \hat{\textbf{w}}_r$& $\hat{\textbf{b}}[\textbf{r}] = 0$ & $\textbf{Q}^{(n)}= 0$, $\textbf{R}^{(n)}= 0$\\
        2 & $\hat{\textbf{w}}_r[\textbf{r}] = \textbf{r}\cdot \grad \hat{\textbf{U}}$& $\hat{\textbf{b}}[\textbf{r}] = 0$ & $\textbf{Q}^{(n)}= 0$, $\textbf{R}^{(n)}= 0$\\
        3 & $\hat{\textbf{w}}_r[\textbf{r}] = \textbf{0}$& $\hat{\textbf{b}}[\textbf{r}] = 0$ & $Q^{(0)}$ \\
4 & $\hat{\textbf{w}}_r[\textbf{r}] = \frac{1}{2}\textbf{r}\textbf{r}\cdot \grad\grad \hat{\textbf{U}}$ & $\hat{\textbf{b}}[\textbf{r}] = 0$ & $\textbf{Q}^{(n)}= 0$, $\textbf{R}^{(n)}= 0$\\ 
        5 & $\hat{\textbf{w}}_r[\textbf{r}] = \textbf{0}$ & $\hat{\textbf{b}}[\textbf{r}] = 0$ & $\textbf{Q}^{(1)}$\\ 
        6 & $\hat{\textbf{w}}_r[\textbf{r}] = \textbf{0}$ & $\hat{\textbf{b}}[\textbf{r}] = 0$ & $\textbf{R}^{(1)}$\\ 
        7 & $\hat{\textbf{w}}_r[\textbf{r}] = 0$ & $\hat{\textbf{b}}[\textbf{r}] = \hat{\textbf{b}} $ & $\textbf{Q}^{(n)}= 0$, $\textbf{R}^{(n)}= 0$\\ 
        \hline
    \end{tabular}
    \end{center}
\end{table}
For any of configurations: 1 through 7, the expressions for $\hat{\boldsymbol{\sigma}}_{i/o}$ and $\hat{\textbf{u}}_{i/o}$ required in~\ref{eq:int_final_step} can be obtained from~\ref{eq:big_solution,eq:big_S,eq:point_source}.

Substituting configuration~1 into \ref{eq:int_final_step} and factoring out $\hat{\textbf{w}}_r$ yields,
\begin{multline}
    \intS[p]{\bm\sigma_{o}\cdot \textbf{n}}
=
    \intS[p]{\textbf{w}_{r}\cdot \mathbb{S}_{o}^{(1)}\cdot \textbf{n}}
    - \intO[i]{ \mathbb{S}_{i}^{(1)} :\grad\textbf{U}}
    - (1-\lambda) \intO[i]{(\mathbb{U}_{i}^{(1)} + \bm\delta) \cdot \grad^2 \textbf{U} }\\ 
    + \intS[p]{(\mathbb{U}_{i/o}^{(1)} + \bm\delta) \cdot  \textbf{b}}
    + \lambda  \intO[i]{(\mathbb{U}_{i}^{(1)} + \bm\delta)\cdot \textbf{f}^{tot}_i} 
    + Re\intO[o]{\mathbb{U}_{o}^{(1)}\cdot \textbf{f}_{o}},
    \label{eq:drag_force}
\end{multline}
which is a general formula for the hydrodynamic drag forces. 
Based on this formula, one can re-derive Faxen law, for example~\citep{stone2001inertial,pozrikidis2011introduction}. 

The general formula for the first moment of force may be obtained by setting configuration~2 into~\ref{eq:int_final_step}, it yields,
\begin{multline}
    \intS[p]{r_j(\bm\sigma_{o}\cdot \textbf{n})_i}
    - \intO[i]{2(\textbf{e}_i)_{ij}}
\overset{i\neq j}{=}
    \intS[p]{(\textbf{w}_{r})_l (\mathbb{S}^{(2)}_{o})_{ijkl} n_k}
    - \intO[i]{(\mathbb{S}^{(2)}_{i})_{ijkl}(\grad\textbf{U})_{kl}}\\
    + (\zeta-1) \intO[i]{(\mathbb{U}_{i}^{(2)} + \textbf{r}\bm\delta)_{ijk} (\grad^2 \textbf{U})_k }
    + \intS[p]{(\mathbb{U}_{i}^{(2)} + \textbf{r}\bm\delta)_{ijk} \text{b}_k}\\
    + \lambda \intO[i]{(\mathbb{U}_{i}^{(2)} + \textbf{r}\bm\delta)_{ijk} (\textbf{f}^{tot}_i)_k} 
    + Re\intO[o]{(\mathbb{U}_{o}^{(2)})_{ijk}(\textbf{f}_{o})_k},
    \label{eq:first_mom}
\end{multline}
It is important to note that to derive~\ref{eq:first_mom} we have factorized the equations by the tensor $(\grad \hat{\textbf{U}})_{ij}$. 
Because $(\nabla \hat{\textbf{U}})_{kk} = \nabla \cdot \hat{\textbf{U}} = 0$, taking the trace of~\ref{eq:first_mom} leads to an incorrect expression, since it would mean factoring by zero. 
Therefore, it is important to note that this equality holds only for $i \neq j$.
Hence, if we note  $M_{ij}$ the whole first moment, then~\ref{eq:first_mom} only provides its traceless part, namely:
    $M_{ij} - 1/3\delta_{ij}M_{kk}$.
Nevertheless, using configuration~3 in~\ref{eq:int_first_step0}, we obtain a formula for the trace of the first moment ($M_{kk}$), which reads, 
\begin{equation}
    \intS[p]{\textbf{r} \cdot  \bm\sigma_{o}\cdot \textbf{n}}
    =
    - 
    Re\intO[o]{\frac{\textbf{r}}{r^3}\cdot \textbf{f}_{o}}. 
    \label{eq:first_mom_Q}
\end{equation}
Indeed, the integral of $\textbf{u}_o \cdot \hat{\boldsymbol{\sigma}}_o$ over the droplet surface on the right-hand side of ~\ref{eq:first_step_out} vanishes because $\textbf{u}_o \cdot \hat{\boldsymbol{\sigma}}_o \propto \textbf{u}_o \cdot \textbf{n}$ at $r=1$, and $\textbf{u}_o$ is divergence-free. 
By adding $\delta_{ij}$ times \ref{eq:first_mom_Q} to the traceless part of \ref{eq:first_mom_Q}, one obtains an expression valid for every pair of indices $i$ and $j$, that is, the complete first-moment tensor.

Using configuration~4 (see \ref{tab:textproblem}), we obtain the general relation for the second moment of the force (in index notation),
\begin{multline}
    \frac{1}{2}\intS[p]{  (\bm\sigma_{o}\cdot \textbf{n})_i r_jr_k}
    - \intO[i]{2(\textbf{e}_i )_{ij} r_k}
\overset{i\neq j,k}{=}
    \intS[p]{(\textbf{w}_{r})_l (\textbf{S}^{(3)}_{o})_{ijklm} n_m}
    - \intO[i]{(\textbf{S}^{(3)}_{o})_{ijklm} (\grad\textbf{U})_{lm}}\\
    - (1-\lambda) \intO[i]{((\textbf{U}^{(3)}_{i})_{ijkl} + \frac{1}{2}r_kr_l\delta_{il})(\grad^2 \textbf{U})_l }
    + \intS[p]{((\textbf{U}^{(3)}_{i})_{ijkl} + \frac{1}{2}r_kr_l\delta_{il}) \text{b}_l}\\
    + \lambda \intO[i]{((\textbf{U}^{(3)}_{i})_{ijkl} + \frac{1}{2}r_kr_l\delta_{il}) (\textbf{f}_{i}^{tot})_l} 
    + Re\intO[o]{(\textbf{U}^{(3)}_{i})_{ijkl}\cdot \textbf{f}_{o}},
    \label{eq:first_formula}
\end{multline}
Where we used the relation: $\intO[i]{(\textbf{u}_i+\textbf{w}_r)}=v_p(\textbf{w}- \textbf{w}) = 0$. 
This formula is valid for a droplet immersed in an otherwise arbitrary background flow with velocity $\textbf{U}$ and an interfacial jump $\textbf{b}$. 
However, it is only valid for $i \neq j,k$. 
Indeed, we have factored out $(\nabla\nabla \hat{\textbf{U}})_{kji}$, and $(\nabla\nabla \hat{\textbf{U}})_{kii} = (\nabla\nabla \hat{\textbf{U}})_{iji} = 0$. 
Consequently, if we denote by $K_{ijk}$ the complete second moment,~\ref{eq:second_mom_text} provides only the traceless part of $K_{ijk}$ under any contraction over $i,j$ or $i,k$. 
Therefore, a separate expression is still required for the trace of~\ref{eq:first_formula}, together with a procedure to combine this trace with the previously undefined traceless part (with respect to $i,j$ and $i,k$) of $K_{ijk}$.
This procedure is detailed in ~\ref{ap:second_mom_isotopic} and yields to   
\begin{equation}
    G_{ijk} = K_{ijk}
    +  \frac{1}{8} (K_{lkl}\delta_{ij}  - 3K_{llk})\delta_{ij}  
    + \frac{1}{8} (K_{llj} -3  K_{ljl})\delta_{ik} 
    \label{eq:def_G}
\end{equation}
One can verify that taking the trace of this expression over $(i,k)$ or $(i,j)$ yields zero. 
To obtain the whole second moment from $G_{ijk}$ (i.e. from~\ref{eq:first_formula}) we use the relation, 
\begin{equation}
    K_{ijk} = G_{ijk}  + \frac{1}{8}(3K_{llk}-K_{lkl}) \delta_{ij} + \frac{1}{8}(3K_{ljl} - K_{llj}) \delta_{ik}, 
    \label{eq:def_K}
\end{equation}
where $\textbf{G}$ is defined as the traceless part of $\textbf{K}$ under contractions over the index pairs $ij$ and $ik$. 
To obtain the traces of the second moment, i.e., $K_{lkl}$ and $K_{llk}$, we can use a procedure similar to the one used for~\ref{eq:first_mom_Q}.
For the moment, we delay this procedure to the next section, where the real problem will be defined in greater detail.

Recall that $\bm\sigma_o$ and $\textbf{e}_i$ are the local stress and shear rate, relative to the bulk stress ($\bm\Sigma$), and bulk shear rate ($\textbf{E}$), respectively. 
Consequently, the left-hand side of~\ref{eq:drag_force,eq:first_mom,eq:first_mom_Q,eq:first_formula} are exactly the terms we seek in~\ref{eq:f_alpha,eq:def_sigma_eff_f}.

Although these formulas are sufficient to close the averaged momentum equations, one may be interested in the expression for the integral of $\textbf{e}_i$ over the droplet volume, independently of the integral of $\textbf{r}\bm\sigma_o\cdot \textbf{n}$, in contrast to~\ref{eq:first_mom} which provides the sum of these two terms.
Indeed, this term can be used for the calculation of the droplet deformation, see for example \citet{fintzi2025averaged}.  
Using configuration~7 of~\ref{tab:textproblem} we obtain:
\begin{multline}
    \intS[p]{(\textbf{u}_{i} + \textbf{w}_r)\textbf{r}}
    =
    \intS[p]{\textbf{w}_{r}\cdot \mathbb{S}_{o}^\text{(2-b)}\cdot \textbf{n}}
    - \intO[i]{\mathbb{S}_{o}^\text{(2-b)} :\grad\textbf{U}}
    + (\zeta-1) \intO[i]{\mathbb{U}_{i}^\text{(2-b)} \cdot \grad^2 \textbf{U} }\\ 
    + \intS[p]{\mathbb{U}_{i}^\text{(2-b)} \cdot  \textbf{b}}
    + \lambda \intO[i]{\mathbb{U}_{i}^\text{(2-b)}\cdot \textbf{f}_{i}^{tot}} 
    + Re\intO[o]{\mathbb{U}_{o}^\text{(2-b)}\cdot \textbf{f}_{o}}.
    \label{eq:internal_e}
\end{multline}
Upon using the divergence theorem on the left-hand side term, one can obtain a formula for the integral of the velocity gradient inside the droplet, whose symmetric part corresponds to the integral of $\textbf{e}_i$ over the droplet volume. 
Hence, the left-hand side of~\ref{eq:internal_e} corresponds rigorously to the integral of $\textbf{e}_i$ over the droplet volume. 

\subsection{Force moments on a droplet immersed in a steady uniform flow at low Reynolds number}
\label{sec:computation_forces}

This subsection considers only a uniform steady state relative motion in the \textit{real} problem, such that $\textbf{w}_r= \textbf{U}-\textbf{w}$ is not a function of space and time.
Furthermore, we assume clean interfaces, such that $\textbf{b}=\textbf{0}$.

\subsubsection{Oseen drag force}

In this section, we show how to obtain the $O(Re)$ correction to the drag force on a translating spherical droplet using the reciprocal theorem. 
Although this result has been known for a long time \citep{taylor1964deformation}, the present derivation allows us to introduce useful notation and to clarify the methodology employed in the subsequent sections. 
Using \ref{eq:drag_force}, we obtain
\begin{equation}
    \intS[p]{\bm\sigma_{o}\cdot \textbf{n}}
=
    \textbf{w}_{r}\cdot \intS[p]{ \mathbb{S}_{o}^{(1)}\cdot \textbf{n}}
+ \zeta Re \intO[i]{(\mathbb{U}_{i}^{(1)} + \bm\delta)\cdot \textbf{f}_{i}} 
    + Re\intO[o]{\mathbb{U}_{o}^{(1)}\cdot \textbf{f}_{o}}. 
    \label{eq:drag_force_application}
\end{equation}
The term involving the divergence of mean stress cancelled out because the latter is a constant vector in this configuration and $\intO[i] {\textbf{w}_r +\textbf{u}_i}=0$. 
The first term on the right-hand side is by definition the Stokes flow contribution to the drag, since $\textbf{w}_{r}\cdot \mathbb{S}_{o}^{(1)}$ represents the stress field of a translating spherical droplet in Stokes flow conditions~\ref{eq:big_solution}. 
Thus, 
\begin{equation}
    - \textbf{w}_{r}\cdot\intS[p]{ \mathbb{S}_{o}^{(1)}\cdot \textbf{n}}
    = - 2 \pi \frac{2+3\lambda}{\lambda+1} \textbf{w}_r
    = \textbf{F}_{s},
    \label{eq:stokes_force}
\end{equation}
where we introduced the vector $\textbf{F}_s$ which represents the Stokes force from the droplet on the continuous phase. 

The real challenge is to compute the last two integrals on the right-hand side of~\ref{eq:drag_force_application}, which require expressions for the forcing terms $\textbf{f}_o$ and $\textbf{f}_i$. 
In a situation where only steady and uniform relative motions are present, we deduce from~\ref{eq:momentum_out,eq:momentum_in} that 
\begin{equation}
    \textbf{f}_{i/o} = 
(\textbf{u}_{i/o} + \textbf{w}_r)\cdot \grad \textbf{u}_{i/o},
    \label{eq:inertial_term}
\end{equation}
where we recall that $\textbf{u}_{i/o}$ are the yet unknown velocity fields inside or outside the translating droplet. 
To obtain the $O(Re)$ correction to the drag force, one only needs to obtain a uniformly valid approximation of $\textbf{f}_{i/o}$ accurate at $O(1)$ in $Re$. 
It is well known that the fields $\textbf{u}_{i/o}$, can be approximated by $\textbf{u}_{i/o} \approx \mathbb{U}^{(1)}_{i/o}\cdot \textbf{w}_r$ only at distance $r < O(Re^{-1})$ \citep{kaplun1957,proudman1957expansions}. 
Otherwise, for $r > O(Re^{-1})$, one may use the velocity field generated by a point force in the Oseen equation, i.e., the solution of \citep{pozrikidis2011introduction}, 
\begin{align}
    -\grad p_{out} + \grad^2 \textbf{u}_{out}
    + \textbf{F}_s\delta(\textbf{r})
    &= 
    Re  \textbf{w}_r \cdot \grad \textbf{u}_{out},
    \label{eq:oseen_eq}
\end{align}
where $p_{out}$ and $\textbf{u}_{out}$ are the outer pressure and velocity fields. 
This equation may be solved in the sense of generalized functions \citep[Chapter 6]{pozrikidis2011introduction}, or with the use of Fourier transforms \citep{candelier2016settling}, it gives $\textbf{u}_{out} =  \mathbb{U}^{(out)}\cdot \textbf{F}_s$, with 
\begin{equation}
    \mathbb{U}^{(out)}
    = \frac{e^X}{4\pi r}\bm\delta
    + \grad \left(
        \frac{e^X-1}{8\pi X}
        (\textbf{w}_r - \textbf{n})
    \right), 
    \label{eq:Oseen_sol}
\end{equation}
where $X = \frac{r Re}{2} (\textbf{w}_r \cdot \textbf{n} - 1)$.
Recall that $\textbf{w}_r$ is the dimensionless relative velocity vector, hence, it is a unit vector. 
A uniformly valid solution for the disturbance velocity field outside the test droplet and for arbitrary $r$ may be obtained by combining~\ref{eq:Oseen_sol,eq:big_solution}.
It yields
\begin{equation}
    \textbf{u}_o 
    = \textbf{u}_{in}
    + Re \textbf{u}_{out}^*,
    \label{eq:patch_vel}
\end{equation} 
where $\textbf{u}_{in} = \mathbb{U}^{(1)}\cdot \textbf{w}_r$, $Re \textbf{u}_{out}^* = [\mathbb{U}^{(out)} - \mathbb{G}]\cdot \textbf{F}_s$ and $\mathbb{G} = (\bm\delta  + \textbf{nn})/(8\pi r)$ denotes the free-space Green's function of Stokes flows. 
Note that expanding~\ref{eq:Oseen_sol} and using the expression of $\mathbb{G}$ one directly gets
\begin{multline}
    \textbf{u}_{out}^* 
    =     \left\{
    \left[\frac{e^X -1}{8\pi X}
    \bm\delta 
    + \frac{e^X-X-1}{16 \pi  X^2}  
    (\textbf{nn} -\bm\delta )\right](\textbf{w}_r\cdot \textbf{n} - 1)\right.\\ \left. 
    + 
    \frac{(X-1) e^X + 1}{X^2 16 \pi} 
    (\textbf{w}_r - \textbf{n})
    (\textbf{w}_r - \textbf{n}) \right\}\cdot \textbf{F}_s. 
    \label{eq:Oseen_sol_star}
\end{multline}
From~\ref{eq:Oseen_sol_star}, one observes that in the limit $X\to 0$, or $r\to 0$ at $Re$ fixed, or equivalently $Re \to 0$ at fixed $r$, the quantity $\textbf{u}_{out}^*$ remains bounded at $O(1)$. 
For $r \to \infty$ or $X\to -\infty$ (since $\textbf{w}_r\cdot \textbf{n}-1 \leq 0$), this function behaves at most as $\sim 1/(Re r)$ depending on the values of $\textbf{w}_r\cdot \textbf{n}$. 
Using~\ref{eq:Oseen_sol_star,eq:inertial_term}, we write $\textbf{f}_o$ as
\begin{equation}
\textbf{f}_o 
= (\textbf{u}_{in}+\textbf{w}_r)\cdot \grad \textbf{u}_{in}
+ Re \textbf{w}_r\cdot \grad \textbf{u}_{out}^\ast
+ \ldots
\end{equation}
where the ellipsis denotes the remaining contributions such as $Re \textbf{u}_{in}\cdot \grad \textbf{u}_{out}$ and $Re^2 \textbf{u}_{out}\cdot \grad \textbf{u}_{out}$, which will be shown to be of order $o(Re)$ in the final result.

The last term on the right-hand side of~\ref{eq:drag_force} can now be evaluated as,  
\begin{align}
    \intO[o]{\mathbb{U}_{o}^{(1)}\cdot \textbf{f}_{o}}
    &=
    \int_{1<|\textbf{r}|<\infty }{
    \mathbb{U}_{o}^{(1)}\cdot 
    (\textbf{u}_{in} + \textbf{w}_r)\cdot \grad \textbf{u}_{in}
    }d^3\textbf{r}, \nonumber\\
    &+ 
    Re \int_{1<|\textbf{r}|<\infty}{
    \mathbb{U}_{o}^{(1)}\cdot  \textbf{w}_r \cdot \grad \textbf{u}_{out}^\ast
    }d^3\textbf{r}
    \label{eq:inertial_term_int}
\end{align}
Although all these integrals could be computed directly, it is useful to proceed to some simplifications. 
Because $\mathbb{U}_o^{(1)}$ and therefore $\textbf{u}_{in}$ are even functions of $\textbf{n}$, and consequently $\grad \textbf{u}_{in}$ an odd function of \textbf{n}, the product,  
$\mathbb{U}_{o}^{(1)}\cdot 
(\textbf{u}_{in} + \bm\delta)\cdot \grad \textbf{u}_{in}$ is odd in \textbf{n}. 
Thus, this term vanishes upon integration over the spherical unit surface centred at the origin. 
Hence, the inner solution does not contribute to the $O(Re)$ correction to the drag force. 
The integral of $ (\mathbb{U}_i^{(1)}+\bm\delta)\cdot \textbf{f}_i$ over $\Omega_i$, in~\ref{eq:drag_force}, vanishes for similar reasons. 

The second integral of~\ref{eq:inertial_term_int} represents the contribution of the outer solution. 
To evaluate the scalings with respect to $Re$ of the integrals involving $\textbf{u}_{out}^\ast$, one must consider separately the inner region where $\textbf{u}_{out}^\ast \sim O(1)$ and the outer region where $\textbf{u}_{out}^\ast \sim O(1/Re r) $. 
For example, the term in the second integral of~\ref{eq:inertial_term_int} scales as $Re O(r^{-2})  r^2 dr$ in the inner region, yielding a contribution of $O(1)$, after integration from $r=1$ to $r=Re^{-1}$, which is of course not negligible. 
However, if we only consider the part of $\mathbb{U}_o^{(1)}$ which decays as $\sim r^{-3}$, then this contribution scales as $Re O(r^{-4})  r^2 dr= Re r^{-2}$, which yields a contribution of $O(Re)$ after integration.
In the outer region now, one sees that the term which scales in $O(r^{-3})$ in $\mathbb{U}_o^{(1)}$ contributes to $Re O(Re^{-1} r^{-5})  r^2 dr= r^{-3}$, which yields a contribution of $O(Re^2)$ after integration between $Re^{-1}$ and $r\to\infty$. 
Consequently, in this integral, one may use the approximation 
\begin{equation}
    \mathbb{U}_o^{(1)} = |\textbf{F}_s| \cdot \mathbb{G} + O(r^{-3}),
    \label{eq:approx_UO}
\end{equation}
which will therefore yield an error of $O(Re^2)$ in the final result.
Using the same reasoning we see that the term $Re \mathbb{U}_o^{(1)} \cdot \textbf{u}_{in}\cdot \grad \textbf{u}_{out}^\ast r^2 dr$ scales as $\sim Re r^{-1}$ in the inner region, which integrates to $O(Re \ln Re)$, and as $\sim r^{-2}$ in the outer region which integrates to $O(Re)$, so this term and the others remain negligible. 
Using these approximations, we only need to evaluate the second integral of~\ref{eq:inertial_term_int} to obtain the drag force, namely 
\begin{align} 
    Re \textbf{F}_s   \cdot \int_{1<r<\infty}{
     \grad \textbf{u}^*_{(out)}\cdot \mathbb{G}
    }  d^3  \textbf{r}
    =
    \textbf{w}_r\frac{(\textbf{F}_s\cdot \textbf{F}_s)}{16\pi} + O(Re).
    \label{eq:ossen_force}
\end{align}
The integral in~\ref{eq:ossen_force} can be evaluated using the Fourier convolution theorem (see \citet{masoud2019reciprocal}), or by direct integration in spherical coordinates, which is the method adopted here, see~\ref{ap:more_deltais_on_symbolics}.
Finally, by using~\ref{eq:ossen_force,eq:stokes_force} in~\ref{eq:drag_force_application} we obtain the drag force on the droplet as, 
\begin{equation}
    \intS[p]{\bm\sigma_{o}\cdot \textbf{n}}
    =
    - \textbf{F}_s+ \frac{Re }{16\pi}(\textbf{F}_s\cdot \textbf{F}_s)\textbf{w}_r
=
    2\pi \frac{2+3\lambda}{\lambda+1}\textbf{w}_r+ 
    \pi \frac{(2+3\lambda)^2}{(\lambda+1)^2}
        \frac{Re }{4} \textbf{w}_r.
    \label{eq:drag_force_application_last}
\end{equation}
We recovered the classic results derived by \citet{taylor1964deformation}.

\subsubsection{First moment of force}

The first moment may be computed directly from~\ref{eq:drag_force,eq:first_mom_Q}. 
The former equation reads, 
\begin{align}
    \label{eq:first_mom_trans}
\intS[p]{r_j(\bm\sigma_{o}\cdot \textbf{n})_i}
        - \intO[i]{2(\textbf{e}_i)_{ij}}
        &\overset{i\neq j}{=}
        \zeta Re \intO[i]{(\mathbb{U}_{i}^{(2)} + \textbf{r}\bm\delta)_{ijk} (\textbf{f}_{i})_k} 
        + Re\intO[o]{(\mathbb{U}_{o}^{(2)})_{ijk}(\textbf{f}_{o})_k},
\end{align}
where we noticed that $\intS[p]{\mathbb{S}^{(2)}\cdot \textbf{n}} = 0$, whence the right-hand side of~\ref{eq:first_mom_Q,eq:first_mom_trans} only contains inertial contributions. 
This is because the Stresslet on a translating droplet in pure Stokes flow is zero \citep{kim2013microhydrodynamics}.
 
We now focus on computing the last inertial term on the right-hand side of \ref{eq:first_mom_trans}, which involves the forcing term $\textbf{f}_o$, still given by \ref{eq:inertial_term}. 
It is found that the outer contribution to this integral is negligible at leading order in $Re$, and in this case the $O(Re)$ inertial effects can be obtained through a regular perturbation procedure.
Indeed, for $r<O(Re^{-1})$, the term $Re \textbf{w}_r\cdot \grad \textbf{u}_{out}^\ast $ scales as $\sim Re r^{-1}$, while $\mathbb{U}_o^{(2)} \sim r^{-2}$. 
Consequently, the integrand associated with the stresslet scales as $O(Re r^{-3})\times r^2$, which when integrated from $1$ to $Re^{-1}$ yields a term of $O(Re \ln Re)$. 
Likewise, in the outer region, $Re \mathbb{U}_o^{(2)} \cdot \textbf{w}_r\cdot \grad \textbf{u}_{out}^\ast  r^2 dr$ scales as $O(r^{-2})$ therefore after integration from $Re^{-1}$ to $\infty$, the outer contribution to the integral is of $O(Re)$. 
The convergence of the volume integrals indicates the regular nature of the inertial contribution \citep{dabade2015}.
Similar arguments are presented in \citet{stone2001inertial} and \citet{raja2010inertial} when computing the inertial correction to the stresslet in shear-dominated flows.
Since the droplet lies entirely within the Stokes region, the Stokes flow field may be used to compute $\textbf{f}_i$, and consequently the inertial terms involving $\textbf{f}_i$ in \ref{eq:first_mom_trans}.

Based on this argument and~\ref{eq:big_solution,eq:inertial_term}, one may re-write the forcing terms as,
\begin{equation}
    \textbf{f}_{i/o}
    =
\textbf{w}_r \cdot
    (\mathbb{U}_{i/o}^{(1)} +\bm\delta)\cdot \grad \mathbb{U}_{i/o}^{(1)}\cdot \textbf{w}_r,
    \label{eq:forcing_inner}
\end{equation}
and compute the last integrals on the right-hand side of~\ref{eq:first_mom_trans,eq:first_mom_Q}, yielding: 
\begin{align}
    \label{eq:first_mom_trans_res_bis}
    \intS[p]{\textbf{r}\bm\sigma_{o}\cdot \textbf{n}}
    - \intO[i]{2\textbf{e}_i}
    &\overset{i\neq j}{=}
    -\pi Re  \frac{63 \lambda^{3} + 150 \lambda^{2} + 112 \lambda + 28}{60 (\lambda+1)^{3}} \textbf{w}_r \textbf{w}_r,
    \\ 
    \intS[p]{\textbf{r}\cdot \bm\sigma_{o}\cdot \textbf{n}}
    &=
    \pi Re  \frac{3 \lambda^2+6\lambda + 4}{12 (\lambda+1)^2} \textbf{w}_r\cdot \textbf{w}_r,
    \label{eq:first_mom_trans_res2}
\end{align}
Note that the integrals only involve tensor products of the position vector $\textbf{r}$ with the distance function $r = |\textbf{r}|$. 
We have used the open-source library \texttt{sympy} of \texttt{Python} to compute these integrals and the subsequent ones in this work.  
Specifically, the integration over the unit sphere is done using the methodology described in~\citet{barthes1973computer,pozrikidis1992boundary}. 
 
To obtain a formula for the whole first moment, we sum the deviatoric part of~\ref{eq:first_mom_trans_res_bis} with~\ref{eq:first_mom_trans_res2} times $\delta_{ij}/3$. 
It gives, 
\begin{align}
    \intS[p]{\textbf{r}\bm\sigma_{o}\cdot \textbf{n}}
    - \intO[i]{2\textbf{e}_i}
    =
    -\pi Re  \frac{63 \lambda^{3} + 150 \lambda^{2} + 112 \lambda + 28}{60 (\lambda+1)^{3}} \textbf{w}_r \textbf{w}_r,\nonumber \\ 
    + \pi Re  \frac{26 \lambda^{3} + 65 \lambda^2+54\lambda + 16}{60 (\lambda+1)^3} (\textbf{w}_r\cdot \textbf{w}_r) \bm\delta. 
    \label{eq:first_mom_trans_res}
\end{align}
Additionally, using~\ref{eq:internal_e,eq:big_solution} and similar arguments for the estimation of $\textbf{f}_{i/o}$ one find:
\begin{equation}
    \label{eq:first_mom_trans_res3}
    \intS[p]{\textbf{u}_{i} \textbf{r}}
    = 
    \intO[i]{\textbf{e}_{i}}
    =
    -\pi Re \frac{12\lambda^2+23\lambda +10}{100(\lambda+1)^3}[\textbf{w}_r\textbf{w}_r-\frac{1}{3} (\textbf{w}_r\cdot \textbf{w}_r )\bm\delta]
\end{equation}
which may be used to compute the droplet deformation in that particular situation following the procedure outlined in \citet{fintzi2025averaged}.

\subsubsection{Second moment of force}

Now, let us turn our attention to the second moment of forces.
Since $\textbf{w}_r$ is uniform and $\textbf{b}=0$, \ref{eq:first_formula} reads
\begin{multline}
    \frac{1}{2}\intS[p]{r_kr_j ( \bm\sigma_{o}\cdot \textbf{n})_i}
    - 2\intO[i]{r_k (\textbf{e}_i)_{ji}}
\overset{i\neq j,k}{=}
    (\textbf{w}_{r})_l\intS[p]{(\mathbb{S}^{(3)}_{o})_{ijklm}n_m}\\
+ \zeta Re \intO[i]{( (\mathbb{U}_{i}^{(3)})_{ijkl} + \frac{1}{2} r_kr_j\delta_{il} ) (\textbf{f}_{i})_l}
    + Re\intO[o]{(\mathbb{U}_{o}^{(3)})_{ijkl}  (\textbf{f}_{o})_l}.
    \label{eq:second_mom_text}
\end{multline}  
This formula gives the second moment of forces (left-hand side of~\ref{eq:second_mom_text}) for a  droplet translating in a uniform flow. 
As noted, previously we have factored out by $(\grad\grad\hat{\textbf{U}})_{kji}$, to derive this formula, hence~\ref{eq:second_mom_text} is only valid when $i\neq j,k$. 
To obtain the traces of the second moment, i.e., $K_{lkl}$ and $K_{llk}$, we use a procedure similar to the one used for~\ref{eq:first_mom_Q} (see~\ref{ap:second_mom_isotopic_sing}). 
Indeed, based on the singularity solutions given by~\ref{eq:pts_source_n} we may show that, 
\begin{align}
    \frac{1}{2}\intS[p]{\textbf{nn} \cdot  \bm\sigma_{o}\cdot \textbf{n}}
    =
    - \frac{1}{2}\intS[p]{ \bm\sigma_{o}\cdot \textbf{n}}
    + 3\textbf{w}_r \intO[i]{}
    + \frac{Re}{2}\intO[o]{(\bm\delta + \textbf{nn})r^{-1}\cdot \textbf{f}_{o}},
    \label{eq:second_mom_text2}
    \\
    \frac{1}{2}\intS[p]{\textbf{nn} \cdot  \bm\sigma_{o}\cdot \textbf{n}}
    - \intS[p]{2 \textbf{r}\cdot\textbf{e}_i}
    =
    \frac{1}{6}\intS[p]{\bm\sigma_{o}\cdot \textbf{n}}
    + 
    \frac{Re}{6}\intO[o]{(\bm\delta - 3\textbf{nn})r^{-3}\cdot \textbf{f}_{o}}.
    \label{eq:second_mom_text3}
\end{align}
It is interesting to note that the drag force term appears explicitly on the right-hand side of these expressions. 
Here, it should be understood that \ref{eq:second_mom_text2} corresponds to $\mathbb{K}_{llk}$, while equation~\ref{eq:second_mom_text3} corresponds to $\mathbb{K}_{lkl}$.

Whether, it is~\ref{eq:second_mom_text,eq:second_mom_text2}, or \ref{eq:second_mom_text3}, the vectors $\textbf{f}_{i/o}$ are still needed, and given by~\ref{eq:inertial_term}, with the approximation given by $\textbf{u}_{i/o}= \textbf{w}_r\cdot \mathbb{U}_{i/o}^{(1)}$ close from the test droplet, and $\textbf{u}_{out}= \textbf{F}_s\cdot \mathbb{U}^{(out)}$ in the Oseen region. 
Additionally, similarly than with~\ref{eq:approx_UO} we introduce the approximation \citep{nadim1991motion}, 
\begin{equation}
    (\mathbb{U}^{(3)}_o)_{ijkl} =  \pi \frac{\lambda}{(\lambda+1)}(\mathbb{G})_{l i}\delta_{jk} + O(r^{-3}). 
    \label{eq:U3approx}
\end{equation}
Hence, the last integrals on the right-hand side of~\ref{eq:second_mom_text,eq:second_mom_text2,eq:second_mom_text3} will be evaluated using the same approach as for~\ref{eq:drag_force_application,eq:first_mom_trans}. 
Using the same symmetry arguments as in the two previous sections, we arrive at the conclusion that in~\ref{eq:second_mom_text,eq:second_mom_text2,eq:second_mom_text3} all the internal contributions integrated over the volume of the test droplet vanish except the second term of~\ref{eq:second_mom_text2}. 
By making use of the same symmetry argument as in the drag force calculation, and of~\ref{eq:U3approx}, we deduce that in~\ref{eq:second_mom_text,eq:second_mom_text2}, only the far field contribution is relevant.
The result of the integral presented in~\ref{eq:ossen_force} is used to compute these two contributions.
In the last integral on the right-hand side of~\ref{eq:second_mom_text3}, only the inner velocity field could in principle contribute, due to the rapid $r^{-3}$ decay of the integrand. 
However, the contribution from the inner velocity field vanishes upon integration, since the final expression is odd in $\textbf{n}$.
Overall, the methodology for computing the integrals requires nothing beyond what has already been presented in the previous two subsections. 
This yields, 
\begin{align}
    \frac{1}{2}\intS[p]{r_kr_j ( \bm\sigma_{o}\cdot \textbf{n})_i}
    - 2\intO[i]{r_k (\textbf{e}_i)_{ji}}
&\overset{i\neq j,k}{=}
\pi \frac{\lambda}{\lambda+1}(\textbf{w}_r)_i \delta_{jk} 
    + Re  \pi \frac{\lambda (2+3\lambda)}{8(\lambda+1)^2} \delta_{jk} (\textbf{w}_r)_i,\\
    \frac{1}{2}\intS[p]{\textbf{nn} \cdot  \bm\sigma_{o}\cdot \textbf{n}}
&=
    \pi\frac{\lambda+2}{\lambda+1}\textbf{w}_r
    + Re \pi \frac{(2+\lambda)(2+3\lambda)}{8(\lambda+1)^2}\textbf{w}_r,   \\
    \frac{1}{2}\intS[p]{\textbf{nn} \cdot  \bm\sigma_{o}\cdot \textbf{n}}
    - \intS[p]{2 \textbf{r}\cdot \textbf{e}_i}
    &=
    \pi \frac{3\lambda +2}{3( \lambda+1) } \textbf{w}_{r}
    +
    \pi \frac{(3\lambda +2)^2}{24( \lambda+1)^2 } \textbf{w}_{r}. 
    \label{eq:second_moment_Trik}
\end{align}
Combining all contributions and using \ref{eq:def_G} and \ref{eq:def_K} yields the expression for the complete second moment of the forces, namely,
\begin{multline}
    \frac{1}{2}\intS[p]{r_kr_j ( \bm\sigma_{o}\cdot \textbf{n})_i}
    - 2\intO[i]{r_k (\textbf{e}_i)_{ji}}
    =
    \frac{\pi }{\lambda+1} [
        \frac{2}{3}\delta_{ij} (\textbf{w}_r)_k 
        + \lambda \delta_{jk} (\textbf{w}_r)_i
    ]\\ 
    +
    \frac{\pi Re(3\lambda+2)}{8(\lambda+1)^2} [
        \frac{2}{3}\delta_{ij} (\textbf{w}_r)_k 
        + \lambda \delta_{jk} (\textbf{w}_r)_i
    ]. 
    \label{eq:second_mom_final}
\end{multline}
One may note that the Stokes-flow contribution derived in~\cite{zhang1997momentum, fintzi2025averaged} is recovered in the first line of~\ref{eq:second_mom_final}.

In~\ref{ap:direct_calculation} we derived the $O(Re)$ inner pressure and velocity fields around a solid sphere in the same configuration, and derived the force moments from these fields, thereby validating the present methodology since we recover the same expressions for the forces moments, at least in the limit $\lambda\to\infty$.

\subsection{Discussion}

The formulas given by~\ref{eq:first_mom_trans_res,eq:second_mom_final} constitute the main results of the paper.
\begin{figure}[h!]
    \centering
\begin{tikzpicture}
\begin{axis}[
    xmode=log,
    xlabel={$\lambda$},
legend pos= south west,
width=10cm,
    height=6cm
]
\addplot[  domain=0.01:100, samples=50]
{ (3*x^2 +6*x +4)/(x +1)^2 /12 };
\addlegendentry{$C_{\text{Iso}}$}
\addplot[ dashed,  domain=0.01:100, samples=50]
{ -(63*x^3 + 150*x^2 +112*x +28)/(60*(x+1)^3) };
\addlegendentry{$C_{\text{Dev}}$}
\end{axis}
\end{tikzpicture}
\caption{Plots of the coefficients of the first moment of force as a function of the viscosity ratio $\lambda$. 
(solid line) Coefficient in front of the isotropic part $C_\text{Iso} = (3\lambda^2 +6\lambda +4)/(\lambda +1)^2 /12$ (see~\ref{eq:first_mom_trans_res2}). 
(dashed line) Coefficient of the deviatoric part of the first moment $C_\text{Dev} = - (63\lambda^3 +150\lambda^2 +112\lambda +28)/ (\lambda +1)^3 /60$ (see~\ref{eq:first_mom_trans_res_bis}). 
}\label{fig:stressletcoeffcient}
\end{figure}
We have plotted in~\ref{fig:stressletcoeffcient} the coefficients of the first moment as functions of the viscosity ratio. 
In this figure, $C_\text{Iso}$ denotes the coefficient associated with the isotropic contribution to the first moment~\eqref{eq:first_mom_trans_res2}, while $C_\text{Dev}$ corresponds to the coefficient of the deviatoric contribution~\eqref{eq:first_mom_trans_res_bis}.
One observes that $C_\text{Iso}$ exhibits only a weak dependence on $\lambda$, whereas $C_\text{Dev}$ is significantly larger for bubbles ($\lambda=0$) than for solid particles ($\lambda \to \infty$). 
Notably, $C_\text{Dev}$ is closely related to droplet deformation~\citep{fintzi2025statistical}, and it is shown in \citep{fintzi2026deformation} how the classical results of \citet{taylor1964deformation} for the inertial deformation of a translating droplet can be recovered. 
One may note the absence of $\zeta$ in these expressions, indicating that the droplet density has no effect on the first and second moments, at least for $Re \ll 1$, $Ca \ll 1$, and in quasi-steady state configuration.
This arises because, upon direct calculation, the first integral on the right-hand side of~\ref{eq:first_mom_trans} vanishes. 
Moreover, as shown previously, the inertial contributions integrated over the volume of the test droplet also vanish in the expressions for the second moment; consequently, the droplet density has no effect on the second moment.
A similar observation was made by \citet{magnaudet2003}, who showed that the inertial lift force acting on a droplet translating parallel to a wall does not depend on the droplet density.
However, \citet{taylor1964deformation} demonstrated that the density ratio enters the force balance through droplet deformation. 
Therefore, one may expect the density ratio to influence the force moments for a slightly deformed droplet.

 \section{Averaged zeroth, first, and second moments of forces}
\label{sec:averaged_moments}

Before presenting the averaged system of equations \ref{eq:dt_phif,eq:div_u,eq:dt_phip,eq:dt_up,eq:dt_uf2} in a (nearly) closed form, we first introduce the averaged and dimensional expressions of the force, first moment, and second moment of the forces. 
Indeed, the results derived in the previous sections still need to be converted back to dimensional form and 
integrated over all particle centre-of-mass velocities ($\textbf{w}$), see~\ref{eq:conditional_average2}.  
Hence, below the vectors $\textbf{w}_r$,$\textbf{U}_p$  and $\textbf{U}$ refer to the dimensional velocity vectors. 

First, note that all stresses appearing on the left-hand side of \ref{eq:drag_force_application_last,eq:first_mom_trans_res,eq:second_mom_final} were nondimensionalized using the stress scale $\mu_f U /a$, with lengths scaled by $a$. 
In addition, recall that the Reynolds number is defined as $(Re = U  \rho_f a / \mu_f)$.
As a result, the viscous contributions in~\ref{eq:drag_force_application_last,eq:second_mom_final} scale linearly with the relative velocity $(\textbf{w}_r = \textbf{U} - \textbf{w})$. 
Therefore, upon averaging over all values of $\textbf{w}$, one simply obtains:
\begin{equation}
    \int_{\mathbb{R}^3} P[\textbf{w}|\textbf{x},t]
    \textbf{w}_r d\textbf{w} = \textbf{U} - \textbf{U}_p = \textbf{U}_r,
    \label{eq:mean_vel}
\end{equation}
where we have introduced $\textbf{U}_r$ as the mean relative velocity between phases. 
In dimensional form the first moment of forces, given by~\ref{eq:first_mom_trans_res}, scale as $\textbf{w}_r\textbf{w}_r$.
Hence, upon averaging overall $\textbf{w}$ one obtain,   
\begin{equation}
    n_p[\textbf{x},t]\int_{\mathbb{R}^3} P[\textbf{w}|\textbf{x},t]
    \textbf{w}_r\textbf{w}_r d\textbf{w}
= 
    n_p \textbf{U}_r \textbf{U}_r
    + \pavg{\textbf{u}_\alpha'\textbf{u}_\alpha'},
    \label{eq:standard_dev1}
\end{equation}
where we recall that $\pavg{\textbf{u}_\alpha'\textbf{u}_\alpha'}$ represents the variance of the droplet center-of-mass velocity, which is responsible for the momentum flux appearing in~\ref{eq:dt_up}. 
We now turn our attention to the Oseen force and to the inertial contribution to the second moment of the force, given by \ref{eq:drag_force_application_last,eq:second_mom_final}. 
In dimensional form, these expressions involve integrals of terms proportional to $|\textbf{w}_r|\textbf{w}_r$.
Since $|\textbf{w}_r|$ is a nonlinear (power-law) function of $\textbf{w}_r$, its average cannot, for an arbitrary distribution $P[\textbf{w}|\textbf{x},t]$, be expressed solely in terms of the mean relative velocity ($\textbf{U}_r$) and $\pavg{\textbf{u}'_\alpha \textbf{u}'_\alpha}$.
Nevertheless, if we assume that $\textbf{w}$ remains close to $\textbf{U}_p$, one may perform a Taylor expansion of $\textbf{w}_r$ about $\textbf{U}_r$, leading to the approximation (see \ref{ap:varience})
\begin{equation}
    n_p  \int_{\mathbb{R}^3} 
    |\textbf{w}_r| (\textbf{w}_r)_k
    P(\textbf{w})
    d\textbf{w}
=
    n_p  
    |\textbf{U}_r|(\textbf{U}_r)_k + 
    n_p (\textbf{R}_p)_k
    +O(\pavg{(\textbf{u}_\alpha')^{(3)}})
    \label{eq:standard_dev2}
\end{equation}
with $\textbf{p} = \textbf{U}_r|\textbf{U}_r|^{-1}$ the unit vector in the direction of $\textbf{U}_r$, and, 
\begin{equation}
    n_p(\textbf{R}_p)_k = 1/2\pavg{\textbf{u}_\alpha'\textbf{u}_\alpha'}_{ij}\left(\delta _{ij} p_k +2p_j \delta_{ik}  -p_ip_jp_k\right). 
\end{equation}
The error generated by this approximation scale as $O(\pavg{(\textbf{u}_\alpha')^{(3)}})$, corresponding to the third-order joint cumulant of the distribution $P[\textbf{w}|\textbf{x},t]$. 
Note that~\ref{eq:standard_dev2} is valid for an arbitrary distribution $P[\textbf{w}|\textbf{x},t]$. 
Consequently, if $P[\textbf{w}|\textbf{x},t]$ is the joint Gaussian distribution this expression is exact, since the joint cumulants of order $n>2$ all vanish~\citep{kardar2007statistical}. 
In the case of bubbly flows, there are few measurements of the probability density function (pdf) of bubble velocities available in the literature \citep{risso2018agitation}. 
One of the only available studies is that of \citet{mercado2010}, conducted for high-Reynolds-number bubbles ($Re \approx 1000$), which showed that the bubble-velocity pdf exhibits non-Gaussian features.
To our knowledge, no equivalent measurements are available at low Reynolds number for bubbly or droplet flows. 
However, measurements of particle velocity in dilute suspensions of solid spheres at low Reynolds number indicate that the velocity pdf has a Gaussian shape \citep{bergougnoux2021dilute}.
In summary, by averaging~\ref{eq:drag_force_application_last,eq:first_mom_trans_res,eq:second_mom_final} over all \textbf{w}, and using \ref{eq:mean_vel,eq:standard_dev1,eq:standard_dev2}, we obtain the ensemble-averaged zeroth, first, and second moments of the forces, which in dimensional form read,
\begin{equation}
    \label{eq:forces_reformulated1_avg}
    \pSavg{\bm\sigma_f^*\cdot \textbf{n}}
=
    \frac{\mu_f}{a^2}  \frac{3(2+3\lambda)}{2(\lambda+1)} \phi \textbf{U}_{r,O},
\end{equation}
\begin{multline}
    \pSavg{\textbf{r}\bm\sigma_f^*\cdot \textbf{n}}
    - \pOavg{2\mu_f \textbf{e}^*_d}
    = \\
    - \rho_f  \frac{63 \lambda^{3} + 150 \lambda^{2} + 112 \lambda + 28}{80 (\lambda+1)^{3}} [\phi \textbf{U}_r \textbf{U}_r + v_p \pavg{\textbf{u}_\alpha'\textbf{u}_\alpha'} ]  \\ 
+ \rho_f   \frac{26\lambda^3 + 65 \lambda^2 + 54\lambda + 16}{80 (\lambda+1)^3} (\phi \textbf{U}_r\cdot \textbf{U}_r+ v_p \pavg{\textbf{u}_\alpha'\cdot \textbf{u}_\alpha'}) \bm\delta
    \label{eq:forces_reformulated2_avg}
\end{multline}
\begin{multline}
    \frac{1}{2}\pSavg{r_kr_j ( \bm\sigma^\ast_f\cdot \textbf{n})_i}
    - \pOavg{2 \mu_f r_k (\textbf{e}^\ast_d)_{ji}}
    =\\
    + \phi \mu_f \frac{ 3}{4(\lambda+1)} [
        \frac{2}{3}\delta_{ij} (\textbf{U}_{r,O})_k 
        + \lambda \delta_{jk} (\textbf{U}_{r,O})_i
    ]
\label{eq:forces_reformulated3_avg}
\end{multline}
where we introduced the definition: 
\begin{equation}
    \textbf{U}_{r,O}
    =
    \textbf{U}_r + 
    \frac{\rho_f a}{\mu_f} \frac{2+3\lambda}{8(\lambda+1)}[\textbf{U}_r|\textbf{U}_r| + \textbf{R}_p]. 
    \label{eq:Ur_oseen}
\end{equation}
As discussed in \citet{lhuillier1996contribution} and ~\citet[Eq. (5.35)]{fintzi2025averaged}, in the averaged momentum equation~\ref{eq:dt_uf2} the second moment of forces, noted $\mathbb{K}_{ijk}$ here~\ref{eq:forces_reformulated3_avg}, appears as, 
\begin{multline}
    K_{i(jk)}
    + K_{j(ik)}
    - K_{k(ij)}
    =
    \phi \mu_f \frac{3\lambda}{4(\lambda+1)} [\delta_{ik} (\textbf{U}_{r,O})_j + \delta_{jk} (\textbf{U}_{r,O})_i]
    + \mu_f \phi  \frac{ 2 - 3\lambda}{4(\lambda+1)}  \delta_{ij} (\textbf{U}_{r,O})_k
\label{eq:symmetry}
\end{multline}
where the $K_{i(jk)}$ represents the symmetric part of the tensor over the indices $_j$ and $_k$, \textit{i.e.} $K_{i(jk)} = 1/2 (K_{ijk}+K_{ikj})$, and so on for $K_{j(ik)}$ and $K_{k(ij)}$. 
Only the permutations of $\textbf{K}$ shown in \ref{eq:symmetry} are physically relevant, since the tensor $\textbf{K}$ appears under the double divergence operator $\partial_k \partial_j$, in~\ref{eq:dt_uf2} \citep{lhuillier1996contribution,fintzi2025averaged}.
This transformation highlights that the associated effective stress is symmetric with respect to the pair of indices $(i,j)$.

 \section{Discussion}
\label{sec:averaged_equations}
In this final section, we summarize the closures derived for the hybrid two-phase flow model and discuss their range of validity. 
We also compare our results with those available in the existing literature. 
Finally, we examine the impact of the proposed closures in the context of realistic pipe flows.

\subsection{Summary and limitations of the proposed closures for the two-fluid formulation}
\label{sec:averaged_summary}

Before presenting the proposed closure within the hybrid Euler–Euler framework, it is important to clarify its range of validity, particularly in the context of inhomogeneous flows.
So far we have considered that the droplet was steadily translating in a uniform flow at small Reynolds numbers.
The question we aim to address is under which conditions the inertial corrections associated with background shear and quadratic velocity contributions can be neglected.
Expanding the ensemble-averaged flow in a Taylor series about the position \textbf{y} of the test droplet yields,  
\begin{equation}
    \textbf{w}_r[\textbf{x}]
    =
    (\textbf{U} - \textbf{w})[\textbf{x}]
    = \textbf{w}_r[\textbf{y},t]
    + \textbf{r}\cdot \textbf{E}[\textbf{y},t]
    + \textbf{rr} : \textbf{Q}[\textbf{y},t]
    + O(\grad\grad\grad \textbf{U}),
    \label{eq:first_def_U}
\end{equation}
where $\textbf{E} = \grad \textbf{U}|_{\textbf{x} =\textbf{y}} $ and $\textbf{Q}=\grad\grad \textbf{U}|_{\textbf{x} =\textbf{y}}$ denote, respectively, the velocity gradient and the quadratic contribution of the flow evaluated at the droplet center-of-mass position. 
Note that these tensors may, in general, depend on time $t$.
Without further assumptions on $\textbf{U}[\textbf{x},t]$, the right-hand side of~\ref{eq:first_def_U} must satisfy the full Navier-Stokes equations, or, in the present framework, the system given by~\ref{eq:dt_phif,eq:div_u,eq:dt_phip,eq:dt_up,eq:dt_uf2}.  
For instance, the flow $\textbf{U} = \textbf{r}\cdot \textbf{E}$ cannot be a steady flow because $\textbf{r}\cdot \textbf{E}$ is not a steady solution of the full Navier-Stokes equations~\citep{raja2010inertial}.

The analysis presented here and detailed in~\ref{ap:scalings} requires that $Re \ll 1$, $Re_E \ll 1$, and $Re_K \ll 1$, where $Re _E = \rho_f E a^2 \mu_f$ and $Re_Q = \rho_f Q a^3 \mu_f$ are the Reynolds numbers associated with the characteristic shear rate $E$ and the characteristic quadratic contribution $Q$, respectively.
From the analysis of~\ref{ap:scalings}, see \ref{eq:scalings_adim}, it follows that inertial corrections associated with shear-induced motion may be neglected when $Re \gg (Re_E)^{1/2}$ while the quadratic contribution becomes negligible when $Re \gg (Re_Q)^{1/3}$.
The first condition corresponds to the requirement that the Oseen length ($a/Re$) remain much smaller than the Saffman length ($a/Re_E^{1/2}$)~\citep{saffman1965,mclaughlin1991inertial}.
The second condition is, \textit{a priori}, less restrictive since~\ref{eq:first_def_U} implies that $Re_E \gg Re_Q$. Consequently, if $Re \gg Re_E^{1/2}$, one generally expects that $Re \gg Re_Q^{1/3}$ also holds.

The analysis of unsteady effects is more involved (\citet{lovalenti1993}, and \ref{ap:scalings}). 
Here, we restrict attention to pure translational motion, while unsteady effects associated with pure shear will be briefly discussed in the next subsection.
In the current setup, we have assumed $S\ll Re$ (where $S = a/(\tau U)$ is the Strouhal number and $\tau$ denotes the timescale) so that unsteady effects remain negligible for distances of the order of the Oseen length or less.
Although the unsteady term may still balance the viscous term at larger distances.
However, estimating the velocity field for distances larger than the Oseen length then requires the use of the outer solution \citep{lovalenti1993}.   
When $S \sim Re$, the unsteady translational contribution cannot be neglected in the Oseen region but can still be neglected in the inner region. 
Hence, the range of validity of the proposed closures depends on whether the inertial correction originates from the inner or the outer region. 
In particular, the stresslet expression remains valid for $S \sim O(1)$, whereas the drag term (and the second moment of force) are only valid in the more restrictive regime where $S \to 0$ since there always exist arbitrarily short times at which unsteady effects are not negligible (see~\ref{ap:scalings}).
As shown by \citet{lovalenti1993}, the long-time behaviour of the history kernel depends on the underlying dynamics of the problem. 
For the configuration of interest here, namely the rise of a spherical solid particle, the history force decays as $t^{-2}$ as the solution approaches the steady Oseen limit. 
Consequently, in this particular case, the relevant timescale constraint becomes $S \ll Re^{1/2}$.
  
Under the assumptions, $Re \ll 1$, $Re \gg (Re _E)^{1/2}$, $Re \gg (Re _Q)^{1/3}$ and $S \ll Re$ the results from the Stokes regime given in \citet{fintzi2025averaged} (see also \citet{zhang1997momentum}) can simply be added with the one given in the previous sections~\ref{eq:forces_reformulated1_avg,eq:forces_reformulated2_avg,eq:forces_reformulated3_avg,eq:symmetry} and yields\footnote{In \citet{fintzi2025averaged} there is a miss print in $\bm\Sigma^f$ in the term $\frac{5\lambda +2}{ \lambda+1 } \textbf{E}$ which is corrected here.  }, 
\begin{align}
    \textbf{F}=&
    \phi
    \frac{\mu_f}{a^2}
    \frac{3(2+3\lambda)}{2(1+\lambda)}\textbf{U}_{r,O}
    + \phi\mu_f  \frac{3\lambda}{4(\lambda +1)} \grad^2 \textbf{U}
\label{eq:drag_final}
    \\
    \bm\Sigma^f 
    =&
     \mu_f \phi \frac{5\lambda +2}{ \lambda+1 } \textbf{E}\nonumber \\
    &- \mu_f \frac{3\lambda}{4(\lambda+1)} [
    \grad(\phi \textbf{U}_{r,O})
    + \grad(\phi \textbf{U}_{r,O})^\dagger]
    + \mu_f \frac{3\lambda - 2}{4(\lambda+1)} \div(\phi \textbf{U}_{r,O})  \bm\delta\nonumber\nonumber \\
    &- \rho_f  \frac{63 \lambda^{3} + 150 \lambda^{2} + 112 \lambda + 28}{80 (\lambda+1)^{3}} [\phi \textbf{U}_r \textbf{U}_r + v_p \pavg{\textbf{u}_\alpha'\textbf{u}_\alpha'} ] \nonumber \\ 
&+ \rho_f   \frac{26\lambda^3 + 65\lambda^2 + 54\lambda + 16}{80 (\lambda+1)^3} (\phi \textbf{U}_r\cdot \textbf{U}_r+ v_p \pavg{\textbf{u}_\alpha'\cdot \textbf{u}_\alpha'}) \bm\delta\nonumber \\
&-\avg{\chi_f\rho_f \textbf{u}_f'\textbf{u}_f'},
    \label{eq:sigma_feffff}
    \\
    \bm\Sigma^p 
    =&-\rho_p v_p\pavg{\textbf{u}_\alpha'\textbf{u}_\alpha'}
    \label{eq:sigma_peffff}.
\end{align}
The system of equations consisting of~\ref{eq:dt_phif,eq:div_u,eq:dt_phip,eq:dt_up,eq:dt_uf2} together with \ref{eq:drag_final,eq:sigma_feffff,eq:sigma_peffff,eq:Ur_oseen} is fully closed, except for the velocity-variance terms $\pavg{\textbf{u}'_\alpha \textbf{u}'_\alpha}$ and $\avg{\chi_f \textbf{u}'_f \textbf{u}'_f}$. 
To determine these quantities, one may either solve transport equations for the variances, as commonly done in turbulence modelling \citep{pope2001turbulent} and kinetic theory \citep{rao2008introduction}, or introduce explicit closures that relate the variances to $\textbf{U}_r$, $\lambda$, $\phi$, and their gradients, and, when relevant, the problem boundaries.
Hints on their modelling using the latter method are provided by Eq. (6.9) and (6.11) of \citet{fintzi2025averaged} and are discussed below. 
Closures for the centre-of-mass velocity variance in the Oseen regime may be found in \citet{koch1993hydrodynamic} and, more recently, in~\citet{bergougnoux2021dilute}. 
In viscous-dominated regimes, experimental evidence suggests that $\avg{|\textbf{u}_\alpha '|^2} = O(\phi^{2/3}|\textbf{U}_r|^2)$ \citep{guazzelli2011fluctuations}.
Hence, in the dilute limit $\phi \ll 1$, the centre-of-mass velocity variance may dominate the contribution arising from the mean flow, since the former scales as $\phi^{2/3}$ whereas the latter scales as $O(\phi)$. 
Consequently, without an appropriate model for the velocity variance, the first-moment closure based solely on the mean relative motion appears to be of limited relevance, at least in the dilute regime where $\phi^{2/3} \gg \phi$. The same observation applies to the drag force and the second moment of force.
Nevertheless, in regimes where pseudo-turbulence is the only source of fluctuations, it is reasonable to expect that the functional form of such closures does not introduce qualitatively new behaviours beyond those already captured by the effective stress tensor $\bm{\Sigma}^f$. 
This follows from symmetry considerations; see \citet{fintzi2025averaged}. 
Consequently, the discussion presented below should remain qualitatively valid even when such closures are included.

Owing to the rigorous averaging procedure employed here, we have shown that the velocity variance of the dispersed phase appears explicitly in the expression for the force. 
To the best of our knowledge, this point has not been raised previously, although the inclusion of relative-velocity variance in drag-force models is common in two-phase flow modeling \citep{simonin1996}. 
While the Oseen drag law is valid only over a limited range of Reynolds numbers \citep{chester1969}, the present results highlight that, when an isolated-particle model is used within an averaged framework, it is necessary to also average over the centre-of-mass velocity of the test particle. 
This procedure naturally introduces higher-order moments of the dispersed-phase velocity distribution.
For instance, the various empirical closures available in the literature for isolated particles \citep{clift2005bubbles}, when averaged over the particle centre-of-mass velocity, would acquire additional contributions proportional to the velocity variance. 
The same remarks apply to the first and second moments of the force.

A second point worth discussing is the presence of a term proportional to $\rho_f \phi \textbf{U}_r \textbf{U}_r$ in the effective stress tensor. 
This term is clearly non-negligible in buoyant droplet suspensions due to the strong relative motion between the phases. 
In particular, it may be of the same order of magnitude as the Newtonian viscous stress proportional to $\mu_f \phi\textbf{E}$ when the Reynolds number associated with the relative motion is sufficiently large. 
However, these two contributions have very different physical effects.
For example, in a vertical established flow with relative velocity only in the channel direction, the divergence of the inertial Stresslet tensor vanishes in the vertical momentum balance. 
However, the Stresslet may have an impact on the horizontal momentum balance which may in turn modify the radial distribution of $\phi$, which may then eventually impact the vertical momentum balance.

In the averaged momentum equation of the continuous phase~\eqref{eq:dt_uf2}, it is the divergence of the stress tensor that governs the dynamics. 
Accordingly, the first moment contributes through the terms $\rho_f \textbf{U}_r \textbf{U}_r \cdot \nabla \phi$ and $\rho_f \phi \nabla \cdot (\textbf{U}_r \textbf{U}_r)$. 
Therefore, in situations where strong gradients of the volume fraction or pronounced spatial variations in $\textbf{U}_r$ are present ---as is the case, for instance, in flows near a wall \citep{cox1971suspended}--- these stress contributions may play an important role.
Although as discussed previously for the vertical channel flow their physical relevance needs to be carefully examined.

In~\ref{eq:sigma_feffff} we clearly see that the second moment of force contribution, at $O(Re)$, has exactly the same functional form as in the Stokes flow regime except that the relative velocity is $\textbf{U}_{r,O}$ as defined in~\ref{eq:Ur_oseen}, instead of $\textbf{U}_r$ in Stokes flow regime.
This is an expected result as the Oseen outer solution basically increases ``the uniform background flow velocity'' seen by the droplet by that amount (i.e.~by the second term of~\ref{eq:Ur_oseen}). 
According to~\ref{eq:second_mom_text2,eq:second_mom_text3} we know that the traces of the second moment is among other contribution exactly proportional to the drag force on particles. 
The way we factorized~\ref{eq:sigma_feffff} with $\textbf{U}_{r,O}$ seems to support this argument to the whole second moment tensor. 
Thus, for future engineering practices, one could use the drag force coefficients available in the literature \citep{clift2005bubbles} to extend the validity of second-force moment to higher $Re$.

The inertial correction to the second moment of the force is proportional to the gradient of a term of the form $a\rho_f \phi\textbf{U}_r |\textbf{U}_r|$. 
Depending on the magnitudes of $\nabla \phi$ and $\nabla \textbf{U}_r$ (or $\nabla |\textbf{U}_r|$), this contribution may be of the same order of magnitude compared to the inertial contribution arising from the first moment. 
However, the scale-separation hypothesis implicitly used in this derivation assumes that these gradients vary over a characteristic length scale ($R$) that is much larger than the droplet radius. 
As a result, the inertial second moment contribution, are of order $O(a/R)$ relative to the inertial contribution of the first moment. 
Consequently, the inertial contribution of the second moment is probably negligible in practical models of suspension rheology.

Overall, the mixture momentum equation obtained by summing~\ref{eq:dt_up,eq:dt_uf2} does not correspond to a Newtonian fluid. 
This non-Newtonian behaviour arises from the Stokes contribution of the second moment, the inertial contribution of the first moment, and, to a lesser extent, the inertial contribution of the second moment. 
These stresses lead to non-Newtonian behaviour for two distinct reasons. 
First, the terms proportional to $\textbf{U}_r$ can arise independently of the mixture velocity $\textbf U$. 
For instance, in one-dimensional sedimentation where $\textbf{U}=0$ everywhere, stresses are still generated due to droplet settling, since $\textbf{U}_r \propto \textbf{g}$. 
This behaviour is non-Newtonian and is rooted in the two-phase flow nature of the present study. 
Second, the dyadic terms $\textbf{U}_r \textbf{U}_r$ and $(\textbf{U}_r \cdot \textbf{U}_r)\bm\delta$ imply that the first moment of momentum contributes to normal stress differences in the momentum equation of the continuous phase. 
This should not be confused with classical non-Newtonian shear-thickening or shear-thinning constitutive laws, which are fundamentally different, as the inertial stress identified here does not depend on the shear rate.

\subsection{Comparison with previous work}

The first-order force moment computed in this work has the same functional form as in the inviscid potential flow limit. 
Indeed, for spherical bubbles embedded in a uniform flow at high Reynolds number the first moment tensor reads \citep{biesheuvel1984two,zhang1994ensemble}, 
\begin{equation}
    a\pavg{\intS{-p\textbf{n} \textbf{n}}}
    = 
    \rho_f   \frac{2}{5} (\phi \textbf{U}_r\cdot \textbf{U}_r+ v_p \pavg{\textbf{u}_\alpha'\cdot \textbf{u}_\alpha'}) \bm\delta
    - \rho_f  \frac{9}{20} [\phi \textbf{U}_r \textbf{U}_r + v_p \pavg{\textbf{u}_\alpha'\textbf{u}_\alpha'} ].   
    \label{eq:first_mom_pot}
\end{equation}
Taking the limit ($\lambda \to 0$) in \ref{eq:forces_reformulated2_avg} yields a coefficient of -7/20 for the deviatoric part and 1/5 for the isotropic part. 
These values may be compared with the inviscid potential-flow coefficients, -9/20 and 2/5, respectively. 
To the best of the authors' knowledge, the second-order moment of the force has not been computed in the potential-flow regime.
Nevertheless, based on the symmetry of the potential flow field generated by the steady motion of a particle, the inviscid steady contribution to the second moment is expected to vanish. 
There is, however, no reason for the viscous potential contribution to the second moment to be zero, although its explicit calculation lies beyond the scope of the present work.

The functional form of the first moment is also identical to that of the particle-fluid-particle stress recently introduced by~\citet{zhang2021stress,zhang2021ensemble}. 
However, it is important to emphasize that this contribution acts on the force $\textbf{F}$ and is therefore absent from the mixture stress $\bm \Sigma ^m = \bm\Sigma^p + \bm\Sigma^f$, unlike the first moment. 
Similarly, the fluid-mediated particle interaction stresses derived in the viscous regime by~\citet{noetinger1989sedimentation,noetinger1989two,zhang2021stress} possess the same functional form as the second moments of force, namely they scale as $\grad (\textbf{U}_r \phi)$, but have a different physical meaning. 
Indeed, the second moment of force contributes directly to $\bm \Sigma ^m$, whereas the above interaction stresses contribute instead to \textbf{F}. 
In summary, the first and second moments of force represent momentum transferred from the dispersed phase to the continuous phase in the form of stresses, while the contributions introduced in the aforementioned studies describe particle interactions mediated by the continuous phase, also expressed through the divergence of a stress tensor.

It is also of interest to compare our results with the pioneering studies of~\citet{stone2001inertial,raja2010inertial}, since the present work relies heavily on their methodology. 
This comparison is worthwhile even though both the flow configurations and the objectives considered here differ from those of their studies.
\citet{raja2010inertial} computed the excess stress due to the presence of the droplets, defined as
\begin{multline}
    \bm\Sigma_\text{Raja}
    = \frac{1}{V}\sum_i^N\int_{S_i}
    [\frac{1}{2}(\bm\sigma_f^0 \textbf{r}
    + \bm\sigma_f^0 \textbf{r}^\dagger)\cdot \textbf{n}
    - (\textbf{nu}^0 + \textbf{u}^0 \textbf{n}) ]dS\\
    - \frac{1}{V}\sum_i^N\int_{V_i}
    \frac{\rho_f}{2}(\textbf{f}_d' \textbf{r}
    + \textbf{r} \textbf{f}_d') dV 
    - \frac{1}{V}\sum_i^N\int_{V}
     \rho_f \textbf{u}' \textbf{u}'dV, 
     \label{eq:raja_stress}
\end{multline}
where $S_i$ and $V_i$ are droplet surfaces and volumes, and $\textbf{f}_d'$ is the acceleration of fluid inside the droplets relative to the  volume-averaged one.
Below, we summarize the main differences between our study and the definition~\ref{eq:raja_stress} given by \citet{raja2010inertial}.  
 
(1), \ref{eq:raja_stress} corresponds to the stress of the mixture, including the contributions arising from the acceleration of the fluid inside the droplets. 
In our framework~\ref{eq:dt_up} is the equation for the averaged centre-of-mass velocity; consequently $\bm\Sigma^p$ in~\ref{eq:dt_up} does not contain the latter contribution and:
\begin{equation}
    \bm\Sigma^f+ \bm\Sigma^p \neq \bm\Sigma_\text{Raja}. 
\end{equation} 
The counterpart is that the advective term of~\ref{eq:dt_up} does not contain acceleration of the fluid inside the droplets since it is written for the centre of mass velocity. 
The differences between our framework and \citet{raja2010inertial}-like frameworks are discussed in length in~\citet{fintzi2025averaged}. 

(2) In \ref{eq:raja_stress}, the inhomogeneous stress contribution, i.e. the second moment of force is not taken into account, in contrast to \ref{eq:def_sigma_eff_f}.
This means that the closure developed by \citet{raja2010inertial} is valid only for homogeneous suspensions, whereas ours extends to weakly non-homogeneous suspensions.

(3)  The definition of the first moment of force in~\ref{eq:raja_stress} is based on $\bm\sigma_f^0$ and $\textbf{u}^0$, which differs from our own expression~\ref{eq:def_sigma_eff_f}, which is based on $\bm\sigma_f^\ast$ and $\textbf{u}^\ast$.  
Using such an expression for the computation of $\bm\Sigma_{raja}$ generates mean stress contributions, see the first line of Eq. (4.6) of~\citet{raja2010inertial}. 
In our case, these contributions are included in the factor of the mean viscous term $\phi_f \div\bm\Sigma$ in~\ref{eq:dt_uf2}. 
For a clear overview of the different formulations, see \citet{fintzi2025averaged}. 

(4) The volume-averaged framework used by \citet{raja2010inertial} and the ensemble-averaged formulation adopted here are equivalent in the limit of viscously dominated dilute suspensions \citep{jackson1997locally,zhang1997momentum}.
Nonetheless, ensemble-averaged equations remove all ``random'' structures of the flows (clusters of droplets, isotropic and random eddies, wakes of particles \ldots), and therefore lead to statistically steady equations provided that the boundary conditions are steady as well \citep{balachandar2024fundamentals}. 
By contrast, volume averaging only filters out fluctuations below the chosen averaging length scale; it does not eliminate larger-scale structures or time-dependent random fluctuations of the flow \citep{balachandar2024fundamentals}.
Hence, the two methodologies are expected to be equivalent for closure terms that depend locally on the mean flow, but not for non-local closures in space or time. 
For instance, the velocity variance is a non-local quantity, as it may depend on the size of the container \citep{guazzelli2011fluctuations} as well as on the elapsed time since the beginning of the experiment \citep{bergougnoux2021dilute}. 
In this case, an ensemble-averaged velocity variance closure is supposed to take into account all structures of the flow, including turbulence generated by sedimentation of clusters of particles or any other macroscopic and random event.
In opposition, the volume-averaged velocity variance model only takes into account the variance within the given volume delimited by the average operator at a given time, this includes pseudo-turbulence due to the wake of particles, but not necessarily macroscopic flow structures that are larger than the volume average operator length scale.
Consequently, the two approaches differ for this type of non-local closure.
Local closures, on the other hand, include the drag force and the first and second moments of force. 
These are local in the sense that the stresses at the surface of a test droplet depend only on derivatives of the ensemble-averaged velocity evaluated at the droplet centre of mass. 
Provided that the test droplet velocity is prescribed, the drag force and related moments can therefore be regarded as local closures. 
For such quantities, one may conjecture that the volume- and ensemble-averaged frameworks are equivalent.

Now, let us delve into more details on the final result \citet{raja2010inertial} obtained, namely
\begin{multline}
\bm\Sigma_\text{Raja} = 2\phi [\frac{5\lambda+2}{2(\lambda+1)}\textbf{E} + \frac{1}{10}\grad^2 \textbf{E}]\\
+ Re_E\phi \frac{2(3\lambda^2 + 3\lambda +1)}{3(\lambda +1)^2}[
    \frac{D\textbf{E}}{Dt} - \frac{2}{3}(\bm\Omega\cdot \textbf{E}+ \textbf{E}\cdot \bm\Omega)
] + O(\phi^2, \phi Re_E , \phi Ca). 
\label{eq:stress_raja_explicit}
\end{multline} 
In the above expression, $Re_E$ represents the Reynolds number based on the shear-rate (corresponding to $Re v_E$ in~\ref{ap:scalings}). 
The main difference between the present work and the studies of \citet{stone2001inertial,raja2010inertial} lies in the origin of inertia effects considered. 
Their analyses focus on the leading inertial effects induced by the ambient shear flow (the first correction in $Re_E$), whereas we consider the leading inertial effects associated with the relative motion between phases (the first correction in $Re$). 
There are also some more subtle differences in the underlying assumptions that they made.
Indeed, they assume that the unsteady flow contribution is of $O(Re_E)$, implying that the characteristic timescale associated with the shearing motion and that associated with the temporal variation of the shear are of the same order. 
In the present work, we instead had to assume that the unsteady contributions associated with translational motion are smaller than $O(Re)$.
Otherwise, unsteady terms would need to be retained in the outer flow region, thereby contributing to the history force (and the second-force moment) in a non-trivial manner \citep{lovalenti1993force,sundberg2026}.  
In the specific calculation of the Stresslet in the limit considered by \citep{raja2010inertial} only the inner field is of importance.
They obtained the unsteady contribution to the stresslet through a regular perturbation of the Stokes flow solution. 
By contrast, evaluating the unsteady contribution to the force, even in the case of a steady background shear, is a considerably more difficult problem~\citep{candelier2013}. 
Because~\citet{raja2010inertial} considered unsteady terms and the vorticity of the background shear, they ended up with a non-material frame indifference expression. 
This is because, in view of~\ref{eq:stress_raja_explicit} the terms proportional to $\bm\Omega$ and the time derivative of \textbf{E} cannot be factored out as an objective time derivative~\citep{gatignol2023thermomecanique}.
In our situation, these problems do not arise as $\textbf{U}_r$ is an objective vector and $|\textbf{U}_r|$ is an objective scalar.  

\subsection{Buoyant motion of a dispersed two-phase flow in a pipe}

To better understand how the derived stresses manifest in practical applications, we now consider the equations of motion in the specific case of a statistically steady flow in a pipe, such as a dilute laminar bubble column or a dilute pneumatic conveying system.
Let us consider a vertical circular pipe of radius $R$ containing a monodisperse suspension of droplets of radius $a$. 
We introduce the local cylindrical coordinate system $(r,\varphi,z)$, where the $z$-axis coincides with the pipe axis, while $(r,\varphi)$ define the polar coordinates in the pipe cross-section (see~\ref{fig:pipe}). 
Following the notation of \citet{happel2012low}, we denote by $(\textbf{i}_r,\textbf{i}_\varphi,\textbf{i}_z)$ the corresponding unit vectors. 
In this last section $r$ denotes $\sqrt{x^2 + y^2}$ in the Cartesian coordinate system, not to be confounded with the distance vector $r$ used up to now. 
In this configuration, gravity is aligned with the pipe axis, such that $\textbf{g} = g\textbf{i}_z$.
\begin{figure}[h!]
    \centering
    \begin{tikzpicture}
\draw  (-1.25,-2) --++ (0,4);
        \draw  (1.25,-2) --++ (0,4);
        \fill[gray!20] (-1,-2) rectangle (1,2);
        \draw[->] (-0.5,0) --++ (-2,0.5)node[left]{Effective medium $(\Omega_{in})$}; 
        \draw[->] (1.1,0) --++ (0.5,0.5)node[right]{Continuous phase only $(\Omega_{out})$};
        \draw[dotted](0,-2.2)--(0 ,2.2); 
        \draw[dashed](0,-2.2)--(0 ,2.2); 
        \draw[->](0,0)--(0,0.5)node[right]{$\textbf{i}_z$}; 
        \draw[->](0,0)--(0.5,0)node[right]{$\textbf{i}_r$}; 
        \draw[->](0,-2.3)--++(1,0)node[below, midway]{$R-a$}; 
        \draw[->](0,-2.4)--++(-1.25,0)node[midway,below]{$R $}; 
\draw[->] (1,-1)--++(0.5,0)node[right]{$\textbf{n} = \textbf{i}_r$};
    \end{tikzpicture}
    \caption{Schematic of a section of the pipe. 
    $\Omega_{out}$ denotes the region of the pipe where no droplet centre of mass can be present due to the impenetrability of the pipe wall.  
    $\Omega_{in}$ denotes the region of the flow where there is a homogeneous mixture of droplets and continuous phase. 
    }
    \label{fig:pipe}
\end{figure}

\subsubsection{General form of the governing equations and closure terms}

Considering statistically steady and fully developed flow conditions, all ensemble-averaged quantities may be assumed to depend solely on the radial coordinate $r$. 
Additionally, writing~\ref{eq:div_u,eq:dt_phip} in this system of coordinates together with the impenetrability conditions 
\begin{align}
\textbf{U} \cdot \textbf{i}_r =&0, \quad \text{at} \quad r=R, \\
\textbf{U}_p \cdot \textbf{i}_r =&0, \quad \text{at} \quad r=R-a, 
\end{align}
gives, 
\begin{align}
    \div \textbf{u} =  \partial_r (r \textbf{U}\cdot \textbf{i}_r) / r = 0, 
    && 
    \pddt \phi  + \div(\textbf{U}_p \phi )
    = \partial_r (r \phi \textbf{U}_p \cdot \textbf{i}_r) / r = 0. 
    \label{eq:div_u_dt_phip_polar}
\end{align} 
From the above boundary conditions and~\ref{eq:div_u_dt_phip_polar} we deduce that the radial component of the velocity vectors $\textbf{U}$, and $\textbf{U}_p$, vanish. 
Hence, we can write
\begin{align}
\textbf{U} = U(r)  \textbf{i}_z,
&& \textbf{U}_{p} = U_{p}(r)  \textbf{i}_z,
&& \textbf{U}_{r} = U_{r}(r)  \textbf{i}_z. 
\label{eq:def_vel_para}
\end{align} 
where we introduced the scalar functions $U,U_p,U_r$ corresponding to the axial components of the velocity vectors. 
Because the velocity vectors are oriented along $\textbf{i}_z$, the advective terms on the left-hand side of~\ref{eq:dt_uf2,eq:dt_up} vanish.

Let us now rewrite the system of equations~\ref{eq:dt_uf2,eq:dt_up} in the ``drift-velocity'' formulation 
The mixture momentum equation can be obtained by adding~\ref{eq:dt_up,eq:dt_uf2} 
\begin{equation}
    \div \bm{\Sigma}
    + \div \bm\Sigma^m 
    + \rho_f \textbf{g} [1 + \phi (\zeta - 1)] 
    = 0,
    \label{eq:dt_uf2_bis}
\end{equation}
where we have used the notation $\bm\Sigma^m = \bm\Sigma^p + \bm\Sigma^f$, and recall that $\zeta$ is the ratio of the density of the dispersed to the continuous phase. 
The dispersed phase ``relative momentum'' equation is obtained by subtracting $\phi$ times~\ref{eq:dt_uf2} from~\ref{eq:dt_up}.
This yields, 
\begin{equation}
    \div \bm\Sigma^p
    - \phi \div \bm\Sigma^m 
    + \textbf{F} 
    + \rho_f \textbf{g} \phi  (\zeta  - 1)(1 - \phi)  
    = 0.
    \label{eq:dt_up_bis}
\end{equation}
Note that the resulting equation does not contain the pressure term $P_f$ but contains the drag force \textbf{F}, which is a function of the relative velocity.
Hence, this equation is to be solved for $\textbf{U}_r$, while~\ref{eq:div_u,eq:dt_uf2_bis} for $P_f$ and \textbf{U}.

According to the discussion in~\ref{sec:averaged_summary} and the present velocity constraints~\eqref{eq:def_vel_para}, the drag force and effective stress tensors may be written as, 
\begin{align}
    \label{eq:force_symmetry}
    \textbf{F} &= F \textbf{i}_z +  L \textbf{i}_r, \\
    \label{eq:particles_stress_symmetry}
    \bm\Sigma^p &= 
    \Sigma^p_{||} \textbf{i}_z\textbf{i}_z
    + \Sigma_\bot^p (\textbf{i}_r\textbf{i}_r+\textbf{i}_\varphi\textbf{i}_\varphi),
    \\
    \label{eq:M_symmetry}
    \textbf{M} &= M_{rz} (\textbf{i}_r \textbf{i}_z+ \textbf{i}_z \textbf{i}_r)
    + M_{||} \textbf{i}_z\textbf{i}_z
    + M_\bot (\textbf{i}_r\textbf{i}_r+\textbf{i}_\varphi\textbf{i}_\varphi),
    \\
    \label{eq:K_symmetry}
    \textbf{K} &= K_\bot ( \textbf{i}_z \bm\delta +  ^\dagger \textbf{i}_z \bm\delta)
    +K_{||} \bm\delta \textbf{i}_z,
\end{align}
where \textbf{M} and \textbf{K} are defined such that $\bm\Sigma^m = \textbf{M} + \div \textbf{K}$. 
Here $F$, $\Sigma^p_{||}$, $\Sigma_\bot^p$, $M_{rz}$, $M_\bot$, $K_\bot$, and $K_{||}$ are scalar functions that can be defined as: 
\begin{align}
\label{eq:def_F}
&F = \frac{\mu_f}{a^2} C_{1,F} U_{r,O} + \mu_f C_{2,F} \frac{1}{r}\partial_r(r\partial_r U),
&& L = \rho_f C_{L} U_r^2,\\
&M_{rz} = \mu_f C_{M} \partial_r U,&& \\
&M_{||} = \rho_f U_r^2 (C_{1,M}+C_{2,M}), &&
M_\bot = \rho_f U_r^2 C_{2,M}, \label{eq:Mbot_def}\\
&\Sigma^p_{||} = \rho_f U_r^2 (C_{1,S}+C_{2,S}), &&
\Sigma_\bot^p = \rho_f U_r^2 C_{2,S},\\
&K_\bot = \mu_f C_{1,K} U_{r,O}, &&
K_{||} = \mu_f C_{2,K} U_{r,O}.  
\label{eq:def_K2}
\end{align} 
Where the $C$ coefficients denotes dimensionless functions of $\phi$, $\lambda$, $Re$, and distance from the wall. 
All terms proportional to $\mu$ are purely viscous contributions (with the exception of the term proportional to $U_{r,O}$, which contains Oseen corrections) and may therefore be derived within the framework of Stokes flow. 
By contrast, the terms proportional to $\rho$ correspond to inertial contributions.
To the best of the authors' knowledge, the complete set of coefficients $C$ is unavailable, even in the asymptotic regime considered here ($\phi \ll 1$, $Re \ll 1$, $Re \gg Re_E$, and $S \ll Re$). 
In the following, we briefly summarize the results currently available in the literature. 

The function $F$ represents the drag force given by~\eqref{eq:drag_final}, together with additional contributions arising from particle-wall interactions. 
These wall-induced corrections may be viscous in origin~\citep{kim2013microhydrodynamics} or inertial~\citep{vasseur1977,magnaudet2003}. 
The analysis of force and force-moments in the presence of walls is difficult because the corresponding coefficients depend on whether the wall is located in the outer region~\citep{vasseur1977} or the inner region~\citep{magnaudet2003}.
The function $L$ denotes the lift force, which is a purely inertial effect induced by the presence of the wall. 
In the Stokes limit, the $O(\phi)$ wall-induced lift vanishes identically as a consequence of the reversibility of the Stokes equations. 
At finite Reynolds number, however, this reversibility is lost, and a droplet translating parallel to a wall may experience a finite lift force; see \citet{magnaudet2003} for a discussion.
The tensor $\textbf{M}$ contains both velocity-variance contributions and the first force moment. 
The components $M_{rz}$ represent off-diagonal contributions associated with purely viscous stresses, corresponding in the unbounded limit to the classical Einstein-viscosity correction. 
One may expect wall effects to modify these coefficients relative to their unbounded-fluid values \citep{brenner1971}.
As discussed previously, we can assume that the velocity-variance tensors have the same tensorial structure as the inertial contribution to the first force moment; their effects are therefore incorporated into the coefficients $M_{||}$ and $M_{\bot}$. 
Significant wall-induced modifications of both particle and fluid velocity variances may also be anticipated~\citep{brenner1999}.
Finally, $\textbf{K}$ denotes the second moment of force derived in~\ref{sec:averaged_equations}. 
Although its value is also expected to depend on the distance from the wall, we are not aware of any study that has computed this quantity in the present context.

To solve~\ref{eq:dt_uf2_bis,eq:dt_up_bis}, appropriate boundary conditions must be specified. 
Near the pipe wall, no droplet centres can be located in the region $r > R-a$ because the droplets are assumed to be non-deformable (recall that the capillary number is assumed to be much smaller than unity) and impenetrable to the wall; see~\ref{fig:pipe}. 
Consequently, the particle number density $n_p(\textbf{x})$, and therefore the volume fraction $\phi(\textbf{x}) = v_p n_p(\textbf{x})$, exhibit a discontinuity at $r=R-a$.
To account for this discontinuity, we write
\begin{equation}
    \phi = \phi_{in} \chi_{in},
    \label{eq:def_phi}
\end{equation}
where $\chi_{in} = H(r - R-a)$, with $H$ denoting the Heaviside function.
The quantity $\phi_{in}$ is a continuous function equal to $\phi$ in the domain $r<R-a$. 
Taking the derivative of~\ref{eq:def_phi} yields
\begin{equation}
    \partial_r \phi(r) 
 = 
    \chi_{in}  \partial_r \phi_{in}
    + \phi_{in} \delta(r - R- a),  
\end{equation}
where $\delta$ is the Dirac delta function. 
More generally, one has~\citep{appel2007mathematics,fintzi2025averaged} 
\begin{equation}
    \grad  \phi  
 = 
    \chi_{in}  \grad \phi_{in}
    - \phi_{in} \textbf{n} \delta_S,
    \label{eq:derivative_phi}
\end{equation}
where \textbf{n} is the unit vector pointing outward the domain $\Omega_{in}$, in the current configuration $\textbf{n} = \textbf{i}_r$ (see~\ref{fig:pipe}), and $\delta_S$ is the surface indicator function defined by~\ref{eq:derivative_phi} and in~\citet{fintzi2025averaged}.

These observations show that ~\ref{eq:dt_uf2_bis,eq:dt_up_bis}, which contain terms proportional to $\nabla\phi$, are not regular equations. 
Indeed, they contain singular contributions proportional to $\phi_{in} \delta_S$. 
Accordingly, ~\ref{eq:dt_uf2_bis,eq:dt_up_bis} must be interpreted in the sense of distributions.
In ~\ref{ap:pipe_flows}, we demonstrate how these singular terms can be transformed into equivalent boundary conditions imposed at the surface $r=R-a$. 

From \ref{eq:dt_uf2_bis,eq:dt_up_bis}, we obtain six independent momentum equations together with four independent boundary conditions at $r=R-a$ (see ~\ref{ap:pipe_flows}). 
In the region $r<R-a$, the momentum balances for the mixture and dispersed phases in the axial direction $\textbf{i}_z$, as well as the corresponding momentum balances in the radial direction $\textbf{i}_r$, are given by
\begin{align}
    \label{eq:vertical_mom_u}
    - \partial_z P_f + \rho_f g [1 + \phi (\zeta - 1)] 
    + \frac{1}{r}[
    \mu_f\partial_r ( r \partial_r U  )
    + \partial_r (rM_{rz})
    + \partial_r (r \partial_r K_\bot)
    ] = 0,  \\
    \label{eq:vertical_mom_ur}
    \rho_f g \phi  (\zeta  - 1)(1 - \phi)  
    - \frac{\phi}{r} [\partial_r (rM_{rz}) + \partial_r (r \partial_r K_\bot)]
    + F = 0,\\
    \label{eq:horizontal_mom_Pf}
    \partial_r  M_\bot  - \partial_r P_f  = 0,  \\
    \label{eq:horizontal_mom_phi}
    \partial_r  \Sigma_\bot^p  -  \phi   \partial_r  M_\bot  + L = 0.
\end{align}
The appearance of $\textbf{M}$ and $\textbf{K}$ in the dispersed-phase momentum balance originates from substituting $\phi\div \bm\Sigma = -\phi \div \bm\Sigma^f + \phi \textbf{F} - \rho_f\phi  \textbf{g}$ (which follows from~\ref{eq:dt_uf2}), into ~\ref{eq:dt_up}.
The physical interpretation is therefore that the mean stress contribution, $\nabla\cdot\bm{\Sigma}$, is modified through the coefficients $M_{rz}$, $M_\bot$, and $K_\bot$ in~\ref{eq:vertical_mom_u,eq:horizontal_mom_Pf}. These modifications, in turn, alter the effective ``buoyancy'' experienced by the dispersed phase.
In the region free of droplet centers of mass, i.e., $r>R-a$, the vertical and horizontal components of the Navier-Stokes equations read,
\begin{align}
    \label{eq:vertical_mom_u_out}
    - \partial_z P_f
    + \rho_f g 
    + \frac{1}{r}[
    \mu_f\partial_r (r \partial_r U  ) 
    ] = 0,\\
    \label{eq:horizontal_mom_u_out}
    - \partial_r P_f = 0.
\end{align}
These equations are completed by boundary conditions at $r=R -a$, namely: continuity of the axial velocity, the jump conditions for the mixture momentum flux in both the $\textbf{i}_z$ and $\textbf{i}_r$ directions, and the jump condition for the particle-phase momentum flux. 
Specifically, these conditions read,
\begin{align}
    \label{eq:BC_slip}
    \mu_f (U_{out}-U_{in}) &= K_\bot, \\ 
    \label{eq:BC_stress}
    \mu_f(\partial_r U_{out} - \partial_r U_{in} ) 
    &= M_{rz} +  \partial_r   K_\bot, \\
    \label{eq:BC_pressure}
    (P_f)_{out} - (P_f)_{in}  &= - M_\bot, \\
    \label{eq:BC_pressure_parts}
 \Sigma_\bot^p &= 0,
\end{align} 
Finally, at the pipe wall ($r=R$), recall that $U=0$. 
The subscript $_{in}$, and $_{out}$, represent limiting values of $U$ and $P_f$ in the region $r<R-a$, and $r>R-a$, respectively. 

The vertical momentum equations~\ref{eq:vertical_mom_u,eq:vertical_mom_ur} are used to find $U$, and $U_r$, respectively. 
The horizontal momentum equation of the mixture~\eqref{eq:horizontal_mom_Pf} is to be solved for the pressure $P_f$, and finally, the horizontal momentum equation of the droplets~\eqref{eq:horizontal_mom_phi} allows us to find $\phi$. 
Recall that $U_r$ and $\phi$ are not always explicitly present in this system of equations but appear in the expression of $F$, $L$, $M_\bot$, $K_\bot$, and $\Sigma_\bot^p$. 
These equations are of course completed by the boundary conditions~\ref{eq:BC_slip,eq:BC_stress,eq:BC_pressure,eq:BC_pressure_parts}, and their counterpart in the domain $r>R-a$, i.e.~\ref{eq:vertical_mom_u_out,eq:horizontal_mom_u_out}. 

These equations clearly illustrate the role played by each stress and force contribution.
In particular, the off-diagonal component of the first moment, $M_{rz}$, acts as an additional viscous stress and contributes to the axial momentum balance~\eqref{eq:vertical_mom_u}, as well as to the stress-jump condition~\eqref{eq:BC_stress}. 
The normal component of the first moment, represented by $M_\bot$, introduces a radial dependence of the fluid pressure $P_f$ through~\eqref{eq:horizontal_mom_Pf}. 
It also generates a pressure jump at $r=R-a$, as described by~\eqref{eq:BC_pressure}.
The second moment of force contributes to the axial momentum balance and likewise appears in the stress-jump condition at $r=R-a$~\eqref{eq:BC_stress}. 
More importantly, one can see that the coefficient $K_\bot$ also induces a discontinuity in the velocity field through the jump condition~\eqref{eq:BC_slip}.
This latter fact has already been identified in~\citet{noetinger1989sedimentation}. 

According to~\ref{eq:vertical_mom_ur}, the relative velocity $U_r$ is primarily governed by $F$ and $\phi\nabla\cdot\bm{\Sigma}^m$. 
Consequently, all force moments influence the vertical momentum balance of the dispersed phase. 
The particle volume fraction $\phi(r)$ must be determined from the radial momentum equation~\eqref{eq:horizontal_mom_phi}. 
As shown by~\ref{eq:horizontal_mom_phi}, the distribution of $\phi$ results from the competition between the gradient of the normal stress, $\partial_r \Sigma_\bot$, the lift force $L$, and, to a lesser extent, the contribution of the dispersed phase to the mixture stress, represented by $-\phi\partial_r M_\bot$.

So in the end, the inertial contribution to the first moment modifies the continuous phase pressure $P_f$, and to a lesser extent, modifies the horizontal momentum equation of the dispersed phase through the buoyancy term $\phi\div \bm\Sigma^m$. 
The action of the second moments is two-fold: it contributes to the vertical momentum equation and, by this means, impacts the vertical momentum equations of the dispersed phase.

\subsubsection{Intrinsic convection in a suspension of rising droplets}

As discussed above, attempting to solve these equations using only the closures given in~\ref{eq:sigma_feffff} is of limited value, since the effect of the first moment of force on the radial pressure gradient has no effect on the velocity, while its contribution to the droplets' phase horizontal momentum equation is negligible at this order in volume fraction. 
Nevertheless, with full knowledge of the above remarks, we revisit the intrinsic convection problem presented in~\citet{geigenmuller1988sedimentation, bruneau1996intrinsic,guazzelli2011,barthes2012microhydrodynamics} and provide insight into the form taken by $\partial_r M_\bot$ in this situation. 

We consider the geometry given by~\ref{fig:pipe} and impose that the net mixture flow rate going through the tube is identically zero, namely, 
\begin{equation}
    2\pi \int_{r=0}^R U(r) r^2 dr = 0. 
    \label{eq:net_flow_rate_zero}
\end{equation}
As mentioned above, the determination of $\phi(r)$ solving~\ref{eq:horizontal_mom_phi} requires at least supplementary closure for $\Sigma_\bot^p$ and $L$ because these contributions are dominant compared to $M_\bot$, however, this is out of the scope of the present work. 
Instead, below we assume that $\phi_{in}(r) = \phi$ is a constant of space and solve for $U_r,U$, and $P_f$ in a section of the pipe.

Any $O(\phi^2)$ are neglected from the above equations consistently with the limit studied here. 
Consequently, we consider only the closure terms presented in~\ref{sec:averaged_summary} and neglect the velocity variance contributions in order to further simplify the problem.
Hence, the constants $C$ can be obtained by identification, by comparing~\ref{sec:averaged_summary} and~\ref{eq:def_F} to~\ref{eq:def_K}.  
Additionally, we consider an additional correction factor for the drag force term, namely, we set $F =\mu_f h C_{F,1}  U_{r,O}$ with  
\begin{equation}
    h(r)
    = 1 
    + \frac{3}{8} \frac{2+3\lambda}{2(\lambda+1)} \kappa 
    + \frac{9}{64} \frac{(2+3\lambda)^2}{4(\lambda+1)^2}  \kappa^2
    + \frac{27}{512} \left(\frac{(2+3\lambda)^3}{8(\lambda+1)^3}  - \frac{\lambda}{8(\lambda +1)}{\lambda}\right) \kappa^3
    + O(\kappa^4), 
    \label{eq:def_h_kappa}
\end{equation}
where $\kappa =  \frac{a}{R} /(1 -  \frac{r}{R})$ is the dimensionless inverse distance to the wall of the pipe. 
This correction factor represents an increase of the drag force coefficient due to the presence of the wall, see~\citet[Eq. (12)]{magnaudet2003}.   
At this stage, no theoretical results nor empirical correlations exist for the second moment of force. 
However, as discussed above, the second moment of forces seems to behave similarly as the drag force term. 
Consequently, we may use the relation $K_\bot = \mu_f h C_{K,1} U_{r,O}$ as well.  
Since it induces a lot of analytical complications and is of limited interest for the discussion that follow, we do not consider the correction factor for the Stresslet term.

We first solve the vertical momentum equation for the slip velocity $U_r$. 
Using~\ref{eq:def_F} in~\ref{eq:vertical_mom_ur} we directly obtain the equations:
\begin{equation}
    U_{r,O} +  a^2 \frac{C_{2,F}  }{C_{1,F} h} \frac{1}{r}\partial_r(r\partial_r U)
    + \frac{\rho_f g \phi  (\zeta  - 1) a^2}{\mu_f C_{1,F} h}
    = 0 
\end{equation}
with $U_{r,O} = U_r + \frac{\rho_f a}{\mu_f}\frac{2+3\lambda}{8(\lambda+1)}|U_r| U_r$.  
Using the Stokes settling velocity, noted $U_s = \frac{\rho_f g \phi  (\zeta  - 1) a^2}{\mu_f C_{1,F}}$, and the length scale $R$, we re-write this equation in dimensionless form as: 
\begin{equation}
    U_{r,O} +  \frac{a^2}{R^2} \frac{C_{2,F}}{ h C_{1,F}} \frac{1}{r}\partial_{r}(r\partial_{r} U)
    + \frac{1}{h}
    = 0.   
    \label{eq:equation_UOr}
\end{equation}
expanding $U_{r,O}$ using~\ref{eq:Ur_oseen} we end up with (considering only $U_r<0$), 
\begin{equation}
    U_r 
    - 
    Ga
    U_r^2 
    +  \frac{a^2}{R^2} \frac{C_{2,F}}{ h C_{1,F}} \frac{1}{r}\partial_r(r\partial_r U)
    + \frac{1}{h}
    = 0. 
    \label{eq:equation_ur_dimensionless}
\end{equation}
In~\ref{eq:equation_UOr,eq:equation_ur_dimensionless} and  in the following, $r,U,U_r$, and, $U_{r,O}$ refer to their dimensionless counterpart, i.e. $r/R, U/U_s, U_r/U_s$, and $U_{r,O}/U_s$, respectively. 
$Ga = \frac{\rho_f^2 g \phi  (\zeta  - 1) a^3}{\mu_f^2 C_{1,F}}\frac{2+3\lambda}{8(\lambda+1)}$ is the Galileo number of the droplets. 
The minus sign in the second term of~\ref{eq:equation_ur_dimensionless} comes from the relation $|U_r| U_r = - U_r^2$ for $U_r<0$. 
This equation can be solved straightforwardly for $U_r$ it yields,
\begin{equation}
U_r(r) = \frac{1- \sqrt{4 (\frac{a^2}{R^2} \frac{C_{2,F}}{h C_{1,F}} \frac{1}{r}\partial_r(r\partial_r U)+1/h) Ga + 1  }}{2 Ga}. 
\label{eq:sol_Ur}
\end{equation} 

In the region $\Omega_{out}$, the pressure is governed by~\ref{eq:horizontal_mom_u_out}, from which we deduce that $P_f(z,r)$ is a constant with respect to $r$, which we arbitrarily set to $0$. 
In the region $\Omega_{in}$, \ref{eq:horizontal_mom_Pf} governs the pressure, using~\ref{eq:Mbot_def} we can write this equation in dimensionless form as: 
\begin{equation}
     \partial_r (  Ga C_{2,M} U_r^2  - P_f) = 0, 
\end{equation}
where $P_f$ is made dimensionless by using the pressure scale $\frac{U_s\mu_f}{a}$.
Integrating this equation once gives 
\begin{equation}
    P_f(r)  = Ga C_{2,M} U_r^2(r). 
    \label{eq:solPf}
\end{equation}
Here recall that $U_r(r)$ is dimensionless and given by~\ref{eq:sol_Ur}. 
Note that the constant of integration in~\ref{eq:solPf} vanishes upon the use of~\ref{eq:BC_pressure} which requires $P_{in} = M_\bot$ at $r=1- a/R$ (recall that $r$ is dimensionless here). 

In a first approach the terms $\partial_r (r \partial_r K_\bot )$ and $\partial_r K_\bot$ are neglected in~\ref{eq:vertical_mom_u,eq:BC_stress}, respectively. 
It is shown below that these terms contribute nothing to the system of equations in the case where $\phi$ is constant in space. 
Using $R$ as length scale and $U_s$ as the velocity scale~\ref{eq:vertical_mom_u,eq:vertical_mom_u_out} can be written in dimensionless form as    
\begin{align}
    \label{eq:vertical_momentum_in_app}
    \Delta P + \mu^{**} 
    + \mu^* \frac{1}{r}\partial_r ( r \partial_r U  ) &= 0,\\
    \label{eq:vertical_momentum_out_app}
    \Delta P
    + \frac{1}{r} \partial_r ( r \partial_r U  ) &= 0,
\end{align}
respectively.
Here $\mu^{**}$ and $\mu^*$ are dimensionless constants defined as,  
$
\mu^{**} 
= 
    \frac{L^2}{a^2} C_{F,1}$, and  
    $\mu^* = 1 + C_M$. 
$\Delta P$ represents the yet unknown dimensionless pressure gradient in the vertical direction, including the dimensionless gravity contribution. Note that because of \ref{eq:solPf,eq:vertical_momentum_in_app}, the vertical pressure gradient, and therefore $\Delta P$, is a constant.
$\mu^*$ represents the dimensionless effective viscosity in the bulk of the suspension and $\mu^{**}$ the droplets' buoyancy contribution to the mixture momentum equation. 
In the current configuration and using~\ref{eq:equation_UOr,eq:def_K2} we have 
\begin{equation}
    \frac{K_\bot}{\mu_f U_s}
 = 
    -C_{1,K}\left(1+
     \frac{a^2}{R^2} \frac{C_{2,F}}{C_{1,F}} \frac{1}{r}\partial_r(r\partial_r U) \right) 
 =
    -C_{1,K}\left(1 
    - \frac{a^2}{R^2} \frac{C_{2,F}}{C_{1,F}} \frac{\Delta P + \mu^{**} }{\mu^*}\right), 
    \label{eq:dimensionless_K}
\end{equation}
where we have used~\ref{eq:vertical_momentum_in_app} for the second equality. 
Note that the terms on the right-hand side are all constant with respect to $r$.
Thus, according to~\ref{eq:dimensionless_K} solving~\ref{eq:vertical_momentum_in_app}, while neglecting $\partial_r K_\bot$,  yields a constant value for $K_\bot$, which is consistent with the assumption that $\partial_r K_\bot = 0$. 
Hence, this validates our first assumption regarding $K_\bot$. 

Now let's solve the system of equations given by~\ref{eq:vertical_momentum_in_app,eq:vertical_momentum_out_app}. 
Applying successive integration from $0$ to $r$ to~\ref{eq:vertical_momentum_in_app} we obtain: 
\begin{align}
U_{in}(r)  &= - \frac{\Delta P + \mu^{**} }{4 \mu^*} r^2  + U_{in}(0)
\end{align}
Similarly, applying successive integration from $r$ to $1$ to~\ref{eq:vertical_momentum_out_app}  (recall that $r$ is dimensionless here, so it varies from 0 to 1) yields 
\begin{align}
U_{out}(r)
    &= -  \frac{\Delta P}{4}(r^2 -  1 + 2 \ln\frac{1}{r})
    - \ln\frac{1}{r} \partial_r U_{out} |_{r=1}, 
\end{align}
where we have used the no-slip boundary condition, namely $U_{out}(1)  = 0$. 
The constants, $U_{in}(0)$, and $\partial_r U_{out} |_{r=1}$, corresponding to the velocity at $r=0$ and the shear rate at $r=1$, respectively, are determined by applying the boundary conditions \eqref{eq:BC_stress,eq:BC_slip}. 
In dimensionless form, they read:  
\begin{align}
U_{out}-U_{in}  
     =K_\bot/(\mu_f U_s),  && 
\partial_r U_{out} - \mu^* \partial_r U_{in}, 
    = 0
\end{align} 
which must be evaluated at $r=1 - a/R$. 
Then the value of $\Delta P$ is found by requiring that the net flow rate through the pipe vanishes, see~\ref{eq:net_flow_rate_zero}.  

\begin{figure}[h!]
    \includegraphics[height=0.24\textwidth]{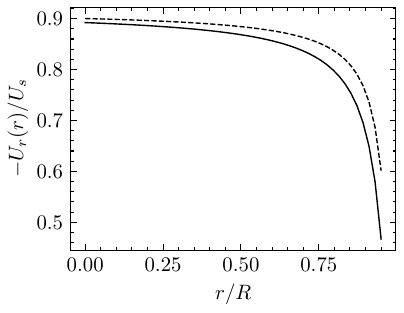}
    \includegraphics[height=0.24\textwidth]{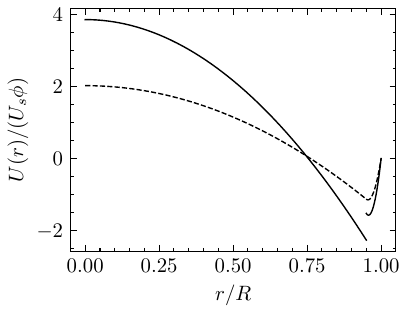}
    \includegraphics[height=0.24\textwidth]{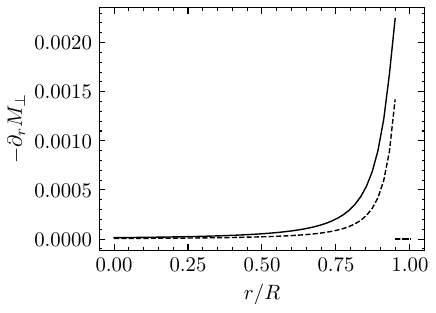}
    \caption{
        (left) Dimensionless droplet velocity relative to the mixture velocity. 
        (middle) Dimensionless mixture velocity induced due to the upward motion of the droplets. 
        (right) Dimensionless lift force due to the first-moment derivative, see~\ref{eq:horizontal_mom_phi}. 
        (solid lines) $\lambda \to\infty $
        (dashed lines) $\lambda = 0$. 
        Value of the parameters: $\phi = 0.01$, $Ga = 0.1$, $a/R = 0.05$. 
    }
    \label{fig:solution}
\end{figure}

Finding $\Delta P, U_{in}(0)$, and $\partial_r U_{out} |_{r=1}$ using these boundary conditions and~\ref{eq:net_flow_rate_zero} leads to a trivial linear system of 3 equations and 3 unknowns.
Nevertheless, the resulting expressions for the velocity and stress fields are somewhat cumbersome; therefore, we directly present the solution in~\ref{fig:solution}.  
We use the parameters $\phi = 0.01$, $Ga = 0.1$, $a/R = 0.05$ and two values of $\lambda$, namely $\lambda \to \infty$ and $\lambda =0$. 
On~\ref{fig:solution}~(left) It is found that the relative velocity is negative indicating that the bubble rise, regarding its magnitude it reaches a maximum of 0.9 at the centre due to the inertial effects ($Ga = 0.1$), and decreases on the sides due to droplet-wall interactions encoded in the function $h(\kappa)$~\eqref{eq:def_h_kappa}. 
In~\ref{fig:solution}~(middle) we observe the mean convective current generated by the upward motion of the droplets, it is found to scale as $U \sim O(U_s \phi)$, as expected in this situation~\citep{bruneau1996intrinsic}. 
$U/U_s$ is negative close to the pipe walls, then decreases until about a droplet radius (namely $a/R = 0.05$), and then it increases until it reaches a maximum at the centre of the pipe. 
One also notes that the velocity jump vanish when $\lambda = 0$ since $C_{1,K}= 3\lambda/(\lambda+1)/4 = 0$ in this case. 
Finally, we display in~\ref{fig:solution}~(right) the values of $\partial_r M_\bot$ given by~\ref{eq:Mbot_def}. 
As $\phi$ is constant $\partial_r M_\bot$ directly follows from the derivative of $U_r^2$.
We can see that it is positive close to the pipe wall and reaches $0$ as the suspension becomes homogeneous at the centre of the pipe.  
Of course the term $\phi \partial_r M_\bot$ is of $O(\phi^2)$ in~\ref{eq:horizontal_mom_phi}  and therefore negligible in the present situation compared to $\Sigma_\bot^p$ or $L$ which are at least $O(\phi^{2/3})$ and $O(\phi)$, respectively. 
Nevertheless, this solution already provides a good qualitative estimate of the behaviour of the first moment of force and its role in the horizontal momentum balance of the droplet phase in this configuration.

\section*{Acknowledgment}

The authors thank H. Stone for sharing his unpublished work \citep{stone2001inertial}, which greatly inspired the present study. 
We also thank him for his comments on an earlier draft of this manuscript. 
During the revision of this work, we came across the PhD thesis of P. M. Lovalenti. 
In the final chapter of his dissertation, he derived the deviatoric part of the first force moment for a solid spherical particle at finite Reynolds number. 
His results are in full agreement with ours.

\bibliography{bib_bulles.bib}
\appendix

\section{Points source, point force, and derivatives}
\label{ap:singularity_sol}
In this appendix we demonstrate how to obtain the solution of a point source, point forces, and higher-order derivatives of these flows. 
First, we recall the useful relation, 
\begin{equation}
    \delta(\textbf{x})
    =-\frac{1}{4\pi}\grad^2(1/r)
    =-\frac{1}{8\pi}\grad^4r.
    \label{eq:usefuk_rel}
\end{equation}
The first equality only holds in the sense of generalized functions, the second equality can be obtained directly by differentiation of $r$. 
These relations are used several times in the demonstration below. 

\paragraph{Point sources :}
We start with the point source solutions, which obey the non-homogeneous Stokes equations: 
\begin{align}
    \div\textbf{u}= 0 
    &&
    \grad^2\textbf{u}=\grad p+ (\textbf{Q}^{(n)}\odot \grad^{(n)})\grad\delta(\textbf{r})
\end{align}
Taking the divergence of the momentum equation gives,
\begin{align}
    \grad^2 p = \frac{1}{4\pi}\grad^{(n)}\grad^2 \grad^2(1/r) \odot\textbf{Q}^{(n)}
    \Longleftrightarrow
    p = \frac{1}{4\pi}\grad^{(n)}\grad^2(1/r)\odot\textbf{Q}^{(n)} = 0 
\end{align}
Hence, the momentum equation reads, 
\begin{equation}
    \grad^2\textbf{u}= -\frac{1}{4\pi } \grad^{(n+1)} \grad^2 (1/r) \odot\textbf{Q}^{(n)}
    \Longleftrightarrow
    \textbf{u}= - \frac{1}{4\pi }\grad^{(n+1)}(1/r)\odot\textbf{Q}^{(n)}
    \label{eq:pts_source}
\end{equation}
Because $\grad\textbf{u}$ is already symmetric and $p=0$ we have, 
\begin{equation}
    \bm\sigma = 2\grad \textbf{u}=- \frac{1}{4\pi }2 \grad^{(n+2)}(1/r)\odot \textbf{Q}^{(n)}. 
\end{equation}
\paragraph{Point force :}
We seek a solution for
\begin{align}
    \div\textbf{u}= 0 
    &&
    \grad^2\textbf{u}=\grad p+ (\textbf{R}^{(n+1)}\odot \grad^{(n)})\delta(\textbf{r})
\end{align}
where $\textbf{R}^{(n+1)}$ is an arbitrary $n+1$ order tensor. 
Taking the divergence of the momentum equation yields, 
\begin{align}
    \grad^2 p = - \grad^{(n+1)} \delta(\textbf{r})\odot \textbf{R}^{(n+1)}
    \Longleftrightarrow
    p= \frac{1}{4\pi} \grad^{(n+1)} (1/r)\odot \textbf{R}^{(n+1)}
\end{align}
Hence, the momentum equation can be rewritten, 
\begin{align}
    \grad^2\textbf{u}
=  \frac{1}{4\pi} \grad^{(n)}[\grad^{(2)} - \bm\delta \grad^2](1/r)
    \odot \textbf{R}^{(n+1)}
    \label{eq:lap_pts_forces}
\end{align}
Using the second equality of~\ref{eq:usefuk_rel} we directly deduce the solution of~\ref{eq:lap_pts_forces} as,
\begin{equation}
    \textbf{u}= 
    \frac{1}{8\pi} \grad^{(n)}[\grad^{(2)} - \bm\delta \grad^2] r\odot \textbf{R}^{(n+1)}. 
    \label{eq:point_force}
\end{equation}
The stress tensor then reads,
\begin{align}
    \bm\sigma
    =
    - p \bm\delta
    + \grad \textbf{u}
    + ^\dagger \grad \textbf{u}
&=
    \frac{1}{8\pi}\grad^{(n)}[2\grad^{(3)} - (\grad\bm\delta + ^\dagger\grad\bm\delta+\bm\delta\grad) \grad^2] r\odot \textbf{R}^{(n+1)}.
\end{align}
Taking $n=0$ one recovers the free space Green function of Stokes flows \citep{pozrikidis2011introduction}. 

\section{Reciprocal relation for the second moment of force}
\label{ap:second_mom_force}

\subsection{Decomposition of a third-order tensor into a traceless tensor and isotopic contributions}
\label{ap:second_mom_isotopic}
The decomposition of $K_{ijk}$ into a deviatoric part and isotopic parts, on only two indices is not a trivial task. 
Following the strategy of \citet{nadim1991motion} we consider an arbitrary third-order tensor, $K_{ijk}$, that represents the second moment of hydrodynamic forces. 
We assume that $\textbf{K}$ can be decomposed into the sum
\begin{equation}
    K_{ijk} = G_{ijk}  + (\textbf{v}^1)_k \delta_{ij} + (\textbf{v}^2)_j \delta_{ik}, 
   \label{eq:defK}
\end{equation}
where $\textbf{v}^1$ and $\textbf{v}^2$ are undetermined vectors, and $\textbf{G}$ is defined as the traceless part of $\textbf{K}$ under contractions over the index pairs $ij$ and $ik$. 
As a result, $\textbf{v}^1$ and $\textbf{v}^2$ represent the isotropic parts of $\textbf{K}$ associated with the index pairs $ij$ and $ik$, respectively. 
To express $\textbf{v}^{1,2}$ in terms of the components of $\textbf{K}$, we successively take the trace of~\ref{eq:defK} over $ij$ and $ik$, which yields the following system of equations
\begin{align}
    K_{llk} - 3(\textbf{v}^1)_i - (\textbf{v}^2)_i &= 0,\\
    K_{lil} - (\textbf{v}^1)_i - 3(\textbf{v}^2)_i &= 0,\\
\end{align}
From which we deduce that, 
\begin{align}
    (\textbf{v}^1) &=  -\frac{1}{8}K_{ljl} + \frac{3}{8}K_{llj}, \\
    (\textbf{v}^2) &=  \frac{3}{8}K_{lkl} - \frac{1}{8}K_{llk}.
\end{align}
Finally, we can define the traceless part of $\textbf{K}$ (on only two pairs of indices), as
\begin{equation}
    G_{ijk} = 
    K_{ijk}
    -\left(\frac{3}{8}\right)K_{ljl}\delta_{ik} - \left(\frac{3}{8}\right)K_{llk}\delta_{ij}  + \left(\frac{1}{8}\right)K_{lkl}\delta_{ij} + \left(\frac{1}{8}\right)K_{llj}\delta_{ik}
    \label{eq:defG}
\end{equation}
Therefore, the formulas derived for the second moment in \ref{eq:first_formula} provide only the tensor $G_{ijk}$ as defined in \ref{eq:defG}, and not the complete second-moment tensor $K_{ijk}$. 
To recover the full second moment, expressions for $\textbf{v}^1$ and $\textbf{v}^2$ are still required. 
These vectors are entirely determined by $K_{lkl}$ and $K_{llk}$, that is, by the traces of the second-moment tensor $K_{ijk}$ over the index pairs $ik$ and $ij$, respectively.

\subsection{Reciprocal theorem for the trace of the second moment}
\label{ap:second_mom_isotopic_sing}
Formulas for $K_{ijk}\delta_{ik}$ and $K_{ijk}\delta_{ij}$, can be obtained using the point source dipole solution (\ref{eq:pts_source} with $n=1$) or the point force solution (\ref{eq:point_force} with $n=0$). 
We start with~\ref{eq:first_step_out} integrated on the exterior of the droplet: 
\begin{equation}
    -\intS[p]{\hat{\textbf{u}}_{o} \cdot  \bm\sigma_{o}\cdot \textbf{n}}
    =
    -\intS[p]{\textbf{u}_{o} \cdot \hat{\bm\sigma}_{o}\cdot \textbf{n}}
    + 
    Re\intO[o]{\hat{\textbf{u}}_{o}\cdot \textbf{f}_{o}}.
    \label{eq:first_step_trace}
\end{equation}
Then we consider the test solutions given by~\ref{eq:pts_source} with $n=1$, and~\ref{eq:point_force} with $n=0$, which read
\begin{align}
    \textbf{u} = \frac{-1}{8\pi r}(\bm\delta + \textbf{nn}),
    && \bm\sigma\cdot \textbf{n} = \frac{6}{8\pi r^{-2}}\textbf{nn},
    \label{eq:first_expr}
    \\
    \textbf{u} = \frac{-1}{4\pi r^3}(3\textbf{nn}- \bm\delta),
    && \bm\sigma\cdot \textbf{n} = \frac{- 6}{4\pi r^4} (\bm\delta - 3\textbf{nn}),
    \label{eq:second_expr}
\end{align}
respectively. 
Inserting~\ref{eq:first_expr} and~\ref{eq:second_expr} in~\ref{eq:first_step_trace}, we get
\begin{align}
    \label{eq:trace1}
    \intS[p]{(\bm\delta + \textbf{nn}) \cdot  \bm\sigma_{o}\cdot \textbf{n}}
    &=
    - 6\intS[p]{\textbf{u}_{o} \cdot \textbf{nn}}
    + 
    Re\intO[o]{(\bm\delta + \textbf{nn})r^{-1}\cdot \textbf{f}_{o}},\\
    \intS[p]{(3\textbf{nn}-\bm\delta) \cdot  \bm\sigma_{o}\cdot \textbf{n}}
    &=
    6\intS[p]{\textbf{u}_{o} \cdot (\bm\delta - 3\textbf{nn})}
    + 
    Re\intO[o]{(\bm\delta - 3\textbf{nn})r^{-3}\cdot \textbf{f}_{o}},
    \label{eq:trace2}
\end{align}
respectively. 
Using the two important identities: 
\begin{equation}
    \intS[p]{\textbf{u}_{o} \cdot \textbf{nn}}
    =
    \intO[i]{\div (\textbf{u}_{o}\textbf{n})}
    = \intO[i]{ \textbf{u}_i}
    = - \textbf{w}_r \intO[i]{},
\end{equation}
and, 
\begin{align*}
    \intS[p]{\textbf{u}_{o} \cdot (\bm\delta - 3\textbf{nn})}
    &=
    - 3 \intS[p]{\textbf{u}_{o}\cdot \textbf{nn}}
    + \intS[p]{\textbf{u}_{o} \textbf{n}\cdot \textbf{n}}\\
    &=
    \intS[p]{\textbf{u}_{o}\cdot \textbf{nn}}
    + \intS[p]{\textbf{u}_{o} \textbf{n}\cdot \textbf{n}}
    - 4 \intS[p]{\textbf{u}_{o}\cdot \textbf{nn}}\\
    &=
    \intO[i]{\grad (\textbf{u}_{i}\cdot \textbf{r})}
    + \intO[i]{\div(\textbf{n}\textbf{u}_{i})}
    - 4 \intO[i]{\div (\textbf{u}_{i}\textbf{n})}\\
    &=
    \intO[i]{\grad \textbf{u}_{i}\cdot \textbf{r}}
    + \intO[i]{\textbf{u}_{i}}
    + 3\intO[i]{\textbf{u}_{i}}
    + \intO[i]{\textbf{r}\cdot \grad\textbf{u}_{i}}
    - 4 \intO[i]{\textbf{u}_{i} }\\
    &=
    \intO[i]{\grad \textbf{u}_{i}\cdot \textbf{r}}
    + \intO[i]{\textbf{r}\cdot \grad\textbf{u}_{i}}
    \\
    &=
    \intO[i]{2 \textbf{r}\cdot \textbf{e}_i},
    \\
\end{align*}
we may reformulate~\ref{eq:trace1,eq:trace2} as,
\begin{align}
    \frac{1}{2}\intS[p]{\textbf{nn} \cdot  \bm\sigma_{o}\cdot \textbf{n}}
    &=
    - \frac{1}{2}\intS[p]{ \bm\sigma_{o}\cdot \textbf{n}}
    + 3\textbf{w}_r \intO[i]{}
    + \frac{Re}{2}\intO[o]{(\bm\delta + \textbf{nn})r^{-1}\cdot \textbf{f}_{o}}
    \label{eq:trace3}
    \\
    \frac{1}{2}\intS[p]{\textbf{nn} \cdot  \bm\sigma_{o}\cdot \textbf{n}}
    &- \intS[p]{2 \textbf{r}\cdot\textbf{e}_i}
    =
    \frac{1}{6}\intS[p]{\bm\sigma_{o}\cdot \textbf{n}}
    + 
    \frac{Re}{6}\intO[o]{(\bm\delta - 3\textbf{nn})r^{-3}\cdot \textbf{f}_{o}}.
    \label{eq:trace4}
\end{align}
The first term on the right-hand side of~\ref{eq:trace3,eq:trace4} is the drag forces acting upon the test droplet, which can be obtained using its own reciprocal relation~\ref{eq:drag_force}. 

To conclude, we have derived~\ref{eq:first_formula} and we explained above that if, $K_{ijk}$ where the whole second moment of forces, then~\ref{eq:first_formula} would represent the tensor $G_{ijk}$ in the decomposition:
\begin{equation}
    K_{ijk}
    = G_{ijk} 
    + \frac{1}{8}(3K_{ljl}
    - K_{llj})\delta_{ik}
    + \frac{1}{8}(3K_{llk} 
    - K_{lkl})\delta_{ij} 
    \label{eq:defK2}
\end{equation}
Hence, to obtain the total second moment of forces, one must then add the traces $K_{llj}$ and $K_{ljl}$ to $G_{ijk}$. 
In the previous subsection, we derived a formula for $K_{ijk}\delta_{ij}$ which is given by~\ref{eq:trace3}, and a second one for $K_{ijk}\delta_{ij}$ given by~\ref{eq:trace4}.
Therefore, combining~\ref{eq:first_formula,eq:trace3,eq:trace4} following the decomposition given by~\ref{eq:defK2}, one directly obtains a formula for the complete second moment of forces.

 \section{Calculation of the $O(Re)$ inner solution around a solid particle}
\label{ap:direct_calculation}
 
Considering the scaling derived in~\ref{ap:scalings} one finds that the $O(1)$ solution for the velocity field in the region $r>Re^{-1}$ is governed by,  
\begin{equation}
    \grad^2 \textbf{u} - \grad p + \textbf{F}_s\delta(\textbf{r}) =  Re \textbf{w}_r\cdot \grad \textbf{u},
\end{equation}
The solution may be written as $\textbf{u} = \mathbb{U}^{(out)}\cdot \textbf{F}_s$ where $\mathbb{U}^{(out)}$ is given by~\ref{eq:Oseen_sol}. 
This expression can equally be written in the form of a Taylor series for small $Re$, namely,
\begin{equation}
    \mathbb{U}^{(out)}
    = 
    \mathbb{G}
    + Re \mathbb{U}^{(out)}_1  + o(Re), 
    \label{eq:ap:Ossen_approx}
\end{equation}
where,
\begin{align}
    \mathbb{G} &= \frac{1}{8\pi r}(\textbf{nn}+\bm\delta),\\
    \mathbb{U}^{(out)}_1 &= 
     \frac{1}{32\pi} [
        (\textbf{w}_r\cdot \textbf{n} - 1)
        (\textbf{nn} +3\bm\delta)
        + (\textbf{w}_r-\textbf{n})(\textbf{w}_r-\textbf{n})
     ].
\end{align}
Here, $\mathbb{G}$ denotes the free-space Green tensor, which, by construction, matches the inner $O(1)$ solution.

The inner fields may be expanded regularly as $(\textbf{u},p) = (\textbf{u}_0,p_0) + Re (\textbf{u}_1,p_1)$. 
According to the scaling assumptions discussed in the body of the text, these fields satisfy the inhomogeneous Stokes equations
\begin{align}
    \div \textbf{u}_0         &= 0 ,
    &&& \div \textbf{u}_1      &= 0,\\
    \grad^2 \textbf{u}_0      &= \grad p_0 ,
    &&& \grad^2 \textbf{u}_1   &= \grad p_1 + (\textbf{w}_r+ \textbf{u}_0)\cdot \grad \textbf{u}_0,
    \label{eq:ap:momentum_order_1}
\end{align}
subject to the boundary conditions,
\begin{align}
    \textbf{u}_0 + \textbf{w}_r = 0, && \textbf{u}_1 = 0 ,
    \label{eq:ap:BCr1}
\end{align}
at the particle surface ($r=1$).
At $r\to\infty$ one has the matching condition,
\begin{align}
    \lim_{r\to\infty}\textbf{u}_0   &= \mathbb{G}\cdot \textbf{F}_s,
    &&& \lim_{r\to\infty}\textbf{u}_1 &= \mathbb{U}^{(out)}_1\cdot \textbf{F}_s,
    \label{eq:ap:BCinfty}
    \\
    \lim_{r\to\infty}p_0 &= 0,
    &&& \lim_{r\to\infty}p_1 &= 0,
\end{align}
where we recall that $\textbf{F}_s = - 6\pi  \textbf{w}_r + o(Re)$ is the $O(1)$ drag force from the test particle to the continuous phase. 
Additionally, note that $p_1$ goes to 0 at $r\to\infty$ because the pressure in the Oseen region is exactly given by $p_0$. 

As discussed in the main text, the order $O(1)$ inner solution may be written as,
\begin{equation}
    \textbf{u}_0 = \mathbb{U}^{(1)}_0\cdot \textbf{w}_r. 
\end{equation}
Then $(\textbf{u}_1,p_1)$ can be decomposed into a homogeneous solution, let say $(\textbf{u}^h_1,p^h_1)$ and a particular solution, noted $(\textbf{u}^p_1,p^p_1)$, which satisfy
\begin{align}
    \grad^2 \textbf{u}_1^h  &= \grad p_1^h,\\
    \label{eq:ap:momentum_order_1_bis}
    \grad^2 \textbf{u}_1^p  &= \grad p_1^p + (\textbf{w}_r + \textbf{u}_0)\cdot \grad \textbf{u}_0,
\end{align}
respectively. 
As clearly explained in~\cite{candelier2023second} and in the additional material accompanying that work, the particular solution can be obtained by using Fourier and inverse Fourier transforms. 
Indeed, since $\textbf{w}_r$ is a constant vector, \ref{eq:ap:momentum_order_1_bis} is an inhomogeneous Stokes equation where the forcing term is a linear combination of terms proportional to some powers of $r$ and n-adics of the position vector $\textbf{r}$, whose Fourier and inverse Fourier transforms are all known~\cite{candelier2023second}. 
If $\mathcal{F}\{\ldots\}$ and $\mathcal{F}^{-1}\{\ldots\}$, are the Fourier and inverse Fourier operators, $\textbf{k}$ the wave vector, with $k = |\textbf{k}|$ its norm, and $\texttt{i}$ the imaginary unit, then the particular solution is given by 
\begin{align}
    \textbf{u}_1^p
    &= 
    \mathcal{F}^{-1}\left\{
        \mathcal{F}\left\{(\textbf{w}_r + \textbf{u}_0)\cdot \grad \textbf{u}_0\right\} \cdot \frac{1}{k^2}(\frac{\textbf{kk}}{k^2} - \bm\delta)
    \right\},\\
    p_1^p
    &= 
    \mathcal{F}^{-1}\left\{
        \mathcal{F}\left\{(\textbf{w}_r + \textbf{u}_0)\cdot \grad \textbf{u}_0\right\} \cdot \texttt{i}\frac{\textbf{k}}{k^2}
    \right\}.
\end{align}
Using the Fourier transform rules provided in~\citet[Additional Material]{candelier2023second} we find: 
\begin{align*}
    \textbf{u}^p_1 &= -\left(\frac{1}{160}\right)\frac{1}{r^{4}}\left(- 30 r^{3} + 45 r^{2} + 8 r + 15\right) \textbf{r}(\textbf{w}_r \cdot \textbf{w}_r) \\
    &- \left(\frac{3}{32}\right)\frac{1}{r^{4}}\left(4 r^{3} - 3 r^{2} + 1\right) \textbf{w}_r(\textbf{w}_r\cdot \textbf{r}) \\
    &+ \left(\frac{3}{80}\right)\frac{1}{r^{6}}\left(- 5 r^{3} + 15 r^{2} + 4 r + 10\right)(\textbf{w}_r\cdot \textbf{r})^2 \textbf{r} + \left(\frac{3}{8}\right) \textbf{w}_r,\\
    p^p_1 &= -\left(\frac{1}{160}\right)\frac{1}{r^{6}}\left(90 r^{4} - 24 r^{3} + 30 r^{2} + 5\right)(\textbf{w}_r\cdot \textbf{w}_r) \\
    &+ \left(\frac{3}{160}\right)\frac{1}{r^{8}}\left(60 r^{4} - 24 r^{3} + 60 r^{2} - 5\right)(\textbf{w}_r\cdot \textbf{r})^2. 
\end{align*}
From $\textbf{u}^p_1$ and $\mathbb{U}_1^{(out)}$ we can now determine an explicit expression of the far fields boundary condition satisfied by $\textbf{u}^h_1$~\eqref{eq:ap:BCinfty}, namely:
\begin{equation}
    \lim_{r\to\infty} \textbf{u}_1^h +  \frac{3}{8}\textbf{w}_r
    = 0. 
\end{equation}
Finally, $\textbf{u}_1^h$, and $p_1^h$ can be found using Lamb's general solution and the boundary condition given by~\ref{eq:ap:BCinfty} and~\ref{eq:ap:BCr1} together with the expression of $\textbf{u}_1^p$. 
We obtain:  
\begin{align*}
    \textbf{u}_1 &= 
    -\left(- \frac{3}{8} + \frac{9}{32 r} + \frac{3}{32 r^{3}}\right) \textbf{w}_r
    - \left(\frac{3}{32}\right)\frac{1}{r^{5}}\left(4 r^{4} - 3 r^{3} + r - 2\right)
    (\textbf{r}\cdot \textbf{w}_r) \textbf{w}_r \\
    &- \left(\frac{3}{32}\right)\frac{1}{r^{7}}\left(2 r^{4} - 6 r^{3} + 3 r^{2} - 4 r + 5\right)(\textbf{r}\cdot \textbf{w}_r)^2 \textbf{r}
    - \left(\frac{9}{32}\right)\frac{1}{r^{5}}\left(r^{2} - 1\right)(\textbf{r}\cdot \textbf{w}_r) \textbf{r} \\
    &+ \left(\frac{3}{32}\right)\frac{1}{r^{5}}\left(2 r^{4} - 3 r^{3} + r^{2} - r + 1\right) (\textbf{w}_r\cdot \textbf{w}_r) \textbf{r}\\
    p_1 &= \left(- \frac{1}{32 r^{6}} -  \frac{3}{16 r^{4}} -  \frac{9}{16 r^{2}} 
    + \frac{7}{16 r^{3}} \right) (\textbf{w}_r\cdot \textbf{w}_r) 
    - \left(\frac{9}{16}\right)\frac{1}{r^{3}} (\textbf{w}_r\cdot \textbf{r})  \\
    &+ \left(\frac{3}{32}\right)\frac{1}{r^{8}}\left(12 r^{4} - 14 r^{3} + 12 r^{2} - 1\right) (\textbf{w}_r\cdot \textbf{r})^2 
\end{align*}
From $\textbf{u}_1$ and $p_1$ we can readily compute the dimensionless stress $\bm\sigma_1 = -\grad p_1 + \grad \textbf{u}_1 + \grad \textbf{u}_1^\dagger$ and therefore get an expression for the drag force, first- and second-order moments of force: 
\begin{align*}
    \int_{\Gamma_p} \bm\sigma_1\cdot \textbf{n} d\Gamma &= \frac{9}{4}\textbf{w}_r,  \\ 
    \int_{\Gamma_p} \textbf{r} (\bm\sigma_1\cdot \textbf{n}) d\Gamma &= -\frac{21}{20}\textbf{w}_r \textbf{w}_r + \frac{13}{30}(\textbf{w}_r \cdot \textbf{w}_r)\bm\delta,\\ 
    \frac{1}{2}\int_{\Gamma_p} \textbf{rr} (\bm\sigma_1\cdot \textbf{n}) d\Gamma &= 
    \frac{3}{8}\bm\delta \textbf{w}_r, 
\end{align*}
We recognize the $O(Re)$ contributions in the limit $\lambda \to\infty$ in~\ref{eq:drag_force_application_last,eq:first_mom_trans_res,eq:second_mom_final}. 

\section{Integral for the Oseen force}
\label{ap:more_deltais_on_symbolics}

In this Appendix we provide some details regarding the computation of the integral: 
\begin{align} 
    Re \textbf{F}_s   \cdot \int_{1<|\wt{\textbf{r}}|<\infty}{
     \grad \textbf{u}^*_{(out)}\cdot \mathbb{G}
    }  d^3  \textbf{r}
    =Re \textbf{F}_s \textbf{F}_s  : \int_{1<|\wt{\textbf{r}}|<\infty}{
     \grad [\mathbb{U}^{(out)} - \mathbb{G}]\cdot \mathbb{G}
    }  d^3  \textbf{r}
\end{align}
In the present configuration it is known that the inertial drag force will be collinear with $\textbf{F}_s$, and that $\textbf{F}_s$ is itself collinear with $\textbf{w}_r$ \citep{kaplun1957,taylor1964deformation}. 
Therefore, we focus in this appendix on the computation of the integral of the scalar: $\grad [\mathbb{U}^{(out)} - \mathbb{G}]\cdot \mathbb{G} \vdots \textbf{w}_r \textbf{w}_r \textbf{w}_r$. 
This is carried out in the spherical coordinate system: $( r,\theta,\omega)$, where $\textbf{w}_r\cdot \textbf{x} = r \cos \theta$. 
Additionally, let us note $\mu = (\textbf{w}_r\cdot \textbf{n} - 1)$, such that $ X = \mu Re r/2$, then we have   
\begin{multline}
    \grad [\mathbb{U}^{(out)} - \mathbb{G}]\cdot \mathbb{G} \vdots \textbf{w}_r \textbf{w}_r \textbf{w}_r  (- r^2 ) d\mu dr d\omega
    = 
    \frac{1}{128 \pi^{2} Re r^{2}} \{
        20 \mu^{2} e^{\frac{\mu Re r}{2}} \\
        - 20 \mu^{2} + \mu Re^{2} r^{2} \left(\mu^{2} + 4 \mu + 4\right) e^{\frac{\mu Re r}{2}} 
        + 40 \mu e^{\frac{\mu Re r}{2}} - 40 \mu \\
        + 2 Re r \left(\mu^{3} + 8 \mu^{2} + 12 \mu + 4\right) e^{\frac{\mu Re r}{2}} 
        + 16 e^{\frac{\mu Re r}{2}} - 16
        \} r^2 d\mu dr d\omega,
\end{multline}
which can be integrated for $Re < \wt r < \infty$, $0<\omega <2\pi$, and $0<\mu<-2$, giving directly the result: 
\begin{equation}
    \frac{\left(Re^{6} e^{Re^{2}} + 3 Re^{4} + 6 Re^{2} - 6 e^{Re^{2}} + 6\right) e^{- Re^{2}}}{12 \pi Re^{8}}
    =
    \frac{1}{16\pi} + O(Re). 
\end{equation}
 \section{Average of the inertial correction}
\label{ap:varience}

In this appendix, we provide a justification for~\ref{eq:standard_dev2}.
We begin by writing, 
\begin{equation}
    n_p[\textbf{x},t] \int_{\mathbb{R}^3} 
    |\textbf{w}_r| \textbf{w}_r
    P(\textbf{w})
    d\textbf{w}_r
    =
    n_p  
    |\textbf{U}_r|\textbf{U}_r
    + n_p  \int_{\mathbb{R}^3} 
     (|\textbf{w}_r | - |\textbf{U}_r|)\textbf{w}_r
    P(\textbf{w})
    d\textbf{w}_r,
    \label{eq:first_step_appendix}
\end{equation}
and focus on the second term on the right-hand side of this equation. 
Provided that the deviation of \textbf{w} around the mean velocity $\textbf{U}_p$ is small, one can use a Taylor expansion of the function $|\textbf{w}_r|$ around $|\textbf{U}_r|$, and write,
\begin{equation}
    |\textbf{w}_r| =   
    |\textbf{U}_r|
    + (\textbf{w}_r - \textbf{U}_r)\cdot \left. \frac{\partial  |\textbf{w}_r|}{\partial \textbf{w}_r} \right|_{\textbf{w}_r = \textbf{U}_r}
    + \frac{1}{2}(\textbf{w}_r - \textbf{U}_r)(\textbf{w}_r - \textbf{U}_r) : \left. \frac{\partial  |\textbf{w}_r|}{\partial \textbf{w}_r\partial \textbf{w}_r}  \right|_{\textbf{w}_r = \textbf{U}_r} + \ldots
\end{equation}
Using the following identities, 
\begin{align}
    \grad|\textbf{x}| =\textbf{x} |\textbf{x}|^{-1}, &&
    \grad\grad|\textbf{x}| = \bm\delta |\textbf{x}|^{-1} - \textbf{xx} |\textbf{x}|^{-3},
\end{align}
 yields, 
\begin{equation}
    |\textbf{w}_r| 
    - |\textbf{U}_r|
    =
    (\textbf{w}_r - \textbf{U}_r)\cdot \textbf{U}_r |\textbf{U}_r|^{-1}
    +\frac{1}{2}(\textbf{w}_r - \textbf{U}_r)(\textbf{w}_r - \textbf{U}_r):
    (\bm\delta |\textbf{U}_r|^{-1} - \textbf{U}_r\textbf{U}_r|\textbf{U}_r|^{-3}). 
\end{equation}
Inserting this expression into~\ref{eq:first_step_appendix}, and integrating overall $\textbf{w}$, while noting that $\textbf{w}_r - \textbf{U}_r = -\textbf{w} + \textbf{U}_p = -\textbf{w}'$ is the fluctuation of the test droplet centre of mass velocity around $\textbf{U}_p$, gives: 
\begin{multline*}
    n_p \int_{\mathbb{R}^3}(|\textbf{w}_r|  - |\textbf{U}_r|) \textbf{w}_r P(\textbf{w}) d\textbf{w} \\
=
    n_p \int_{\mathbb{R}^3}
    [
    -\textbf{w}' \cdot \textbf{U}_r |\textbf{U}_r|^{-1}
    +\frac{1}{2}\textbf{w}' \textbf{w}' :
    (\bm\delta |\textbf{U}_r|^{-1} - \textbf{U}_r\textbf{U}_r|\textbf{U}_r|^{-3})
    ] \textbf{U}_r P(\textbf{w}) d\textbf{w}\\
    +
    n_p \int_{\mathbb{R}^3}
    [
    +\textbf{w}' \textbf{w}' \cdot \textbf{U}_r |\textbf{U}_r|^{-1}
    -\frac{1}{2}\textbf{w}' \textbf{w}' \textbf{w}' :
    (\bm\delta |\textbf{U}_r|^{-1} - \textbf{U}_r\textbf{U}_r|\textbf{U}_r|^{-3})
    ]  P(\textbf{w}) d\textbf{w}.
\end{multline*}
Finally, by using the relation,
\begin{align*}
    n_p \int_{\mathbb{R}^3}\textbf{w}'\textbf{w}' P(\textbf{w}) d\textbf{w} 
    &=\pavg{\textbf{u}_\alpha'\textbf{u}_\alpha'}, 
\end{align*}
and neglecting the triple covariance terms, we get, 
\begin{align}
    &n_p  \int_{\mathbb{R}^3} 
    |\textbf{w}_r| (\textbf{w}_r)_k
    P(\textbf{w})
    d\textbf{w}
     \nonumber\\
    &= n_p  
    |\textbf{U}_r|(\textbf{U}_r)_k + \frac{1}{2}
    \pavg{\textbf{u}_\alpha'\textbf{u}_\alpha'}_{ij}
    \left(\delta _{ij} p_k +2p_j \delta_{ik}  -p_ip_jp_k\right)
    +O(\pavg{(\textbf{u}_\alpha')^{(3)}}),
\end{align}
with $\textbf{p} = \textbf{U}_r|\textbf{U}_r|^{-1}$  which proves~\ref{eq:standard_dev2}. 
\section{Validity of the closures}
\label{ap:scalings} 

This appendix provides a detailed description of the scalings regarding the inertial terms present on the right-hand side of~\ref{eq:momentum_out}. 
In this section we only consider a solid isolated test sphere, whence we drop the subscripts ``$_o$'' to enhance readability, as we consider only the region outside the test particle.  
Making \ref{eq:first_def_U} dimensionless using the velocity scale $U=|\textbf{w}_r|$, and the droplet radius $a$, yields: 
\begin{equation}
    \textbf{w}_r
    = \textbf{w}_r
    + v_E\textbf{r}\cdot \textbf{E}
    + v_Q\textbf{rr} : \textbf{Q},
\end{equation}
where $v_E =  \frac{a E}{U}$, and $v_Q = \frac{a^2 Q}{U}$, with $E = |\textbf{E}|$ and $Q=|\textbf{Q}|$.

When using regular as well as singular perturbation approaches, one estimates $\textbf{f}$ in the area close to the test droplet, based on the $O(1)$ Stokes flow solution of the velocity field around the test droplet immersed in the velocity field $\textbf{w}_r[\textbf{x}]$. 
According to~\ref{eq:big_solution} the latter may be written: 
\begin{equation}
    \textbf{u}  
    = 
    \mathbb{U}^{(1)}\cdot \textbf{w}_r
    + v_E\mathbb{U}^{(2)} : \textbf{E}
    + v_Q\mathbb{U}^{(3)} \vdots \textbf{Q},
    \label{eq:ap:assumptions}
\end{equation}
where we recall that the $\mathbb{U}^{(i)}$ are tensors that satisfy Stokes equations and are only functions of the dimensionless distance vector $\textbf{r} = \textbf{x}- \textbf{y}$ and its norm $r=|\textbf{r}|$. 
Recall that from the Stokes flow solution one has: $\mathbb{U}^{(1)}\sim r^{-1}$, $\mathbb{U}^{(2)}\sim r^{-2}$, and $\mathbb{U}^{(3)}\sim r^{-1}$. 

Let us assume that~\ref{eq:ap:assumptions} is a good estimate to evaluate the advective term at $O(Re)$. 
Injecting~\ref{eq:ap:assumptions} into the inertia term, $\textbf{f}$, one gets 
\begin{align}
    \label{eq:ap:full_forcing}
    Re \textbf{f}
    &= Re [
    S  \pddt \textbf{u}_U
    + S_E v_E \pddt \textbf{u}_E
    + S_Q v_Q \pddt \textbf{u}_Q] \nonumber  \\
    &+ Re  (\textbf{u}_U + \textbf{w}_r) \cdot \grad \textbf{u}_U \nonumber \\
    &+ Re v_E^2 [\textbf{u}_E \cdot \textbf{E}
    + (\textbf{u}_E+\textbf{r}\cdot \textbf{E}) \cdot \grad \textbf{u}_E]  \nonumber \\
    &+ Re v_Q^2 [
        2 \textbf{u}_Q \cdot (\textbf{Q}\cdot \textbf{r})
    + (\textbf{u}_Q+\textbf{rr}:\textbf{Q}) \cdot \grad \textbf{u}_Q
    ] \nonumber \\
    &+ Re v_E [
        \textbf{u}_U \cdot \textbf{E}
        + (\textbf{u}_U+\textbf{w}_r) \cdot \grad \textbf{u}_E 
        + (\textbf{u}_E+\textbf{r} \cdot \textbf{E}) \cdot \grad \textbf{u}_U 
    ] \nonumber \\
    &+ Re v_Q [
        2 \textbf{u}_U \cdot (\textbf{Q}\cdot \textbf{r})
        + (\textbf{u}_U + \textbf{w}_r) \cdot \grad \textbf{u}_Q 
        + (\textbf{u}_Q +\textbf{rr}: \textbf{Q}) \cdot \grad \textbf{u}_U 
    ] \nonumber \\
    &+ Re v_Q v_E [\textbf{u}_Q \cdot \textbf{E}
    + (\textbf{u}_Q+\textbf{rr}:\textbf{Q}) \cdot \grad \textbf{u}_E
    + 2\textbf{u}_E \cdot (\textbf{r}\cdot \textbf{Q})
    + (\textbf{u}_E+\textbf{r}\cdot \textbf{E}) \cdot \grad \textbf{u}_Q]. 
\end{align}
We have introduced on the first line the time timescale ratios:
\begin{align*}
    S    = \frac{a}{\tau    U}
    && S_E =  \frac{a}{\tau_E  U}
    && S_Q =  \frac{a}{\tau_Q  U}
\end{align*}
which, together with the coefficients $v_E$ and $v_Q$, compare the time scales estimated from the scaling $a/U$ to the time scales of variation of the macroscopic flows, noted $\tau,\tau_E$, and $\tau_Q$. 
In other words $S,S_E$, and $S_Q$ correspond to the ratios at which $\textbf{w}_r$, $\textbf{E}$, and $\textbf{Q}$ vary with respect to time, respectively, compared to the scale $a/U$. 

To estimate $Re\textbf{f}$ in the body of the text we have neglected all unsteady terms, as well as all terms involving $\textbf{E}$, or $\textbf{Q}$ at $O(Re)$ in the region close to the test droplet.
This required that all terms in~\ref{eq:ap:full_forcing} must be negligible compared to the advecting term on the second line which is of $O(Re)$ close to the test droplet. 
At $r=O(1)$ the velocity fields $\textbf{u}_U,\textbf{u}_E$, and $\textbf{u}_Q$ are of $O(1)$ and so are their corresponding gradients. 
In view of~\ref{eq:ap:full_forcing}, the underlying assumptions taken in this study is therefore 
\begin{align}
    Re S \ll Re,
    && Re v_E S_E \ll Re,
    && Re v_Q S_Q \ll Re,
    && Re v_E  \ll  Re,  
    && Re v_Q  \ll  Re, 
    &&
    \ldots
    \label{eq:first_contrains}
\end{align}
As expected, following these scaling arguments one can see that the only term of $O(Re)$ in~\ref{eq:ap:full_forcing} is: 
\begin{equation}
    Re \textbf{f} = Re (\textbf{u}_U + \textbf{w}_r) \cdot \grad \textbf{u}_U + o(Re). 
\end{equation}
Nevertheless, as pointed out in the pioneering studies of~\citet{kaplun1957,proudman1957expansions}, although the right-hand side of the momentum equation ($Re\textbf{f}$), is of $O(Re)$ and decreases as $r\to\infty$, the left-hand side of the momentum equation ($\div \bm\sigma$), also decays, though not necessarily at the same rate. 
As a consequence, the inertial contribution ($Re \textbf{f}$) may become of $O(1)$ compared to the viscous contribution ($\div\bm\sigma$) despite the prefactor $Re$ in $Re \textbf{f}$. 
If the inertial terms in $Re \textbf{f}$, evaluated using the presumed $O(1)$ solution ($\textbf{u}_U$,$\textbf{u}_E$, and $\textbf{u}_Q$) turns out to be of same order than the viscous stress at some distance $r$, then the predictions provided by $\textbf{u}_U$,$\textbf{u}_E$, and $\textbf{u}_Q$ are erroneous in that region.

Now let us estimate the dominant contributions appearing in~\ref{eq:ap:full_forcing} at the Oseen length scale $r=O(Re^{-1})$, it read:
{\footnotesize
\begin{align*}
                                                                                                                        &&  r = O(Re^{-1})      \\\grad^2 \textbf{u}_U                                              =              O(r^{-3})                &&  O(        Re^3)     \\\grad^2 \textbf{u}_E                                              =              O(r^{-4})                &&  O(        Re^4)     \\&&                      \\Re S   \pddt \textbf{u}_U                                              =Re S     \; O(r^{-1})             &&  O(S  Re^2)     \\Re S_E v_E  \pddt \textbf{u}_E                                           =Re S_E v_E \; O(r^{-2})             &&  O(S_E v_E  Re^3) \\Re S_Q v_Q  \pddt \textbf{u}_Q                                           =Re S_Q v_Q \; O(r^{-3})             &&  O(S_Q v_Q  Re^4) \\&&                      \\Re  \textbf{w}_r \cdot \grad \textbf{u}_U                                   =Re\; O(r^{-2})                         &&  O(Re^3)             \\&&                      \\Re v_E^2 \textbf{u}_E \cdot \textbf{E}                                      = Re v_E^2  \; O(r^{-2})                &&  O(v_E^2 Re^3)       \\Re v_E^2 \textbf{r}\cdot \textbf{E}  \cdot \grad \textbf{u}_E               = Re v_E^2  \; O(r^{-2})                &&  O(v_E^2 Re^3)       \\&&                      \\&&                      \\Re v_E \textbf{u}_U \cdot \textbf{E}                                        =Re v_E \; O(r^{-1})                    &&  O(v_E Re^2)         \\Re v_E \textbf{r}\cdot \textbf{E} \cdot \grad \textbf{u}_U                  =Re v_E \; O(r^{-1})                    &&  O(v_E Re^2)         \\&&                      \\Re v_Q 2\textbf{u}_U \cdot (\textbf{Q}\cdot \textbf{r})                     =Re v_Q\; O(1)                          &&  O(v_Q Re)           \\Re v_Q \textbf{rr}: \textbf{Q} \cdot \grad \textbf{u}_U                     =Re v_Q\; O(1)                          &&  O(v_Q Re)           \\&&                      \\\end{align*}
}
In the Oseen region $r=O(Re^{-1})$, we observe that the viscous term scales as $O(Re^3)$, hence it appears of the same order as the terms $\textbf{w}_r\cdot \grad \textbf{u}_U$, while the other terms may be negligible in the Oseen region provided that: 
\begin{align*}
    v_Q Re\ll Re^3, 
    && v_E Re^2\ll Re^3, 
    && S Re^2 \ll Re^3, 
    && \ldots
\end{align*}
The other conditions indicated by the ellipsis are satisfied provided the above conditions hold. 
In others words the dimensionless group, $v_E,v_Q,S$, must satisfy: 
\begin{align}
    && v_Q \ll Re^2, 
    && v_E \ll Re,
    && S \ll Re, 
    \label{eq:scalings_adim}
\end{align}
so that the terms proportional to $v_Q,S$, and $v_E$, be negligible at $O(1)$ in the Oseen region. 
Note that because $Re\ll 1$ the constraints for $v_E$, $v_Q$, and $S$ are actually a lot more restrictive than the previous scaling required for $r=O(1)$ given in~\ref{eq:first_contrains}. 
Note that the scalings proposed here are only approximate, since the outer velocity field $\textbf{u}_{out}$ is not isotropic~\citet[Appendix B]{lovalenti1993}.
In particular, different scalings may apply along the direction of $\textbf{w}_r$ and in the directions normal to $\textbf{w}_r$.

Here, we had to assume $S\ll Re$ so that unsteady effects are negligible for $r=O(Re^{-1})$ and below.
However, note that the unsteady term can still balance the viscous term for even larger $r$, but care must be taken to estimate \textbf{u} for $r>O(Re^{-1})$ since one must use $\textbf{u}_{out}$ instead of $\textbf{u}_U$ in this region.  
Indeed, in the Oseen region, the dominant term is $Re \textbf{w}_r \cdot \grad \textbf{u}_{out}$ and it balances the unsteady term at $r = O(S^{-1})$.
Thus, above this length scale we must consider unsteadiness into the computation, and this new solution will have an effect on the drag force~\citep{lovalenti1993}.
For shorter time scales $S = O(Re)\ll 1$, the unsteady term related to translation motion cannot be neglected in the Oseen region in view of the above scalings. 
If $S = O(1)$, meaning that the actual timescale of variation of $\textbf{w}_r[t]$ is of the same order as the timescale $a/|\textbf{w}_r|$, one sees that the inner region contribution is still $O(Re)$ greater than the viscous term so that $\textbf{u}_U$ still provides a good approximation close to the particle.
However, for these timescales, the unsteady term and the viscous term are equivalent at a length scale $r = O(Re^{-1/2})$, which is smaller than the Oseen length \citep[appendix B]{lovalenti1993}. 
Finally, for very short time scales when $S = O(Re^{-1})$, the unsteady and viscous terms become comparable within the inner region. 
In that case, the estimate of $\mathbf{u}$ entering $\mathbf{f}$ must be obtained from the regular perturbation of the unsteady Stokes equations in the inner region. 
These issues are discussed in more detail by~\citet{lovalenti1993force,lovalenti1993}.

\section{From singular to regular system of equations}
\label{ap:pipe_flows}

The goal of this appendix is to transform~\ref{eq:dt_uf2_bis,eq:dt_up_bis} which contains singular terms at $r=R-a$ into sets of regular equation valid in the centre of the pipe ($\Omega_{in}$), in the region close to the pipe wall ($\Omega_{out}$), and at the surface between these two regions ($\Gamma$). 
To achieve that, we use the same framework as the one used to derive the ``two-fluid'' and ``one-fluid'' formulations of two-phase flows~\citep{fintzi2025averaged,drew1983mathematical}; except that here one phase is represented by the non-vanishing value of $\phi(\textbf{x})$ while the other represents the fluid empty of the droplet's centre of mass.

We introduce the ``domain-indicator'' functions $\chi_{in}$ and $\chi_{out}$, which equal to $1$ in $\Omega_{in}$ and $\Omega_{out}$, respectively and zero otherwise. 
In~\ref{eq:dt_up_bis,eq:dt_uf2_bis} the terms $\textbf{F},\bm\Sigma^p,\textbf{M}$ and \textbf{K} can be, by definition, factorized by $\chi_{in}$ because their contribution is identically null for areas where there are no droplet centres of mass. 
For example, we introduce the definition
\begin{align}
    \textbf{F}\to \chi_{in} \textbf{F},
\end{align}
where the ``new'' vector $\textbf{F}$ on the right-hand side of this equation is equal to $\textbf{F}$ in the domain $\Omega_{in}$ and is a continuous function of \textbf{x} everywhere in the domain $\Omega = \Omega_{in} \cup \Omega_{out}$. 
Same definitions apply for $\bm\Sigma^p,\textbf{M}$ and $\textbf{K}$. 

Then, the forcing term on the right-hand side of~\ref{eq:dt_uf2_bis} can be written as 
\begin{equation}
    \div (\chi_{in} \textbf{M})
    + \grad\grad (\chi_{in} \textbf{K})
 = 
    \chi_{in} (\div \textbf{M}+ \grad\grad :\textbf{K})
    + \grad \chi_{in} \cdot [
        \textbf{M}
        + 2 \div \textbf{K}
    ]
    + \grad\grad \chi_{in} : \textbf{K}
    \label{eq:forcing_step1}
\end{equation}
Similar expressions holds for $\div(\chi_{in} \bm\Sigma^p)$ in~\ref{eq:dt_up_bis}. 
The derivative of the phase indicator function $\chi_{in}$ is~\citep{fintzi2025averaged}
\begin{equation}
    \grad \chi_{in} = - \textbf{n}\delta_S,
\end{equation}
where \textbf{n} is the unit vector pointing outward the domain $\Omega_{in}$, in the current configuration $\textbf{n} = \textbf{i}_r$ (see~\ref{fig:pipe}), and $\delta_S$ is the surface indicator function defined by its action on a test function, see in~\citet{appel2007mathematics,fintzi2025averaged}. 
The gradient of this expression reads, 
\begin{equation}
    \grad \grad \chi_{in} = 
    - \grad (\textbf{n}\delta_S)
 = 
    - \grad \textbf{n} \delta_S 
    - \textbf{n}  \textbf{n}\delta_S^{(1)}, 
\end{equation}
where we used the relation $\grad \delta_S = \delta_S^{(1)}\textbf{n}$\footnote{
 The gradient of $\delta_S$ may be reformulated as follows,
\begin{equation}
    \grad \delta_S
 =\gradI \delta_S
    + \textbf{n} (\textbf{n}\cdot \grad )\delta_S
= \textbf{n} \delta_S^{(1)}
    \label{eq:grad_delta_S}
\end{equation} 
where we introduced the notations: $\gradI = (\bm\delta - \textbf{nn}) \cdot \grad$, and $\textbf{n}\cdot \grad\delta_S = \delta_S^{(1)}$. 
Second equality of~\ref{eq:grad_delta_S} is derived using the relation~\citep{fintzi2025averaged},  
\begin{equation*}
    \gradI\delta_S  
 = \delta_S \textbf{n}\cdot \grad \textbf{N} 
 = \delta_S \partial_r \textbf{i}_r = 0.
\end{equation*}
}.  
Finally, using these two relations~\ref{eq:forcing_step1} can be written as
\begin{multline}
    \div (\chi_{in} \textbf{M})
    + \grad\grad : (\chi_{in} \textbf{K})\\
 = 
    \chi_{in} (\div \textbf{M} + \grad\grad : \textbf{K})
    - \delta_S [\textbf{n}\cdot \textbf{M}
    + (\grad \textbf{n} + 2 \textbf{n}\grad ): \textbf{K} ]
    - \textbf{n}  \textbf{n}: \textbf{K} \delta_S^{(1)} 
    \label{eq:source_term}
\end{multline}
It is now evident that~\ref{eq:dt_uf2_bis,eq:dt_up_bis} contain singular terms of first and second order localized at $r=R-a$ due to the presence of $\delta_S$, and $\delta_S^{(1)}$ in this equation. 

\subsection{Method of solutions}

To balance the singular terms in~\ref{eq:dt_uf2_bis}, the Newtonian contribution $\div \bm\Sigma$ must equally contain a singular part of opposite sign.
Hence, we must seek the unknowns: $\textbf{U}$ and $P_f$, as distributions, not as functions, therefore we express the unknown velocity and pressure fields as, 
\begin{align}
    \label{eq:def_P}
 P  &= \chi_{out} P_{out} + \chi_{in} P_{in} + \delta_S P_S, \\
    \textbf{U}  &= \chi_{out} \textbf{U}_{out} + \chi_{in} \textbf{U}_{in}, 
    \label{eq:def_U}
\end{align}
where $P_{out}, P_{in},P_S, \textbf{U}_{out}$, and $\textbf{U}_{in}$, are defined as continuous functions in $\Omega$ which equal $P$, and $\textbf{U}$ in their respective domain denoted by the subscript $_{in}$, $_{out}$, and $_S$, for ``in'', ``out'', and ``at the surface location''.  
Using these two expressions and by direct differentiation, one deduces that the Newtonian stress can be written: 
\begin{align}
    \bm\Sigma
    &= \chi_{out} \bm\Sigma_{out}  + \chi_{in} \bm\Sigma_{in} + \delta_S \bm\Sigma_S \\
    \bm\Sigma_{out} &= - P_{out} + \mu_f (\grad \textbf{U}_{out}+ \grad \textbf{U}_{out}^\dagger)\\
    \bm\Sigma_{in} &= - P_{in} + \mu_f (\grad \textbf{U}_{out}+ \grad \textbf{U}_{out}^\dagger)\\
    \bm\Sigma_S &=  - P_S 
    + \mu_f (\textbf{U}_{out} - \textbf{U}_{in}) \textbf{n} 
    + \mu_f \textbf{n} (\textbf{U}_{out} - \textbf{U}_{in})    
    \label{eq:def_sigma_s}
\end{align}
Taking the divergence of each component of this stress reads, 
\begin{equation}
    \div \bm\Sigma
 =
    \chi_{out} \div\bm\Sigma_{out}
    + \chi_{in} \div\bm\Sigma_{in}
    + \delta_S [\textbf{n}\cdot (\bm\Sigma_{out} - \bm\Sigma_{in})+ \div \bm\Sigma_S]
    + (\textbf{n}\cdot \bm\Sigma_S ) \delta_S^{(1)}
    \label{eq:div_sigma_sing}
\end{equation}
where we used once again the property $\grad \delta_S = \delta_S^{(1)} \textbf{n}$. 
Under these assumptions, the stress possesses a singular part proportional to the normal stress jump at $r=R-a$, and to the divergence of the yet unknown interfacial stress $\bm\Sigma_S$. 
Additionally, the normal stress $(\textbf{n}\cdot \bm\Sigma_S )$ is a singular contribution of second order, i.e., proportional to $\delta_S^{(1)}$, that remains to be determined. 

Using~\ref{eq:def_U} into~\ref{eq:div_u} the continuity equation reads, 
\begin{equation}
    \chi_{out} \div \textbf{U}_{out}  
    + \chi_{in} \div \textbf{U}_{in}  
    - \delta_S (\textbf{U}_{out} - \textbf{U}_{in})\cdot \textbf{n} = 0
    \label{eq:div_U_distribution}
\end{equation}
Likewise injecting~\ref{eq:div_sigma_sing,eq:source_term} into~\ref{eq:dt_uf2_bis}, gives: 
\begin{multline}
    \chi_{out} \div\bm\Sigma_{out}
    + \chi_{in} (\div\bm\Sigma_{in}+ \rho_f \textbf{g} [1 + \phi (\zeta - 1)] +\div\textbf{M}+\grad\grad : \textbf{K})\\
    + \delta_S [
        \textbf{n}\cdot ( \bm\Sigma_{out} - \bm\Sigma_{in} - \textbf{M})
        - (\grad \textbf{n} + 2\textbf{n}\grad ): \textbf{K}
    + \div \bm\Sigma_S]\\
    + (\textbf{n}\cdot \bm\Sigma_S - \textbf{nn}: \textbf{K} ) \delta_S^{(1)} = 0
    \label{eq:div_dt_u_deneralized}
\end{multline}
In~\ref{eq:div_U_distribution,eq:div_dt_u_deneralized}, terms proportional to $\chi_{in}$,$\chi_{out}$, $\delta_S$, and $\delta^{(1)}_S$, are independent hence one have to satisfy, 
\begin{align}
    \label{eq:div_u_o}
    \div \textbf{U}_{out} = 0, \\
    \label{eq:div_u_i}
    \div\bm\Sigma_{out} + \rho_f \textbf{g} = 0, \\
    \label{eq:div_S_o}
    \div \textbf{U}_{in} = 0, \\
    \label{eq:div_S_i}
    \div\bm\Sigma_{in}+ \rho_f \textbf{g} [1 + \phi (\zeta - 1)] + \div\textbf{M}+\grad\grad : \textbf{K} = 0, 
\end{align}
where it is implied that the first two equations must be satisfied in $\Omega_{out}$, while the second and third equations have to be satisfied in the domain $\Omega_{in}$. 
At the location $r=R-a$, one has the jump conditions  
\begin{align}
    \label{eq:vel_jump_2}
    (\textbf{U}_{out} - \textbf{U}_{in})\cdot \textbf{n}= 0,\\
    \label{eq:stress_jump_1}
    \textbf{n}\cdot ( \bm\Sigma_{out} - \bm\Sigma_{in} - \textbf{M})
    - (\grad \textbf{n} + 2\textbf{n}\grad ): \textbf{K}
    + \div \bm\Sigma_S = 0,\\
    (\textbf{n}\cdot \bm\Sigma_S - \textbf{nn}: \textbf{K} ) \delta_S^{(1)} = 0,
    \label{eq:stress_jump_2}
\end{align}
which can also be interpreted as interface equations which have to be solved for the unknown surface pressure: $P_S$ introduced in~\ref{eq:def_P}.
These equations are completed by the no-slip boundary condition, namely $\textbf{U}=0$ at the wall ($r=R$). 
 
From~\ref{eq:stress_jump_2,eq:def_sigma_s,eq:vel_jump_2}, one directly deduce that 
\begin{align}
    \label{eq:sol_VS}
\mu_f (\textbf{U}_{out} - \textbf{U}_{in}) 
 = \textbf{nn}: \textbf{K}_{||}   
    \\
 P_S 
 = - \textbf{nn}: \textbf{K}\cdot \textbf{n}
    \label{eq:sol_PS}
\end{align}
where $\textbf{K}_{||} = \textbf{K}\cdot (\bm\delta - \textbf{nn})$. 
At this stage, it is very interesting to note that~\ref{eq:sol_VS} actually represents a tangential velocity jump for $\textbf{U}$ due to the presence of the second moment of force \textbf{K}. 
From~\ref{eq:sol_PS,eq:sol_VS} one has
\begin{equation}
    \bm\Sigma_S 
 = (\textbf{nn}: \textbf{K}\cdot \textbf{n}) \bm\delta
    + (\textbf{nn}: \textbf{K}_{||}) \textbf{n}
    + \textbf{n} (\textbf{nn}: \textbf{K}_{||}). 
    \label{eq:sol_Ss}
\end{equation}
This expression can then be used in \ref{eq:stress_jump_1}, which gives an explicit expression for the stress jump at $r=R-a$. 

To obtain the droplets phase equation~\eqref{eq:dt_up_bis} we have multiplied~\ref{eq:dt_uf2_bis} by $\phi$ and substituted that results to~\ref{eq:dt_up}, hence~\ref{eq:dt_up_bis} is only valid in $\Omega_{in}$. 
However,~\ref{eq:dt_up} as derived in~\citet{fintzi2025averaged} is valid for all \textbf{x} in $\Omega$. 
The term $\div (\chi_{in} \bm\Sigma^p)$ present in~\ref{eq:dt_up} expands to 
\begin{equation}
    \div (\chi_{in} \bm\Sigma^p) = \chi_{in} \div \bm\Sigma^p + \delta_S \textbf{n} \cdot \bm\Sigma^p. 
\end{equation}
Because that is the only singular contribution to the droplet phase equation, we deduce the boundary condition: 
\begin{equation}
    \bm\Sigma^p\cdot \textbf{n}= 0,
    \label{eq:stress_S_n}
\end{equation}
at $r=R-a$. 
Note that we have not considered direct contact force between wall and droplets, in which case we would have to add a contribution of the form $\delta_S \textbf{F}_{wall\to droplet}$, because this force would be non-zero only on the surface indicated by $r=R-a$ if we consider undeformable spherical droplets. 
In that respect, this boundary condition is consistent with the recent study of \citet{dupuy2024investigation} if one includes kinetic and contact stresses to $\bm\Sigma^p$. 
In the domain $\Omega_{in}$, the droplets phase equations simply read 
\begin{equation}
    \rho_f \textbf{g} \phi  (\zeta  - 1)(1 - \phi)  
    + \div \bm\Sigma^p
    - \phi (\div \textbf{M}+ \grad\grad :\textbf{K})
    + \textbf{F}
 =0 . 
    \label{eq:dt_up_ter}
\end{equation}

The objective of the next section is now to re-write~\ref{eq:div_u_i,eq:div_u_o,eq:div_S_i,eq:div_S_o,eq:dt_up_ter} together with the boundary condition~\ref{eq:stress_jump_1,eq:sol_VS,eq:sol_Ss,eq:stress_S_n}, in the specific case of established pipe flow.

\subsection{Cylindrical coordinates}

Now we write the system of equations as scalar equations in the cylindrical coordinate system. 
In view of the steady-state form given by~\ref{eq:def_vel_para,eq:force_symmetry,eq:particles_stress_symmetry,eq:M_symmetry,eq:K_symmetry} the divergence of the Newtonian stress tensor $\bm\Sigma_{in/out}$, of the first moment \textbf{M}, and of the second moment \textbf{K} read:  
\begin{align}
    \div\bm\Sigma_{in/out} &= 
    - \textbf{i}_r \partial_r P_{in/out} 
    + \textbf{i}_z \{ - \partial_z P_{in/out} + \frac{\mu_f}{r}\partial_r [r \partial_r U_{in/out} ] \}, \\
    \div \textbf{M}
    &=  
    \textbf{i}_z \frac{1}{r}\partial_r (rM_{rz})
    + \textbf{i}_r \partial_r  M_\bot , \\
    \div \bm\Sigma^p
    &=  
    \textbf{i}_r \partial_r  \Sigma^p_\bot , \\
    \div \textbf{K}
    &=  
    \partial_r K_\bot (\textbf{i}_r\textbf{i}_z +\textbf{i}_z\textbf{i}_r ), \\ 
    \grad\grad : \textbf{K}
    &=  
    \frac{1}{r}\partial_r (r\partial_r K_\bot)\textbf{i}_z, 
\end{align}
respectively. 
We directly conclude that the momentum equation~\ref{eq:dt_uf2_bis} written in the domain $\Omega_{out}$ and $\Omega_{in}$ reads 
\begin{align}
    \textbf{i}_z\{- \partial_z P_{in} + \rho_f g [1 + \phi (\zeta - 1)] 
    + \frac{1}{r}[
    \mu_f\partial_r ( r \partial_r U_{in}  )
    + \partial_r (rM_{rz})
    + \partial_r (r \partial_r K_\bot)
    ] \} \nonumber \\
    + \textbf{i}_r \{
    \partial_r M_\bot
    - \partial_r P_{in} 
    \}
    \label{eq:ap:dt_u_in}
 = 0, \\
    \label{eq:ap:dt_u_{out}ut}
\textbf{i}_r \{
    - \partial_r P_{out}
    \}
    + \textbf{i}_z\{- \partial_z P_{out} +\rho_f g 
    + \frac{1}{r} \mu_f\partial_r (r \partial_r U_{out}  )
    \}
 = 0,
\end{align}
respectively. 
Likewise, the droplets phase equation~\ref{eq:dt_up_ter} can be written
\begin{multline}
    \textbf{i}_z \{\rho_f g \phi  (\zeta  - 1)(1 - \phi)  
    - \frac{\phi}{r} [\partial_r (rM_{rz}) + \partial_r (r \partial_r K_\bot)]
    + F\} \\
    + \textbf{i}_r \{\partial_r  \Sigma^p_\bot 
    -  \phi   \partial_r  M_\bot 
    + L
    \}
 =0,
    \label{eq:ap:dt_up_{in}n}
\end{multline}
in the domain $\Omega_{in}$. 

To express the boundary conditions in cylindrical polar coordinates, we start by noting that: 
\begin{align}
    \bm\Sigma_{in/out}\cdot \textbf{n}
    &= - P_{in/out} \textbf{i}_r
    + \mu_f \partial_r U_{o/i}  \textbf{i}_z,\\
    \textbf{M}\cdot \textbf{n}
    &= M_{rz} \textbf{i}_z 
    + M_\bot \textbf{i}_r. 
\end{align}
Likewise, from the definition of \textbf{K}~\eqref{eq:K_symmetry} we directly get, 
\begin{align}
 n_{in} n_jn_k (\textbf{K})_{ijk} 
    = 
    0
    &&
 n_jn_k (\textbf{K}_{||})_{ijk}
    = 
 K_\bot  \textbf{i}_z  
    &&
    \bm\Sigma_S 
    = 
 K_\bot
    (\textbf{i}_z \textbf{i}_r
    + \textbf{i}_r \textbf{i}_z)
\end{align}
where we used the property $\textbf{i}_r \cdot \textbf{i}_z = 0 $. 
From which we deduce the relations: 
\begin{align*}
    \div\bm\Sigma_S 
    = 
    \textbf{i}_z \frac{1}{r}[
        \partial_r (r    K_\bot)  
    ],&&
    (\grad \textbf{n})_{jk} (\textbf{K})_{ijk} = 
    \frac{1}{r} K_\bot  \textbf{i}_z,
\end{align*}
and
\begin{equation}
    \textbf{n}\cdot (\div \textbf{K})
 = 
    \textbf{i}_r\cdot [
      \textbf{i}_z \grad K_\bot           
    + \textbf{i}_z \grad K_\bot ^\dagger  
    +   \bm\delta \partial_z (K_{||})]
 = 
    \textbf{i}_r \cdot \grad K_\bot \textbf{i}_z           
 = 
    \partial_r K_\bot \textbf{i}_z.         
\end{equation}
We have used the relation $\grad \textbf{n} = \grad \textbf{i}_r = \textbf{i}_\varphi\textbf{i}_\varphi/ r$ in this last equation.   
Then according to~\ref{eq:stress_jump_2,eq:stress_S_n} and~\ref{eq:sol_VS} and the last few remarks we write 
\begin{align}
    \label{eq:ap:BC_u}
    \mu_f (\textbf{U}_{out} - \textbf{U}_{in}) 
    -K_\bot \textbf{i}_z &=0,  
    \\
    \label{eq:ap:BC_stress}
[\mu_f\partial_r U_{out} - \mu_f\partial_r U_{in} - M_{rz} - \partial_r K_\bot  ]\textbf{i}_z 
    -(P_{out} - P_{in} + M_\bot) \textbf{i}_r &= 0, \\
 \Sigma^p_\bot \textbf{i}_r &= 0.
    \label{eq:ap:BC_stress_up}
\end{align}
From~\ref{eq:ap:BC_u,eq:ap:BC_stress,eq:ap:BC_stress_up,eq:ap:dt_up_{in}n,eq:ap:dt_u_in,eq:ap:dt_u_{out}ut} we directly deduce~\ref{eq:vertical_mom_u,eq:vertical_mom_u_out,eq:vertical_mom_ur,eq:horizontal_mom_Pf,eq:horizontal_mom_u_out,eq:BC_slip,eq:BC_stress,eq:BC_pressure,eq:BC_pressure_parts,eq:horizontal_mom_phi} of the main text.

\end{document}